\documentclass[aps,prc,superscriptaddress,longbibliography, showpacs,floatfix,letterpaper,preprintnumbers,nofootinbib,notitlepage,reprint]{revtex4-2}
\usepackage{subcaption, caption}
\usepackage[colorlinks=true, linkcolor=blue, citecolor=blue, urlcolor=blue]{hyperref}
\usepackage{amssymb}
\usepackage{color}
\usepackage{graphicx} 
\usepackage{bm}
\usepackage{amsmath,mathrsfs}
\usepackage{braket}
\usepackage{xfrac}
\usepackage{soul}
\usepackage{lineno}
\usepackage{slashed}
\usepackage{appendix}
\usepackage{tabularx,enumitem,paralist}
\usepackage{cancel} 
\usepackage{xcolor} 

\newcommand{\polylog}[1]{\operatorname{Li}_{#1}}
\DeclareMathOperator{\arcsinh}{arcsinh}

\usepackage{float} 
\usepackage{orcidlink}

\begin{document}
\title{Closed-form expressions for tree-level gluon-gluon scattering: a framework for obtaining theoretical systematic uncertainties for jet-medium Monte Carlo simulations}

\author{Lukas Opitz\,\orcidlink{0009-0008-0131-4307}}
\affiliation{Department of Physics, University of Regina, Regina, Saskatchewan S4S 0A2, Canada}

\author{Hemanth Regi\,\orcidlink{0009-0001-8564-159X}}
\affiliation{Department of Physics, University of Regina, Regina, Saskatchewan S4S 0A2, Canada}

\author{Gojko Vujanovic\,\orcidlink{0000-0001-5397-6662}}
\email[Corresponding author: ]{gojko.vujanovic@uregina.ca}
\affiliation{Department of Physics, University of Regina, Regina, Saskatchewan S4S 0A2, Canada}

\date{\today}

\begin{abstract}
Modern Bayesian theory-to-data comparisons for jet-medium interactions, such as [\href{https://journals.aps.org/prc/10.1103/PhysRevC.111.054913}{Phys. Rev. C 111,054913 (2025)}], are lacking the careful accounting of theoretical systematic uncertainties included within their uncertainty budget. Tree-level gluon-gluon scattering is revisited to establish a framework capable of quantifying theoretical systematic uncertainties to be used in Bayesian jet-medium constraints. The behavior of tree-level scattering in thermal QCD is examined in detail, finding deviations away from the commonly used approximations. Deviations in the scattering rate affect jet-medium transport coefficients at the partonic level, leading to a more intricate kinematic dependence for $\hat q$ and $\hat e$ than, say, the well-known logarithmic behavior. These deviations away from well-known behavior are used to estimate theoretical systematic uncertainties in Bayesian analysis.
\end{abstract}

\maketitle
\section{Introduction}

Since the discovery that the QCD coupling decays with increasing energy scale \cite{Gross:1973id,Politzer:1973fx}, the goal of colliding nuclei at relativistic energies has been to study nuclear interactions at the most fundamental, quark and gluon, level while also producing a quark-gluon plasma (QGP) \cite{Bjorken:1982qr, Rafelski:1982pu}. For a more detailed historical review of how the physics of the QGP has been explored, see Ref.~\cite{Busza:2025gpg}. 

Many observables have been used to extract the transport coefficients of the QGP bulk, from soft hadron production \cite{Bernhard:2019bmu,JETSCAPE:2020shq,Nijs:2020ors,Heffernan:2023gye} to electromagnetic probes~\cite{Paquet:2015lta,Vujanovic:2019yih,Gale:2021emg}. The physics of jets, and in particular that of jet quenching in the QGP, has long served as a way to test our understanding of the strong force \cite{Baier:2002tc,JETSCAPE:2022jer,JETSCAPE:2023ikg,JETSCAPE:2024cqe} at finite temperature. Jet-medium interactions probed by high-energy heavy-ion collisions give a regime where perturbation theory, at finite temperature, can be tested in detail. As jets are multi-scale objects, their shower also reaches a region of phase space where perturbative approaches fail, and lattice QCD calculations become a way of investigating QGP properties via jet-medium interactions. Lattice QCD calculations \cite{Kumar:2020wvb} show how the temperature dependence of $\hat q$, the transport coefficient governing transverse diffusion processes of jets in the nuclear medium \cite{Baier:2002tc} such as the QGP \cite{JETSCAPE:2021ehl,JETSCAPE:2022jer,JETSCAPE:2023ikg,JETSCAPE:2024cqe}, changes when going from perturbative to non-perturbative regimes. Moreover, in the non-perturbative regime, $\hat q$ possesses a very weak energy dependence \cite{Kumar:2020wvb} in contrast to perturbative calculations. In addition to temperature and energy dependence of $\hat q$, the latter is also sensitive to the virtuality dependence of partons in the jet, as explored in Ref.~\cite{Kumar:2025rsa}. Our work focuses on jet-medium tomography, presenting an in-depth discussion of how tree-level parton-parton scattering, and the associated jet-medium transport coefficients, are computed at leading order, and complements the results shown in Ref.~\cite{short_paper}.

With the advent of Bayesian model-to-data comparisons, our quantitative understanding of the properties of the QGP have been heightened, with detailed exploration of Bayesian constraints on jet-medium transport coefficients \cite{JETSCAPE:2021ehl,JETSCAPE:2024cqe,Kumar:2025asj,Kumar:2025egh,Sirimanna:2022zje,Modarresi-Yazdi:2024vfh,Mehtar-Tani:2022zwf}. Indeed, the simulation of the nuclear medium itself has been separated into various stages, namely the pre-hydrodynamic, the hydrodynamic, and hadronic transport approaches being commonly used \cite{Gale:2021emg}, while jets are treated as multi-scale objects \cite{JETSCAPE:2021ehl,JETSCAPE:2024cqe}. A topic of great importance that resulted from these detailed Bayesian model-to-data comparisons pertains to the careful accounting of theoretical systematic uncertainties present within model calculations. 

Currently, Bayesian constraints on transport coefficients of the QGP bulk are arguably better understood than constraints of jet-medium transport coefficients, given the efforts put towards accounting for theoretical systematic uncertainty \cite{JETSCAPE:2020shq,Paquet:2023rfd,Jaiswal:2025deb,Jaiswal:2025hyp} in bulk simulations of the nuclear medium. The goal of this work is to provide a better accounting of theoretical systematic uncertainty for tree-level jet-medium scatterings and the associated transport coefficients. Transport coefficients are encoding processes of jet drag and diffusion, encoded in $\hat e$ and $\hat q$, respectively, which in turn allow to characterize the medium itself. The transverse diffusion direction in this case is orthogonal to the jet parton propagation direction. For a scattering where $(a,1)$ labels jet partons while $(b,2)$ labels the medium partons, the transverse momentum diffusion coefficient $\hat q$ is given by
\begin{align}
\hat{q}&= \frac{1}{2E_a}\left[\prod_{i=b,1,2}\int\frac{d^3 p_i}{\left(2\pi\right)^3 2E_i}\right] \int \frac{d^4 q}{(2\pi)^3} f_b\left(p_b\right) \left[1\pm f_2\left(p_2\right)\right]\nonumber\\
&\times\left(2\pi\right)^4\delta^{(4)}\left(p_a-p_1-q\right)\delta^{(4)}\left(p_b+q-p_2\right) q^2_\perp \overline{\left|\mathcal{M}_{a,b\to1,2}\right|^2}\nonumber\\
&=\frac{\langle q^2_\perp \rangle_L}{L}.
\label{eq:qhat}
\end{align}
In Eq.~(\ref{eq:qhat}), $f_b(p_b)$ and $f_2(p_2)$ correspond to the distribution of particles in the QGP, before and after scattering, respectively, which, in thermal equilibrium, are Bose-Einstein or Fermi-Dirac distributions. Finally $\overline{\left|\mathcal{M}_{a,b\to1,2}\right|^2}$ is the scattering matrix element. 

The high-energy limit, the dimensionless quantity $\hat{q}/T^3$ for $gg\to gg$ scattering at leading logarithm accuracy is given by \cite{JETSCAPE:2021ehl,JETSCAPE:2022jer,JETSCAPE:2023ikg,JETSCAPE:2024cqe}
\begin{align}
\frac{\hat{q}}{T^3}&=\frac{9\alpha^{2}_s\zeta(3)}{\pi^4}\ln\left[\frac{2E_a T}{m^2_D}\right]+O\left(e^{-E_a/T}\right),
\label{eq:qhat_leading-log}
\end{align}
where $E_a$ is the incoming jet parton energy, $m^2_D=6\pi\alpha_s T^2$ is the square of the Debye screening mass for a three quark-flavor QGP at temperature $T$, and $\zeta(3)$ is Ap\'ery's constant \cite{Cohl:2014drm}. The goal in this work is to complement \cite{short_paper} by providing a detailed analytical exploration of the scattering rate and transport coefficients beyond the leading logarithmic accuracy, to more precisely examine the theoretical systematic uncertainty, of particular usefulness to future Bayesian endeavors. To avoid infrared singularities, the cut-off scale of our calculation will be set to the Debye mass. Intricate infrared physics giving rise to the Debye screening mass that modify tree-level $2\to 2$ scatterings are left for future work.   

This paper is organized as follows: Section~\ref{sec:rate} discusses how the scattering rate is computed and a subset of important prior approximations, Section~\ref{sec:qhat_ehat} computes $\hat q$ and $\hat e$, contrasting our results to prior approximations, while Section~\ref{sec:conclusion} provides concluding remarks.
\section{Scattering rate}\label{sec:rate}
The differential rate of a relativistic $2\to2$ scattering process $a+b\to1+2$ is given by
\begin{align}
\frac{d^3R}{d^3p_a}&=\frac{1}{\left(2\pi\right)^32E_a}\int\frac{d^3p_b\, f_\pm(p_b)}{\left(2\pi\right)^32E_b}\frac{d^3p_1}{\left(2\pi\right)^32E_1}\frac{d^3p_2}{\left(2\pi\right)^32E_2}\nonumber\\
&\times\left[1\mp f_\pm\left(p_2\right)\right]\overline{|\mathcal{M}|^2}\left(2\pi\right)^4\delta^{(4)}\left(p_a+p_b-p_1-p_2\right)\nonumber\\
&=\frac{1}{16\left(2\pi\right)^8E_a}\int\frac{d^3p_b\;d^3p_1\;d^3p_2}{E_bE_1E_2}
\label{eq:scatt_rate}\\
&\times f_\pm(p_b)\left[1\mp f_\pm\left(p_2\right)\right]\overline{|\mathcal{M}|^2}\delta^{(4)}\left(p_a+p_b-p_1-p_2\right)\nonumber
\end{align}
where $\delta^{(4)}$ is the usual momentum conserving $\delta$-function, and the distribution functions of the thermalized medium particles are given by either the Bose-Einstein $(f_-)$ or the Fermi-Dirac $(f_+)$ distributions. For a medium with temperature $T$, and flow velocity $u^\mu$
\begin{alignat}{2}
	f_-(p)&=\frac{1}{e^{(u\cdot p-\mu)/T}-1},
	\, f_+(p)&=\frac{1}{e^{(u\cdot p-\mu)/T}+1},
\end{alignat}
with $\left(1+f_-\right)$ Bose-enhancement and $\left(1-f_+\right)$ Fermi-suppression factors for the final-state medium particles. 
\subsection{Performing the tree-level rate integrals}
Five out of the nine integral in Eq.~(\ref{eq:scatt_rate}) can be performed exactly without specifying the matrix element. This is done by introducing Mandelstam variables through Dirac $\delta$-functions, with details of such an approach presented in Ref.~\cite{Kapusta:2006pm}. Following integration,
the scattering rate becomes
\begin{align}
\frac{d^3 R}{d^3 p_a}&=\frac{1}{16\left(2\pi\right)^7|\vec{p}_a|E_a}\int_0^\infty dE_b\int_0^{E_a+E_b} dE_2\nonumber\\
&\times \int_{t_{\min}}^{t_{\max}}dt\int_{s_-}^{s_+}ds\frac{f_\pm(p_b)\left[1\mp f_\pm\left(p_2\right)\right]\overline{|\mathcal{M}|^2}(s,t)}{\sqrt{\Lambda(s,t)}}\nonumber\\
\Lambda(s,t)&=\left(t^2-c_ut\right)s-a_st^2-\left(a_t-t\right)s^2,
\end{align}
where
\begin{align}
\begin{split}
a_s&=\left(E_a+E_b\right)^2,\, a_t=\left(E_b-E_2\right)^2,\\
c_u&=2[E_aE_2+E_b\left(E_a+E_b-E_2\right)],
\end{split}
\label{eq:abc_stu}
\end{align}
which is a useful representation if integrating first over Mandelstam $s$ is prioritized. Equivalently, one inverts the order of integration using
\begin{align}
\frac{d^3R}{d^3p_a}&=\frac{1}{16\left(2\pi\right)^7|\vec{p}_a|E_a}\int_0^\infty dE_b\int_0^{E_a+E_b}dE_2\\
&\times
\int_{s_{\min}}^{s_{\max}}ds\int_{t_-}^{t_+}dt\frac{f_\pm(p_b)\left[1\mp f_\pm(p_2)\right]\overline{|\mathcal{M}|^2}(s,t)}{\sqrt{\Lambda(s,t)}},\nonumber\\
\Lambda(s,t)&=\left(s^2-c_us\right)t-a_ts^2-\left(a_s-s\right)t^2.\nonumber
\end{align}
As far as the integration boundaries are concerned,
\begin{align}
\begin{split}
s_\pm&=\frac{t^2-c_ut\pm\sqrt{\left(t^2-c_ut\right)^2-4(a_t-t)\left(a_st^2-t\right)}}{2(a_t-t)},\\
t_\pm&=\frac{s^2-c_us\pm\sqrt{\left(s^2-c_us\right)^2-4(a_s-s)\left(a_ts^2-s\right)}}{2(a_s-s)},
\end{split}
\label{eq:bdy_Lambda}
\end{align}
while
\begin{align}
\begin{split}
t_{\max}&=0, \,t_{\min}=-4\min(E_aE_1,E_bE_2)\equiv\lfloor t \rfloor,\\
s_{\min}&=0, s_{\max}=4\min(E_aE_b,E_1E_2)\equiv\lfloor s\rfloor.
\end{split}
\label{eq:bdy_st}
\end{align}
The boundaries in Eq.~(\ref{eq:bdy_Lambda}) stem from solid angle integral limits, which are equivalent to finding the roots of the $\Lambda(s,t)$. Boundaries in Eq.~(\ref{eq:bdy_st}) are obtained by kinematic analysis, specifically those affecting $E_b$ and $E_2$. 

\begin{figure}
\centering
\begin{subfigure}{0.28\linewidth}
	\includegraphics[width=\textwidth]{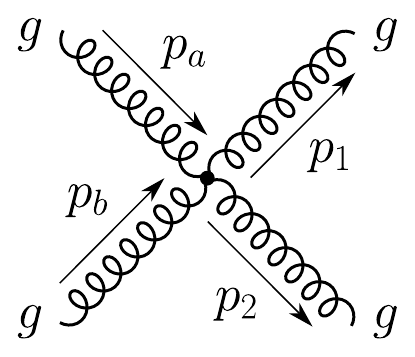}
    \caption{four-point (vertex) channel}
\end{subfigure}
\begin{subfigure}{0.45\linewidth}
	\includegraphics[width=\linewidth]{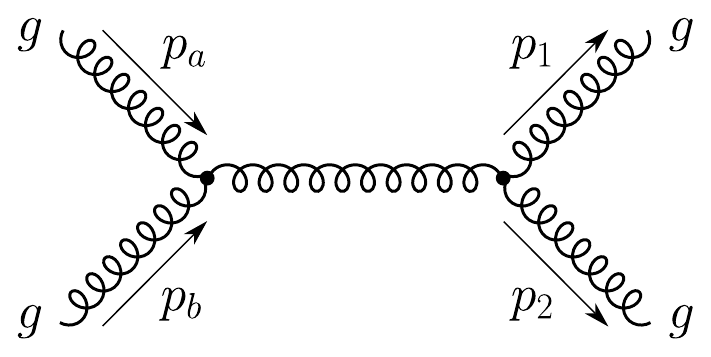}
    \caption{$s$-channel}
\end{subfigure}
\begin{subfigure}{0.45\linewidth}
	\includegraphics[width=\linewidth]{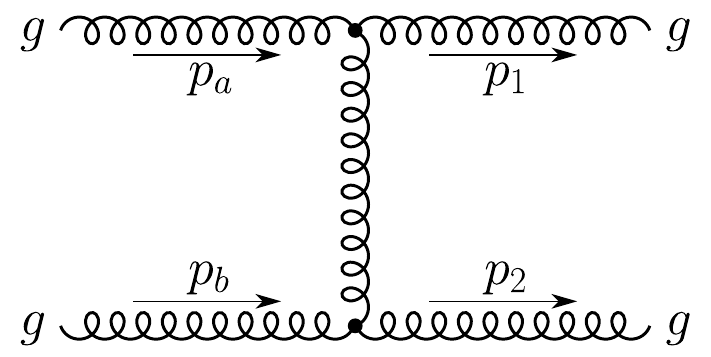}
    \caption{$t$-channel}
\end{subfigure}
\begin{subfigure}{0.45\linewidth}
	\includegraphics[width=\linewidth]{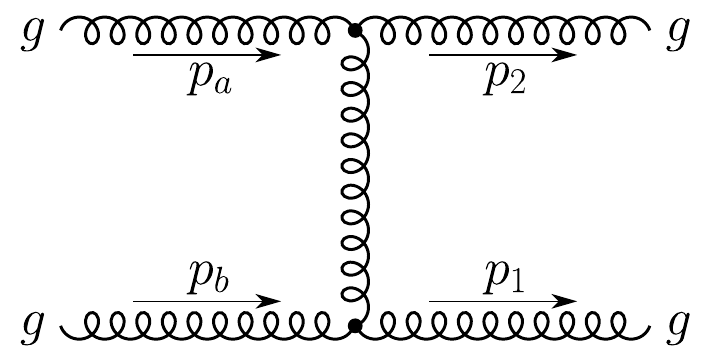}
    \caption{$u$-channel}
\end{subfigure}
    \caption{$gg\to gg$ tree-level diagrams}
\label{fig:Feynman}
\end{figure}

To illustrate our approach, gluon-gluon scattering is considered, with four channels contributing at tree-level depicted in Fig.~\ref{fig:Feynman}.
The matrix element is given by
\begin{align}
\overline{|\mathcal{M}|^2}=288\alpha_s^2\left(2\pi\right)^2\left[3-\frac{tu}{s^2}-\frac{su}{t^2}-\frac{st}{u^2}\right].
\label{eq:matrix}
\end{align}
This matrix element exhibits a divergence as $t\to 0$, which is partially screened by including Hard Thermal Loops (HTL) \cite{Pisarski:1989cs, Braaten:1989kk, Braaten:1989kk, Braaten:1989mz, Braaten:1990az}. While including HTL resummations will be considered in an upcoming study, the focus herein is to revisit the high-energy behavior of $gg\to gg$ processes analytically, while excising the divergence as $t\to 0$ and $u\to 0$ by cutting off the infrared singularity using the Debye mass: $m^2_D=6\pi\alpha_s T^2$. Thus, the maximal value of $t,u$ is reduced to $t_{\max}=-m_D^2=u_{\max}$ for divergent denominators. Performing the integrals over $s$ and $t$ in the local fluid rest frame $u^\mu=(1,\vec 0)$ gives
\begin{align}
\frac{d^3R}{d^3p_a}&=\frac{9\alpha_s^2}{4\pi^4E_a^2}\int_0^\infty \frac{dE_b}{e^{\beta E_b}-1}\int_0^{E_a+E_b} \frac{dE_2}{1-e^{-\beta E_2}}\nonumber\\
&\qquad\qquad\qquad\times\left[\mathbf{I}+\mathbf{II}+\mathbf{III}+\mathbf{IV}\right],
\label{eq:rate_full}
\end{align}
where
\begin{align}
\mathbf I&=3\min[E_a,E_b,E_a+E_b-E_2,E_2],\nonumber\\
\mathbf{II}&=-\frac{E_aE_b\left(E_a+E_b-E_2\right)E_2}{\left(E_a+E_b\right)^3},\nonumber\\
\mathbf{III}&=\frac{E_aE_b\left(E_a+E_b-E_2\right)E_2}{\left[\left(E_b-E_2\right)^2+m_D^2\right]^\frac{3}{2}},\nonumber\\
\mathbf{IV}&=\frac{E_aE_b\left(E_a+E_b-E_2\right)E_2}{\left[\left(E_a-E_2\right)^2+m_D^2\right]^\frac{3}{2}},
\label{eq:rate_I_IV}
\end{align}
where $\beta=1/T$, with $T$ being the temperature. The terms ${\bf I}$ through ${\bf IV}$ match the four terms in Eq.~(\ref{eq:matrix}) in the order written one-to-one, respectively, and are associated to the diagrams depicted in Fig.~\ref{fig:Feynman} in the same order. More details leading to Eqs.~(\ref{eq:rate_full}, \ref{eq:rate_I_IV}) are in Appendix~\ref{sec:Mandelstam_int}. The term in $\mathbf{I}$ can be evaluated analytically giving 
\begin{align}
\begin{split}
\frac{d^3R_c}{d^3p_a}&=\frac{9\alpha_s^2}{4\pi^4E_a^2}\int_0^\infty \frac{dE_b}{e^{\beta E_b}-1}\int_0^{E_a+E_b} \frac{dE_2}{1-e^{-\beta E_2}}\\
&\times3\min[E_a,E_b,E_a+E_b-E_2,E_2]\\
&=\frac{27\alpha_s^2T}{4\pi^4}\bigg\{\frac{\pi^2}{6x_a}+\frac{\zeta(3)}{x^2_a}+\frac{1}{2x_a}\ln^2\left(1-e^{-x_a}\right)\\
&\qquad\qquad\quad+\frac{1}{x^2_a}\Big[\polylog{3}\left(1-e^{-x_a}\right)-\polylog{3}\left(e^{-x_a}\right)\\
&\qquad\qquad\qquad\quad -\polylog{2}\left(1-e^{-x_a}\right)\ln\left(1-e^{-x_a}\right)\Big]\bigg\},
\label{eq:rate_c_analytic}
\end{split}
\end{align}
where $x_a=\beta E_a$, while the polylogarithm $\polylog{n}$ is given by \cite{Cohl:2014drm}
\begin{align}
\polylog{n}\left(e^{x_a}\right)&\equiv \frac{1}{\Gamma(n)}\int^\infty_0 dt \frac{t^{n-1}}{e^{t-x_a}-1} \,\,\,\forall\,\, {\rm Re}(n)>0,\nonumber\\
\Gamma(n)&\equiv \int^\infty_0 dt\, t^{n-1} \,e^{-t}\,\,\,\forall\,\, {\rm Re}(n)>0.
\label{eq:PolyLog_Gamma}
\end{align}
For this term, one can analytically appreciate the effect of neglecting the Bose-enhancement in the rate. Neglecting Bose-enhancement gives 
\begin{align}
\frac{d^3R_c}{d^3p_a}\simeq\frac{9\alpha_s^2T}{8\pi^2x_a}+O\left(e^{-x_2}\right),
\label{eq:rate_c_no_Bose_enh}
\end{align}
where $x_2=\beta E_2$. While losing exponentially suppressed terms is to be expected from the omission of Bose-enhancement, the secondary effect of not including Bose-enhancement is the lack of $x_a^{-2}$ terms, with solely the leading $x_a^{-1}$ term being recovered in the approximate result. Indeed, for $x_a\gg 1$, $\polylog{n}(1-e^{-x_a})\to \zeta(n)$, where $\zeta$ is the Riemann $\zeta$ function, and thus $x^{-2}_a$ contributions are present in the full result. Therefore, omitting Bose-enhancement changes the power counting of the final answer (which is not immediately appreciated from dropping terms under the integral), and care should be taken from here onward. 

The $s$-channel with Bose-enhancement, stemming from integrating term-$\mathbf{II}$, can be expressed as
\begin{align}
\frac{d^3R_s}{d^3p_a}&=-\frac{9\alpha_s^2}{4\pi^4E_a^2}\int_0^\infty \frac{dE_b}{e^{\beta E_b}-1}\int_0^{E_a+E_b} \frac{dE_2}{1-e^{-\beta E_2}}\nonumber\\
&\times\frac{E_aE_b\left(E_a+E_b-E_2\right)E_2}{\left(E_a+E_b\right)^3},\nonumber\\
&=-\frac{9\alpha_s^2T}{4\pi^4x_a}\int_0^\infty\frac{dx_b}{e^{x_b}-1}\bigg\{\frac{x_b}{6}-\frac{2\zeta(3)x_b}{\left(x_a+x_b\right)^3}\nonumber\\
&\qquad\qquad\quad+\frac{\pi^2x_b}{6\left(x_a+x_b\right)^2}+\frac{2x_b\polylog{3}\left[e^{-\left(x_a+x_b\right)}\right]}{\left(x_a+x_b\right)^3}\nonumber\\
&\qquad\qquad\quad+\frac{x_b\polylog{2}\left[e^{-\left(x_a+x_b\right)}\right]}{\left(x_a+x_b\right)^2}\bigg\}\label{eq:rate_s},\\
&\frac{d^3R_s}{d^3p_a}\simeq-\frac{\alpha_s^2T}{16\pi^2x_a}+O\left(e^{-x_2}\right),\label{eq:rate_s_approx}
\end{align}
where $x_b=\beta E_b$, while the last line neglects Bose-enhancement. In doing so, not only are exponentially suppressed terms neglected, but some power-law decaying terms are as well. We can capture power-law terms via a binomial expansion of the denominators $(1+x_b/x_a)^{-n}$, for $x_a\gg x_b$. Start by considering 
\begin{align}
\frac{d^3\tilde{R}_s}{d^3p_a}&=-\frac{9\alpha_s^2T}{4\pi^4x_a}\int_0^\infty\frac{dx_b}{e^{x_b}-1}\bigg\{\frac{x_b}{6}-\frac{2\zeta(3)x_b}{\left(x_a+x_b\right)^3}\nonumber\\
&\qquad\qquad\quad+\frac{\pi^2x_b}{6\left(x_a+x_b\right)^2}\bigg\}
\label{eq:rate_s_noLi},
\end{align}
where fourth and fifth term in Eq.~(\ref{eq:rate_s}) are ignored. The expansion of the binomial second and third terms in Eq.~(\ref{eq:rate_s_noLi}) are, in general, given by 
\begin{align}
\begin{split}
    &\int_0^\infty\frac{dx_b}{e^{x_b}-1}\frac{x_b}{\left(x_a+x_b\right)^n}\\
    &=\sum_{k=0}^\infty\left(-1\right)^k\frac{(n+k-1)!}{(n-1)!k!x_a^{n+k}}\int_0^\infty\frac{dx_b}{e^{x_b}-1}x_b^{k+1}\\
    &=\sum_{k=0}^\infty\left(-1\right)^k\frac{(n+k-1)!(k+1)}{(n-1)!x_a^{n+k}}\zeta(k+2),
    \label{eq:binomial_zeta_series}
\end{split}
\end{align}
where we switched the order of the integral and sum, resulting in a series of Riemann $\zeta$-functions. For the specific case considered herein, the relevant powers are $n=2,3$, which, when incorporated within the rate, yield
\begin{align}
\frac{d^3 \tilde R_s}{d^3 p_a}&\simeq \left.\frac{d^3R_s^{(N)}}{d^3p_a}\right|_\zeta+O\left(e^{-x_a}\right)\\
\left.\frac{d^3R_s^{(N)}}{d^3p_a}\right|_\zeta&=-\frac{9\alpha_s^2T}{4\pi^4x_a}\Bigg[\frac{\pi^2}{36}-\frac{\pi^2}{6x_a}\sum_{n=1}^N\left(-1\right)^n\frac{n!\,n\,\zeta\left(n+1\right)}{x_a^n}\nonumber\\
&\qquad\qquad\quad+\frac{\zeta(3)}{x_a^2}\sum_{n=1}^N\left(-1\right)^n\frac{(n+1)!\,n\,\zeta\left(n+1\right)}{x_a^n}\Bigg].
\label{eq:s-ch_approx}
\end{align}
Note that Eq.~(\ref{eq:s-ch_approx}) is an asymptotic series, as shown in Appendix~\ref{apnd:zeta_series}. Alternatively, one can expand the Bose distribution in Eq.~(\ref{eq:rate_s_noLi}) as a geometric series 
\begin{align}
\frac{1}{e^{x_b}-1}=\frac{e^{-x_b}}{1-e^{-x_b}}=e^{-x_b}\sum^\infty_{k=0} e^{-k x_b}=\sum^\infty_{k=1} e^{-k x_b},\label{eq:BE_geom}
\end{align}
which, in turn, requires evaluating integrals of the form
\begin{align}
&\int_0^\infty dx_b\frac{x_be^{-kx_b}}{\left(x_a+x_b\right)^n}=e^{kx_a}k^{n-2}\int_{kx_a}^\infty dt\left[\frac{e^{-t}}{t^{n-1}}-kx_a\frac{e^{-t}}{t^n}\right]\nonumber\\
&=e^{kx_a}k^{n-2}\left[\Gamma(2-n,kx_a)-kx_a\Gamma(1-n,kx_a)\right],
\end{align}
where $t=k(x_a+x_b)$, while the upper incomplete gamma function \cite{Cohl:2014drm} is defined  via
\begin{align}
\Gamma(s,x)\equiv \int^\infty_x dt\, t^{s-1}e^{-t}.
\end{align}
As the upper incomplete gamma function only admits non-positive integer $s$ values herein, we rewrite it as \cite{Cohl:2014drm}
\begin{align}
    \Gamma(-n,x)=\frac{\left(-1\right)^{n+1}}{n!}\left[\mathrm{Ei}(-x)+e^{-x}\sum_{k=0}^{n-1}\frac{\left(-1\right)^k k!}{x^{k+1}}\right],
\end{align}
where the exponential integral is given by
\begin{align}
{\rm Ei}(z)&\equiv-\int^\infty_{-z} dt\, \frac{e^{-t}}{t}=-\Gamma(0,-z)=-E_1(-z)\nonumber\\
{\rm E}_1(z)&\equiv\int^\infty_z \frac{e^{-t}}{t},
\end{align}
where ${\rm E}_1$  can also be found in Ref.~\cite{Cohl:2014drm}. These results are combined to yield a convergent series (see details in Appendix~\ref{apnd:zeta_series})
\begin{align}
\label{eq:Ei-expansion}
\left.\frac{d^3R_s^{(N)}}{d^3p_a}\right|_{\rm Ei}&\simeq-\frac{9\alpha_s^2T}{4\pi^4x_a}\Bigg\{\frac{\pi^2}{36}\nonumber\\
&-\frac{\pi^2}{6}\sum_{n=1}^N\left[1+(1+nx_a)e^{nx_a}\mathrm{Ei}(-nx_a)\right]\nonumber\\
&-\zeta(3)\sum_{n=1}^N\Bigg[\frac{1}{x_a}+n\\
&\qquad\qquad\quad+n(2+nx_a)e^{nx_a}\mathrm{Ei}(-nx_a)\Bigg]\Bigg\}\nonumber\\
&+O\left(e^{-x_a}\right).\nonumber
\end{align}
\begin{figure}[h]
    \centering
	\includegraphics[width=\linewidth]{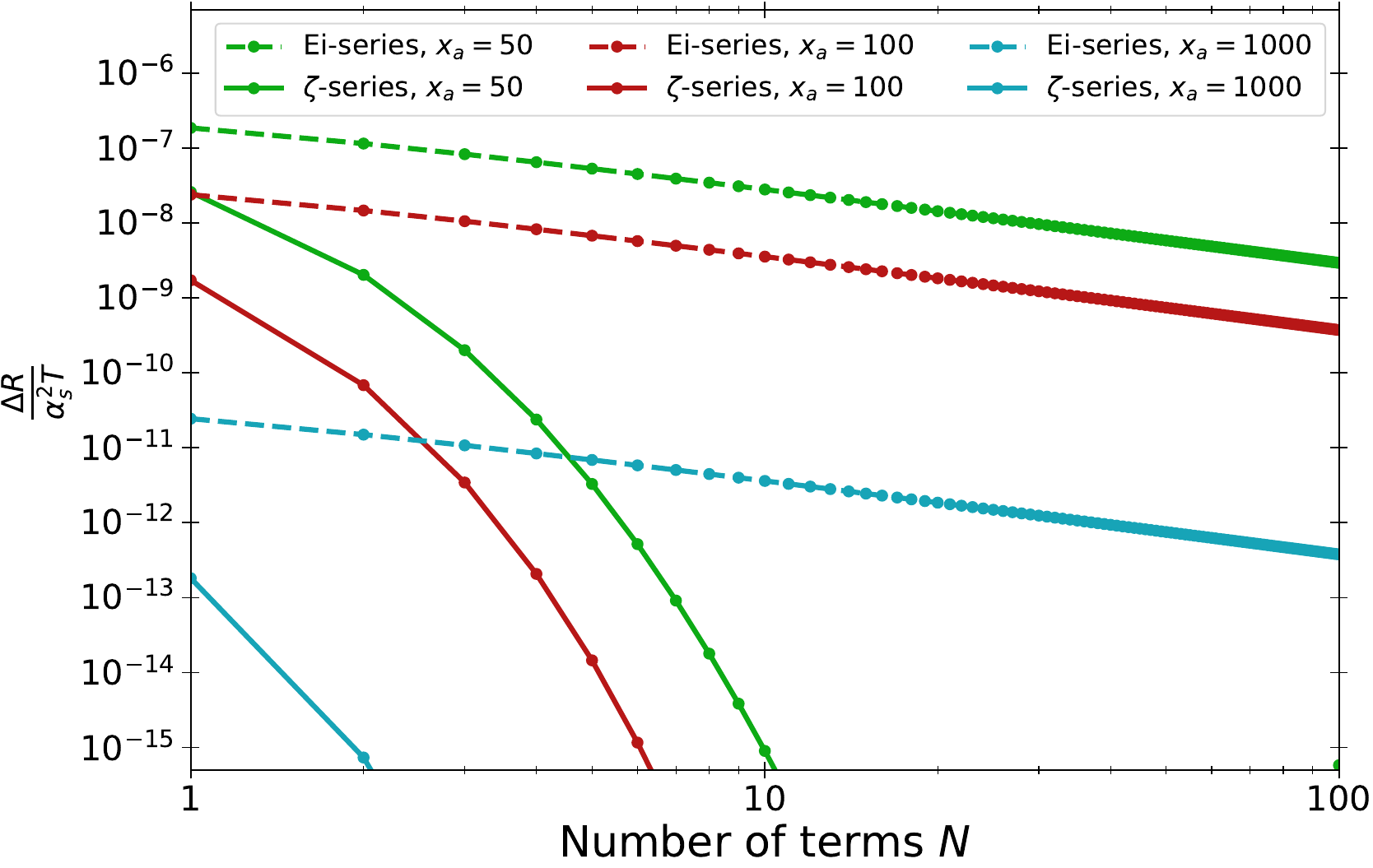}
    \caption{Differences of both series expansions and the full result, as the number of terms is increased, for several fixed values of $x_a$ chosen.}
    \label{fig:Zeta-Ei}
\end{figure}
When comparing the convergence rate of the two expansions at fixed $x_a$, the binomial series ($\zeta$-series) is approaching the true value faster for smaller $N$ than the geometric series (Ei-series) of the thermal distribution function, as shown in Fig.~\ref{fig:Zeta-Ei}, even though the $\zeta$-series is asymptotic, and thus doesn't converge to the original integral as $N\to\infty$. Depicted in Fig.~\ref{fig:Zeta-Ei} is the absolute difference $\Delta R$ between the full, numerical result in Eq.~(\ref{eq:rate_s}), and either of the two expansions $\frac{d^3 R^{(N)}_s}{d^3 p_a}$, be it the the Ei-series or the $\zeta$-series, rescaled by $\alpha^2_sT$. Specifically, 
\begin{align}
\label{eq:DeltaR_N}
\frac{\Delta R(x_a,N)}{\alpha^2_s T}\equiv \frac{\left| \frac{d^3R_s}{d^3p_a}-\frac{d^3R^{(N)}_s}{d^3p_a} \right|}{\alpha^2_s T},
\end{align}
where $N$ accounts for an increasing number of terms in the series, while $x_a$ is held fixed. The $x_a\gg x_b$ approximation holds well since typical parton energies of a jet are larger than typical medium parton energies, and thus the $\zeta$-function expansion provides a good approximation to $\frac{d^3 R_s}{d^3 p_a}$ for Monte Carlo simulations of jet-medium scattering. Moreover, truncating the series at a given number of terms allows to determine the theoretical systematic uncertainty within the above-suggested Monte Carlo simulation, by computing the difference between a series of $n+1$ terms compared to one that only contains $n$ terms. 

For sufficiently large $x_a$, the region $x_b>x_a$ does not contribute significantly to the overall integral on the left hand side of Eq.~(\ref{eq:binomial_zeta_series}), owing to the exponential suppression of large $x_b$ values (see details in Appendix~\ref{apnd:zeta_series}), thus the $\zeta $-function expansion provides a very good approximation to the full integral. For completeness, a binomial expansion for $\polylog{2}$ and $\polylog{3}$ terms in the $s$-channel, namely for the $\frac{d^3R_s}{d^3 p_a}-\frac{d^3 \tilde R_s}{d^3 p_a}$, can be found in Appendix~\ref{apnd:multi_Li_series}. However, Fig.~\ref{fig:Zeta-Ei} already shows that the $\polylog{2}$ and $\polylog{3}$ contributions are small for $x_a\geq 50$.

For the $t$-channel, neither of the energy integrals can be obtained using well-known functions. A closed form using solely functions in Eq.~(\ref{eq:thermo_int}) can be obtained whence Bose-enhancement is neglected. Thus,
\begin{align}
\frac{d^3R_t}{d^3p_a}&=\frac{9\alpha_s^2}{4\pi^4E_a^2}\int_0^\infty \frac{dE_b}{e^{\beta E_b}-1}\int_0^{E_a+E_b} \frac{dE_2}{1-e^{-\beta E_2}}\nonumber\\
&\qquad\qquad\quad\times\frac{E_aE_b\left(E_a+E_b-E_2\right)E_2}{\left[\left(E_b-E_2\right)^2+m_D^2\right]^\frac{3}{2}},\nonumber\\
\frac{d^3R_t}{d^3p_a}&\simeq\frac{9\alpha_s^2T}{4\pi^4}\int_0^\infty\frac{dx_b}{e^{x_b}-1}\Bigg[\frac{x_ax_b\sqrt{x_b^2+z^2}+x_b^2\sqrt{x_a^2+z^2}}{z^2x_a}\nonumber\\
&\qquad\qquad-\frac{x_b\arcsinh\left(\frac{x_a}{z}\right)}{x_a}-\frac{x_b\arcsinh\left(\frac{x_b}{z}\right)}{x_a}\Bigg]\nonumber\\
&+O(e^{-x_2}),
\end{align}
where, from the first to the second line, $\left[1-e^{-\beta E_2}\right]^{-1}$ was expanded as a geometric series, and solely the leading term is kept. Performing the integral over $x_b$ gives
\begin{align}
\begin{split}
\frac{d^3R_t}{d^3p_a}&\simeq \frac{9\alpha_s^2T}{4\pi^4}\Bigg[\frac{A^-_1(z)}{z^2}+\frac{2\zeta(3)\sqrt{x^2_a+z^2}}{z^2 x_a}\\
&\qquad\qquad-\frac{\pi^2\arcsinh\left(\frac{x_a}{z}\right)}{6x_a}-\frac{B^-_1(z)}{x_a}\Bigg]\\
&\quad+O(e^{-x_2}),    
\end{split}
\end{align}
where $z=\beta m_D$, while $A^-_1(z)$ and $B^-_1(z)$ are defined in Eq.~(\ref{eq:thermo_int}). The $u$-channel contribution produces the same approximate result as the $t$-channel one, hence their sum gives
\begin{align}
\frac{d^3R_{t+u}}{d^3p_a}&\simeq \frac{9\alpha_s^2T}{\pi^4}\Bigg[\frac{A^-_1(z)}{2z^2}+\frac{\zeta(3)\sqrt{x^2_a+z^2}}{z^2 x_a}\label{eq:rate_tu_approx}\\
&\qquad\qquad-\frac{\pi^2\arcsinh\left(\frac{x_a}{z}\right)}{12x_a}-\frac{B^-_1(z)}{2x_a}\Bigg]+O(e^{-x_2}).\nonumber
\end{align}
The sum of all four channels, neglecting Bose-enhancement, yields
\begin{align}
\frac{d^3R}{d^3p_a}&\simeq \frac{17\alpha_s^2T}{16\pi^2x_a}+\frac{9\alpha_s^2T}{\pi^4}\Bigg[\frac{A^-_1(z)}{2z^2}+\frac{\zeta(3)\sqrt{x^2_a+z^2}}{z^2 x_a}\nonumber\\
&\qquad-\frac{\pi^2\arcsinh\left(\frac{x_a}{z}\right)}{12x_a}-\frac{B^-_1(z)}{2x_a}\Bigg]+O\left(e^{-x_2}\right).
\end{align}
Expanding the Bose-Einstein distribution within $A^{-}_1(z)$ and $B^{-}_1(z)$ using Eq.~(\ref{eq:BE_geom}) gives
\begin{align}
\tilde{A}^-_1(z,N)&=\sum^{N}_{n=1}\int_0^\infty dx \,x \sqrt{x^2+z^2}\,e^{-nx}\nonumber\\
&=\sum^{N}_{n=1}\frac{\pi z}{2n^2}\left[2K_1\left(nz\right)-nzK_0\left(nz\right)\right],\nonumber\\
\tilde{B}^-_1(z,N)&=\sum^{N}_{n=1}\int_0^\infty dx\, x\arcsinh\left(\frac{x}{z}\right)e^{-nx}\nonumber\\
&=\sum^{N}_{n=1}\frac{\pi}{2n^2}\left[K_0\left(nz\right)-nzK_{-1}\left(nz\right)\right],
\label{eq:Struve-Expansion}
\end{align}
where $K_\alpha$ is the Struve function of the second kind \cite{Cohl:2014drm}, given by
\begin{align}
    K_\alpha(x)\equiv\frac{2}{\Gamma\left(\alpha+\frac{1}{2}\right)\sqrt{\pi}}\left(\frac{x}{2}\right)^\alpha\int_0^\infty dt\left(1+t^2\right)^{\alpha-\frac{1}{2}}e^{-xt}.
\end{align}
\begin{figure}[H]
    \centering
	\includegraphics[width=\linewidth]{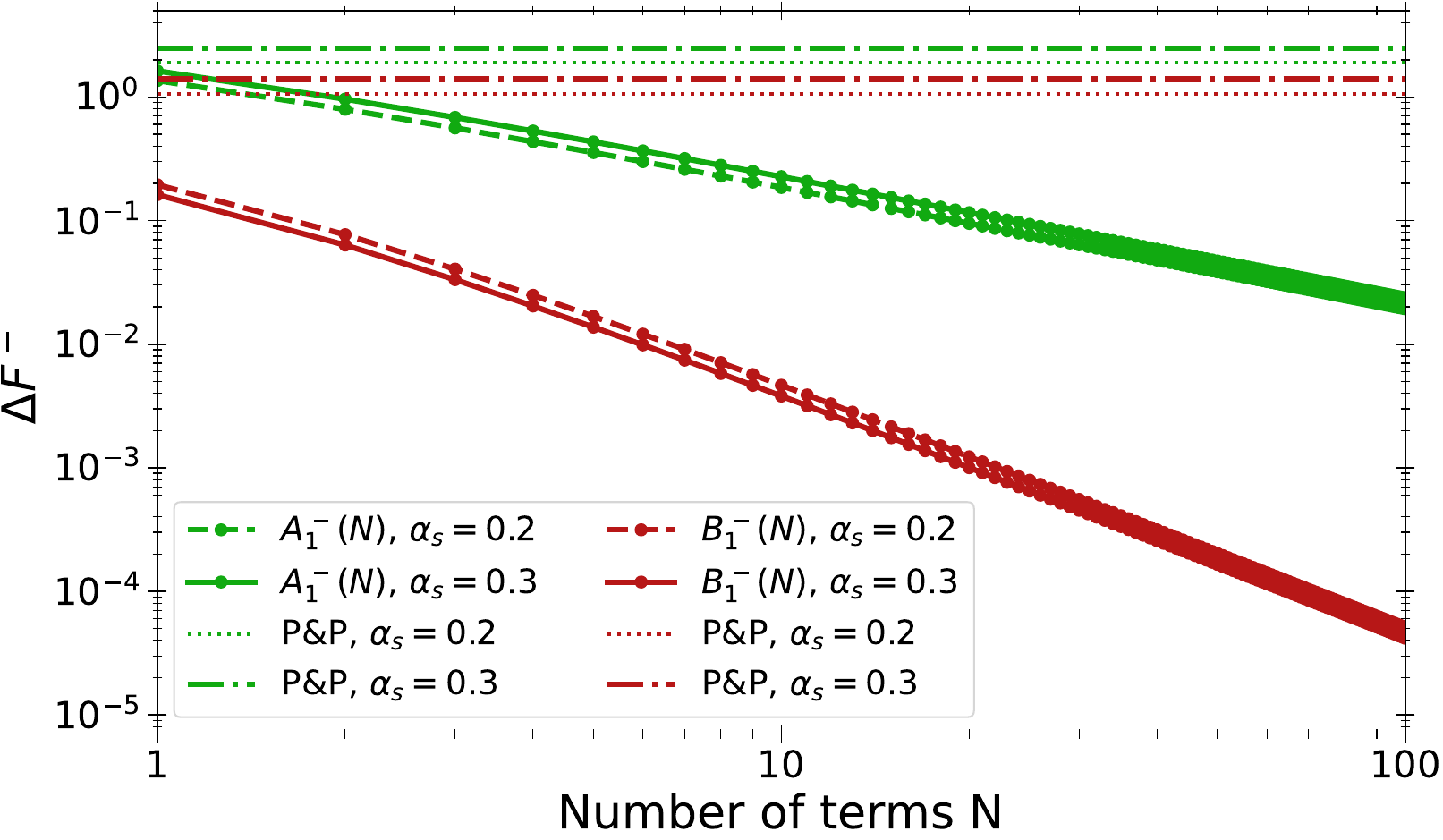}
    \caption{Differences of both Struve expansions and their full result, as the number of terms is increased. For comparison, the difference of the first term from the Peign\'e \& Peshier expansion to the full result (``P\&P'') are given as constant lines. Both evaluated for two different $z(\alpha_s)$, where $\alpha_s$ is taken to be a constant.}
    \label{fig:Struve}
\end{figure}
The $K_\alpha$-series convergence with $N$ is depicted in Fig.~\ref{fig:Struve}. Shown along the $y$-axis is the absolute difference, 
\begin{align}
    \Delta F^-(z,N)\equiv\left|F^-_1(z)-\tilde{F}_1^-(z,N)\right|,
\end{align}
where here $F^-_1\in\{A_1^-,B_1^-\}$ is one of the two thermodynamic integrals, defined in Eq.~(\ref{eq:thermo_int}), which appear in the rate's $t$- and $u$-channel, while $\tilde{F}_1^-$ is the sum of the first $N$ coefficients of the expansion in Eq.~(\ref{eq:Struve-Expansion}).

One notices that the convergence rate of the series in Eq.~(\ref{eq:Struve-Expansion}) is similar to that in Eq.~(\ref{eq:Ei-expansion}), being slower than Eq.~(\ref{eq:s-ch_approx}). With these expansions at hand, the $gg\to gg$ scattering rate becomes 
\begin{align}
\frac{d^3R_{t+u}}{d^3p_a}&\simeq \frac{9\alpha_s^2T}{\pi^4}\Bigg[\frac{\zeta(3)\sqrt{x^2_a+z^2}}{z^2 x_a}-\frac{\pi^2\arcsinh\left(\frac{x_a}{z}\right)}{12x_a}\nonumber\\
&\qquad\qquad+\sum_{n=1}^\infty\frac{\pi}{4n^2z}\left[2K_1\left(nz\right)-nzK_0\left(nz\right)\right]\nonumber\\
&\qquad\qquad-\sum_{n=1}^\infty\frac{\pi}{4n^2x_a}\left[K_0\left(nz\right)-nzK_{-1}\left(nz\right)\right]\Bigg]\nonumber\\
&\qquad\qquad+O(e^{-x_2}).
\end{align}
Instead of expanding the Bose-Einstein distribution in $A^{-}_1(z)$ and $B^-_1(z)$, one can expand,   
\begin{align}
\arcsinh\left(x/z\right)&\simeq \ln\left(2x/z\right)+O\left(z^2x^{-2}\right)\nonumber\\
\sqrt{1+z^2x^{-2}}&\simeq 1+O\left(z^2x^{-2}\right),
\label{eq:approx_arcsinh_sqrt}
\end{align}
assuming $x\gg z$, as this is the kinematic region where tree-level matrix elements are valid. This was done by Peign\'e \& Peshier \cite{Peigne:2007sd,Peigne:2008nd}, and they have kept solely the first term in Eq.~(\ref{eq:approx_arcsinh_sqrt}). Extending Eq.~(\ref{eq:approx_arcsinh_sqrt}) to next to leading order (NLO) and inserting it within $A^-_1(z)$ and $B^-_1(z)$ generates issues at NLO, as one encounters poles in the series, as explored in Ref.~\cite{short_paper}. These issues do not occur in the $s$-channel's binomial approximation explored, see Eq.~(\ref{eq:s-ch_approx}). Expanding instead in the region $x\ll z$, and taking $A^-_1(z)$ as an example, gives the asymptotic series
\begin{align}
    A^-_1(z)&=\sum^{\infty}_{n=0}\frac{\left(-1\right)^{n+1}(2n)!}{4^n\left(n!\right)^2(2n-1)}\int_0^\infty\frac{dx}{(e^x-1)}\left(\frac{x}{z}\right)^{2n+1}\nonumber\\
    &=\sum^{\infty}_{n=0}\frac{\left(-1\right)^{n+1}\left(2n\right)!\left(2n+1\right)!\zeta(2n+2)}{4^n\left(n!\right)^2\left(2n-1\right)z^{2n+1}}.
    \label{eq:divergent_A1_series}
\end{align}
Indeed, when evaluating this expansion numerically, one finds that the magnitude of successive terms to be growing boundlessly, which is shown via d'Alembert's ratio test. Calling the coefficients in the series above $a_n$, one gets
\begin{align}
    \lim_{n\to\infty}\left|\frac{a_{n+1}}{a_n}\right|=\lim_{n\to\infty}\left[\frac{(2n+3)(2n-1)\zeta(2n+4)}{\zeta(2n+2)z^2}\right]=\infty,
\end{align}
thus the infinite series diverges for any finite $z$. A similar result can be found for the series associated with $B^-_1(z)$. Therefore, we will use the full integral representation of $A^-_1(z)$ and $B^-_1(z)$.

The Peign\'e \& Peshier result stemming from the first term in the series shown in Eq.~(\ref{eq:approx_arcsinh_sqrt}) is depicted as horizontal lines in Fig.~\ref{fig:Struve}, and also compares their result to $K_\alpha$-series in Eq.~(\ref{eq:Struve-Expansion}). 

As can be seen in Fig.~\ref{fig:Struve}, the Peign\'e \& Peshier approximation for $A^-_1$ is off by a factor $\sim 2$ compared to the first term in the $K_\alpha$-expansion. Adding more terms in the Struve-expansion decreases the error on $A^-_1$, while all higher-order terms in the Peign\'e \& Peshier approximation diverge. On the other hand, expanding the arcsinh according to Eq.~(\ref{eq:approx_arcsinh_sqrt}) within $B_1^-$ gives a much cruder approximation of $B^-_1(z)$. This can mostly be attributed to the fact that approximating the inverse hyperbolic sine as a logarithm produces a negative region close to $x_b\gtrsim0$, while the $\arcsinh(x)>0$ for $x>0$. Approximating the arcsinh with a logarithm actually gives a wrong sign of the result, namely
\begin{align}
    \int_0^\infty\frac{dx_b}{e^{x_b}-1}x_b\ln\left(\frac{2x_b}{z}\right)\approx -0.53,
\end{align}
for $z(\alpha_s=0.3)$, while integrating numerically $B^-_1(z)$ gives $B_1^-[z(\alpha_s=0.3)]\approx0.87$. Expanding instead arcsinh to first order $\arcsinh(x)\simeq x +O(x^3)$ (which still gives a divergent infinite series), and keeping solely the first term, yields a better approximation of $B_1^-[z(\alpha_s=0.3)]$ of about $1.01$, after which, subsequent partial sums start oscillating rapidly between increasing values of alternating sign.

\subsection{Determining the $z$ cut-off}
\begin{figure}[H]
    \centering
	\includegraphics[width=\linewidth]{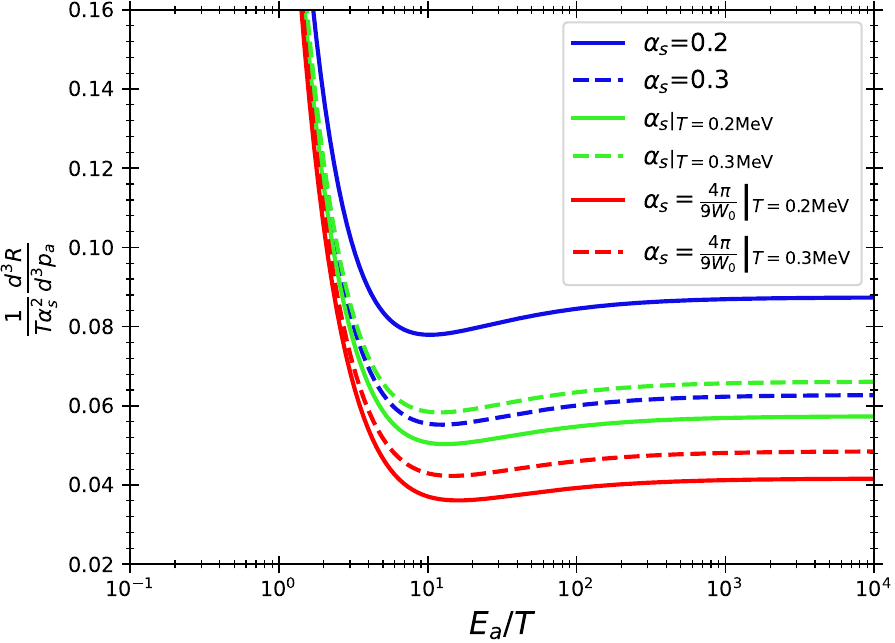}
    \caption{Full scattering rate for various ways of evaluating $\alpha_s$ (see text for details).}
    \label{fig:RunningAlpha}
\end{figure}
While the coupling strength $\alpha_s$ only enters directly as an overall normalization factor, the $t$- and $u$-terms of the rate obviously depend on the cut-off used, which herein is chosen to be $m_D$. The leading order Debye mass is related to the strong coupling, via
\begin{align}
    m_D^2=4\pi\alpha_s\left(1+\frac{N_f}{6}\right)T^2=6\pi\alpha_sT^2
    \label{eq:Debye_mass}
\end{align}
for a system at zero chemical potential and $N_f=3$ flavors. One way to choose the $z$ cut-off is therefore to simply fix $\alpha_s$ to a value. Two, reasonably chosen, fixed values for $\alpha_s$ are depicted by blue lines in Fig.~\ref{fig:RunningAlpha}. More realistically, the coupling runs as a function of the scale of the exchanged momentum, which is chosen to be Mandelstam $t$ herein. The leading order running of $\alpha_s$ is given by
\begin{align}
	\alpha_s(t)=\frac{12\pi}{(33-2N_f)\ln\left[\frac{t}{\Lambda^2_{QCD}}\right]},
\label{eq:alpha_s_2piT}
\end{align}
where $N_f$ is allowed to change with Mandelstam $t$ in the above expression. Besides $t$, in the medium, there is another scale to consider: one associated with the temperature of the medium. Thus, $\alpha_s$ can run with the propagator's four-momentum, or the temperature. If the temperature is chosen as the relevant scale, then the strong coupling is $\alpha_s(t=4\pi^2 T^2)$. Such a choice generates the pale green lines in Fig.~\ref{fig:RunningAlpha}. The last option, depicted by red lines in Fig.~\ref{fig:RunningAlpha}, stems from the requirement $t\leq -m^2_D$, which generates a self-consistent equation for $t$, as the strong coupling runs with the Mandelstam $t$ exchanged in the scattering~\cite{Peigne:2007sd,Peigne:2008nd}. Requiring that $t_{\min}$ satisfies $t=-6\pi\alpha_s(t)T^2$ yields
\begin{align}
	m_D^2=\frac{8\pi^2T^2}{3W_0\left(\frac{8\pi^2T^2}{3\Lambda^2_3}\right)}
\label{eq:mD_w_running}
\end{align}
where $\Lambda_3$ (the QCD scale for three lightest quark flavors) is obtained by requiring that the running strong coupling $\alpha_s(t)$ is continuous across all scales probed by Mandelstam $t$, with no discontinuities as $N_f$ changes with $t$. So, starting from $\Lambda_5$ (the QCD scale at the five quark flavors), which is determined from the measured value of $\alpha_s$ at scale of the $Z$-boson mass, one obtains $\Lambda_4$, and $\Lambda_3$, by requiring no discontinuities in $\alpha_s$ across the kinematic boundary where $N_f$ changes from $N_f=4$ to $N_f=3$. From $m^2_D$ in Eq.~(\ref{eq:mD_w_running}), one obtains
\begin{align}
	\alpha_s=\frac{4\pi}{9W_0\left(\frac{8\pi^2T^2}{3\Lambda^2_3}\right)}
\label{eq:alpha_s_run_W0}
\end{align}
where $W_0$ includes the positive region of the Lambert $W$-function \cite{Cohl:2014drm}, defined as the function that solves the transcendental equation
\begin{align}
    W(x)e^{W(x)}\equiv x,
\end{align}
and whose Taylor series about $x=0$ gives
\begin{align}
W_0(x)=\sum^\infty_{n=1} \frac{(-n)^{n-1}}{n!} x^n.
\end{align}
\subsection{Exploring the individual channels of the rate}
\begin{figure}[h]
    \centering
	\includegraphics[width=\linewidth]{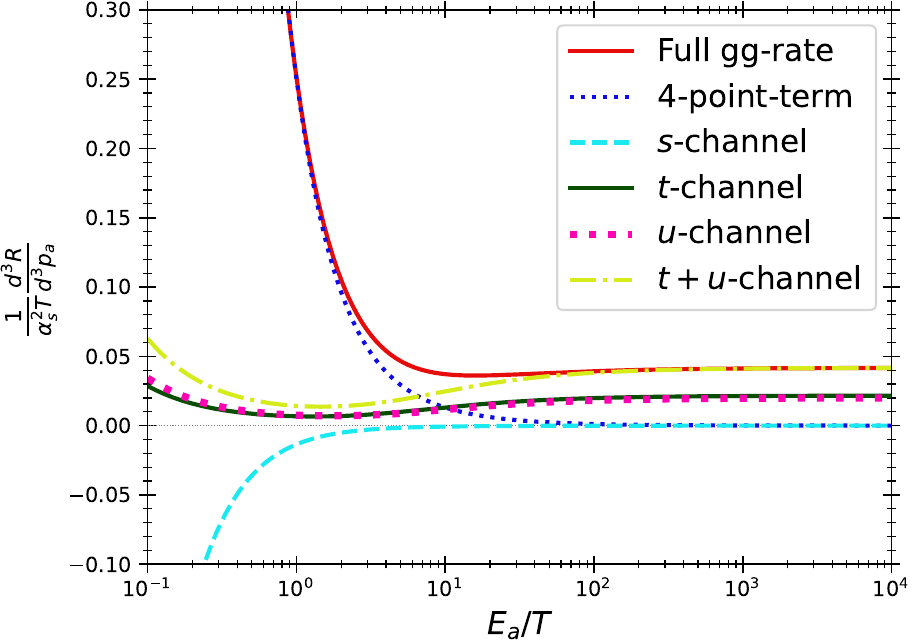}
    \caption{Individual channel contributions to the tree-level $gg\to gg$ scattering rate, with $m_D$ being evaluated at $T=0.2$ GeV and satisfying Eq.~(\ref{eq:mD_w_running}).}
    \label{fig:rate_channel_decomposition}
\end{figure}

The individual channel's contribution to the overall scattering rate at tree-level is depicted in Fig.~\ref{fig:rate_channel_decomposition}. While the scattering rate is dominated by the $t$- and $u$-channel contributions for $x_a>100$, Fig.~\ref{fig:rate_channel_decomposition} shows that the four-point vertex contribution to the scattering rate, i.e. $\frac{d^3 R_c}{d^3 p_a}$, starts to contribute already in the $10<x_a<100$ region, far above the region where HTL effects are expected to be significant~\cite{Shen:2014nfa}.

\subsection{A good approximation to the full tree-level rate}
\begin{figure}[h]
    \centering
	\includegraphics[width=\linewidth]{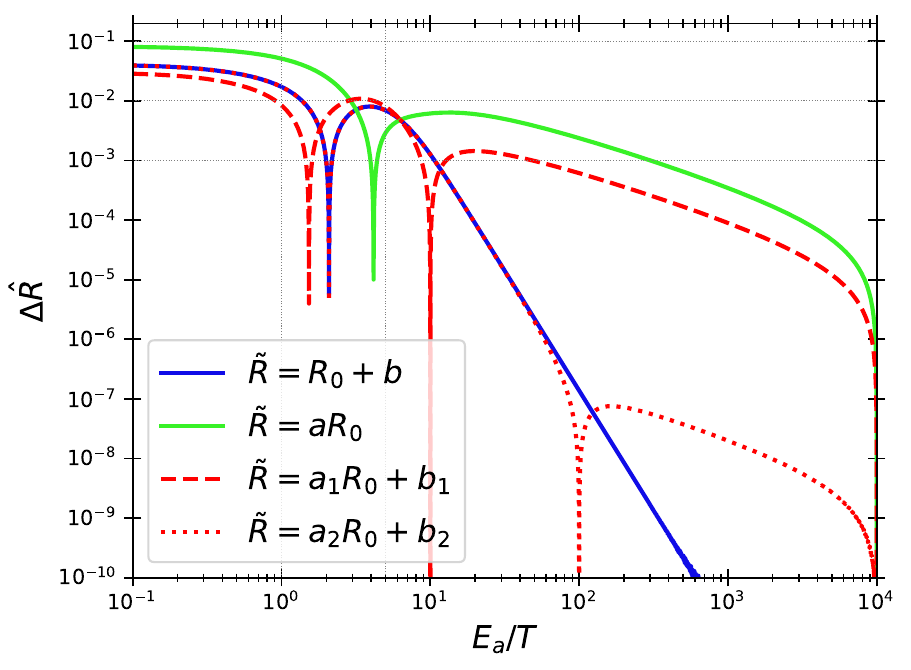}
    \caption{Using Eq.~(\ref{eq:rescaled_approx_rate}), with various values of $a$ and $b$ (c.f. text) and Eq.~(\ref{eq:rate_c_analytic}) to approximate the full result in Eqs.~(\ref{eq:rate_full}, \ref{eq:rate_I_IV}) obtained numerically. $m_D$ is taken to follow Eq.~(\ref{eq:mD_w_running}) with $T=0.2$ GeV.}
    \label{fig:rate_fit}
\end{figure}

Figure~\ref{fig:rate_fit} investigates ways of re-using the rates in Eqs.~(\ref{eq:rate_s_approx}, \ref{eq:rate_tu_approx}) to obtain a better approximation of the full rate in Eqs.~(\ref{eq:rate_full}, \ref{eq:rate_I_IV}). We define the approximate rate as
\begin{align}
\frac{d^3R_{\rm approx}}{d^3p_a}&\equiv\frac{9\alpha_s^2T}{\pi^4}\Bigg[\frac{A^-_1(z)}{2z^2}+\frac{\zeta(3)\sqrt{x^2_a+z^2}}{z^2 x_a}\nonumber\\
&\quad-\frac{\pi^2\arcsinh\left(\frac{x_a}{z}\right)}{12x_a}-\frac{B^-_1(z)}{2x_a}\Bigg] -\frac{\alpha_s^2T}{16\pi^2x_a}\nonumber\\
&+\frac{d^3R_c}{d^3p_a},
\label{eq:rate_approx_stu_w_c_exact}
\end{align}
where $\frac{d^3 R_c}{d^3 p_a}$ is given by Eq.~(\ref{eq:rate_c_analytic}). This expression is then linearly transformed according to
\begin{align}
\frac{d^3\tilde{R}}{d^3p_a}\equiv a\left[\frac{d^3R_{\rm approx}}{d^3p_a}\right]+b
\label{eq:rescaled_approx_rate}
\end{align}
in order to construct the relative difference
\begin{align}
    \Delta\hat{R}\equiv\left|\frac{d^3R}{d^3p_a}-\frac{d^3\tilde{R}}{d^3p_a}\right|\left(\frac{d^3R}{d^3p_a}\right)^{-1}
\end{align}
between the full, Bose-enhanced, result, which is evaluated numerically, and a linear transformation of the approximate rate expression. The relative difference is depicted in Fig.~\ref{fig:rate_fit}, where different values of the parameters $a$ and $b$ in this linearly transformed, approximate rate are compared. Note that not using Eq.~(\ref{eq:rate_c_analytic}) inside the $\frac{d^3 R_{\rm approx}}{d^3 p_a}$ gives a larger $\Delta \hat R$.

In all cases, both rates (the linearly transformed approximate result as well as the full result) are required to be equal in the UV-region at $x_a=10^{4}$. This determines the scaling factor $a$ for the green line (where $b=0$), and the translation coefficient $b$ for the blue line (where $a=0$). The red lines have one more free parameter, so one more point can be fixed to match the full rate. Shown in Fig.~\ref{fig:rate_fit} are two examples where the additional anchoring point is at $x_a=10$ and $x_a=100$ for the dashed and dotted red lines, respectively, while the other point remains at $x_a=10^{4}$.

Examining the various configurations, it appears that using a full linear transformation (c.f. red lines, with two free parameters) solely improves upon the simple rescaling of the approximate rate, i.e. the green line. Instead, simply shifting the approximate rate by a constant, as depicted by the blue line in Fig.~\ref{fig:rate_fit}, seems to be the best approximation.\footnote{Note that the additional dips around $1<x_a<5$ correspond to a change in sign of the difference $\frac{d^3R}{d^3p_a}-\frac{d^3\tilde{R}}{d^3p_a}$.} Figure~\ref{fig:rate_fit} is especially useful for the development of Monte Carlo simulations of jets in the QGP, as it allows to re-use Eq.~(\ref{eq:rate_approx_stu_w_c_exact}), a closed-form expression with respect to $x_a$, to develop a numerically efficient scheme for evaluating $gg\to gg$ scattering rates to below $10^{-3}$ accuracy in the Monte Carlo relevant range: $x_a>10$. Theoretical systematic uncertainties can also be estimated by comparing any approximate result with the full result. 

Finally, we have numerically verified that adding the first few $O(e^{-n x_2})$ corrections to $\frac{d^3 R_{\rm approx}}{d^3 p_a}$, stemming from higher order terms in the geometric series expansion of $\left[1-e^{-x_2}\right]^{-1}$, does not improve $\Delta \hat R$. Moreover, there is no well-established analytic expression for these $O(e^{-n x_2})$ corrections.

\subsection{Comparison to other approximate approaches in the literature}
Several approximations to tree-level leading order $gg\to gg$ scattering are present in the literature. We compare our results to the approximate approaches of Braaten \& Thoma \cite{Braaten:1991jj,Braaten:1991we}, Peign\'e \& Peshier \cite{Peigne:2007sd,Peigne:2008nd}, as well as the leading logarithm.
\subsubsection{The leading logarithm approximation}
To obtain the leading logarithm approximation, it is easiest to start from
\begin{align}
\frac{d^3R}{d^3p_a} &= \frac{1}{\left(2\pi\right)^8 2E_a}\int d^4q\, \frac{d^{3}p_b}{2E_b}d^4p_1\, \delta\left(p^2_1\right)\Theta(E_1) \nonumber\\
&\times \int d^4p_2\, \delta\left(p^2_2\right)\Theta(E_2) f_\pm(E_b)\left[1\mp f_\pm(E_2)\right] \overline{|\mathcal{M}|^2}\nonumber\\ 
&\qquad\times \delta^{(4)}\left(p_a - p_1 - q\right)\,\delta^{(4)}\left(q - p_2 + p_b\right),\nonumber\\
&=\frac{1}{\left(2\pi\right)^8 2E_a}\int d^4q\, \frac{d^3p_b}{2E_b}f\left(E_b\right)\left[1\pm f\left(E_{b} +\omega\right)\right]\nonumber\\
&\qquad\qquad\qquad \times \overline{|\mathcal{M}|^2}\Theta\left(E_{a} - \omega\right)\Theta\left(E_{b} + \omega\right)\nonumber\\
&\qquad\qquad\qquad \times\delta\left(\left(p_a - q\right)^2\right)\delta\left(\left(q + p_b\right)^2\right),
\end{align}
where $q^\mu=(\omega,\vec{q})$ denotes the four-momentum of the propagator. Expressing $\left(p_a-q\right)^2$ in the forward scattering approximation yields
\begin{align}
\left(p_a-q\right)^2&= q^2 - 2E_a \omega + 2 \lvert\vec{p}_a\rvert \lvert \vec{q}\rvert \cos\theta_{aq}\nonumber\\
&=\omega^2-q^2_\perp-q^2_z - 2E_a \omega + 2 \left\vert\vec{p}_a\right|q_z\nonumber\\
&\simeq -q^2_\perp-2E_a\omega+2 \left\vert\vec{p}_a\right|q_z
\end{align}
and thus
\begin{align}
\frac{d^3R}{d^3p_a} &\simeq \frac{1}{\left(2\pi\right)^8 2E_a}\int d^4q\, \frac{d^3p_b}{2E_b}f\left(E_b\right)\left[1\pm f\left(E_{b} +\omega\right)\right]\nonumber \\
&\qquad \qquad \qquad \times\overline{|\mathcal{M}|^2}\, \Theta\left(E_{a} - \omega\right)\Theta\left(E_{b} + \omega\right)\nonumber\\
&\qquad \qquad \qquad \times\frac{\delta\left(\omega-\omega^*\right)}{2E_a} \delta\left(\left(q + p_b\right)^2\right),
\end{align}
where
\begin{align}
\omega^*=\frac{-q^2_\perp+2\lvert \vec p_a\rvert q_z}{2E_a}.
\end{align}
After integrating over $\omega$ one obtains
\begin{align}
&\frac{d^3 R}{d^3 p_a}\simeq \frac{1}{\left(2\pi\right)^8 \left(2E_a\right)^2}\int d^3q\, \frac{d^3p_b}{2E_b}f\left(E_b\right)\left[1\pm f\left(E_{b} + \omega_*\right)\right]\nonumber\\
& \qquad \qquad \times\overline{|\mathcal{M}|^2}\, \Theta\left(E_{a} - \omega^*\right)\Theta\left(E_{b} + \omega^*\right)\delta\left(\left(q + p_b\right)^2\right).
\end{align}
Using the forward scattering approximation on $\left(p_b+q\right)^2$, one obtains
\begin{align}
\left(p_b+q\right)^2&\simeq-q^2_\perp\,+\\
& 2\left\{E_bq_z\left[1-\cos\left(\theta_{ba}\right)\right]-q_\perp\lvert\vec{p}_b\rvert\sin\left(\theta_{ba}\right) \cos\left(\Delta\varphi\right)\right\}\nonumber
\end{align}
with $\Delta\varphi=\varphi_{qa}-\varphi_{ba}$, where the azimuthal angles are measured relative to $p_a$, as highlighted by the indices on $\varphi$ on the right hand side. This allows to write the rate as
\begin{align}
\frac{d^3R}{d^3p_a} &\simeq \frac{1}{\left(2\pi\right)^8 \left(2E_a\right)^2}\int d^3q\, \frac{d^3p_b}{2E_b}f\left(E_b\right)\left[1\pm f\left(E_{b} + \omega^*\right)\right]\nonumber \\
&\quad \times\overline{|\mathcal{M}|^2}\, \Theta\left(E_{a} - \omega^*\right)\Theta\left(E_{b} + \omega^*\right) \frac{\delta\left(q_z-q^*_z\right)}{2E_b\lvert1-\cos(\theta_{ba})\rvert},
\end{align}
where 
\begin{align}
q^*_z=\frac{q^2_\perp+2q_\perp\lvert \vec p_b\rvert\sin\left(\theta_{ba}\right)\cos\left(\Delta\varphi\right)}{2E_b\left[1-\cos\left(\theta_{ab}\right)\right]}.
\end{align}
Using solely the first term in Eq.~(\ref{eq:BE_geom}), the $gg\to gg$ scattering rate simplifies down to
\begin{align}
\frac{d^3 R}{d^3 p_a} &\simeq \frac{1}{\left(2\pi\right)^6 16E_a^2}\int dq_{\perp} q_{\perp}  dE_b f(E_b) \frac{d[\cos(\theta_{ab})]\overline{|\mathcal{M}|^2}}{\lvert 1-\cos\left(\theta_{ab}\right)\rvert}.
\end{align}
In the $t\to 0$ forward scattering limit, approximating the $gg\to gg$ matrix elements amounts to setting $u\simeq-s$ such that solely the $t$- and $u$-channels contribute in the leading logarithm approximation. In fact, the $t$-channel and $u$-channel contributions are identical, yielding 
\begin{align}
\overline{|\mathcal{M}_{gg\to gg}|^2}&\simeq 288\alpha_s^2\left(2\pi\right)^2 \left[\frac{2\left\{2E_aE_b\left[1-\cos\left(\theta_{ba}\right)\right]\right\}^2}{q^4_\perp}\right].
\end{align}
To ensure that $\hat q\sim\ln\left[\frac{2E_aT}{m^2_D}\right]$ is obtained (as will be shown below), the integration limits in the rate are  
\begin{align}
\begin{split}
\frac{d^3 R}{d^3 p_a} &\simeq \frac{288\alpha^2_s}{\left(2\pi\right)^4}\int^{\sqrt{2E_aT}}_{m_D} \frac{dq_{\perp} q_{\perp}}{q^4_\perp} \int^\infty_0 dE_b E^2_b f(E_b)\\
&\qquad\qquad\times\int^1_{-1}d\left[\cos\left(\theta_{ab}\right)\right]\frac{1-\cos\left(\theta_{ab}\right)}{2},\\
\frac{d^3 R}{d^3 p_a}&\simeq\frac{9\alpha^2_s T}{\pi^4}\left[\frac{2\zeta(3)}{z^2}-\frac{\zeta(3)}{x_a}\right].    
\end{split}
\end{align}

\subsubsection{The Peign\'e \& Peshier approximation and the Braaten \& Thoma limit}
Starting from the initial expression for the scattering rate again
\begin{align}
\frac{d^3R}{d^3p_a} &= \frac{1}{2E_a\left(2\pi\right)^{12}}\int \frac{d^3p_b}{2E_b}\frac{d^3p_1}{2E_1}\frac{d^3p_2}{2E_2}f_\pm(p_b)\left[1 \mp f_\pm(p_2)\right]\nonumber\\
&\qquad\qquad\qquad\times\left(2\pi\right)^4 \delta^{(4)}\left(p_a +p_b - p_1 - p_2\right)\overline{\left\vert\mathcal{M}\right\vert^2},
\end{align}
and following the approach in Refs.~\cite{Peigne:2007sd, Peigne:2008nd}, the energy $\omega=E_a-E_1$ of the propagator, as well as Mandelstam variables $t=(p_a-p_1)^2$, $s=(p_a+p_b)^2$, are introduced via
\begin{align}
\begin{split}
1&=\int dt\delta\left\{t+2\left[E_aE_1-\cos\left(\theta_{1a}\right)\right]\right\},\\
1&=\int ds\delta\left\{s-2\left[E_aE_b-\cos\left(\theta_{ba}\right)\right]\right\},\\
1&=\int d\omega\delta\left(\omega-E_a+E_1\right).   
\end{split}
\label{eq:stomega_delta_fct}
\end{align}
These relations enable to write the scattering rate as
\begin{align}
\begin{split}
\frac{d^3R}{d^3p_a} &= \frac{1}{16E_a\left(2\pi\right)^{8}}\int \frac{d^3p_b}{E_b}\frac{d^3p_1}{E_1}\frac{d^3p_2}{E_2}dsdtd\omega f_\pm(p_b)\\
&\times \left[1 \mp f_\pm(p_2)\right]\overline{\left\vert\mathcal{M}\right\vert^2}\delta^{(4)}\left(p_a +p_b - p_1 - p_2\right)\\
&\times \delta(\omega-E_a+E_1) \delta\left\{t+2\left[E_aE_1-\cos\left(\theta_{1a}\right)\right]\right\}\\
&\times \delta\left\{s-2\left[E_aE_b-\cos\left(\theta_{ba}\right)\right]\right\}.  
\end{split}
\end{align}
Following multiple integrations, one obtains 
\begin{align}
\frac{d^3R}{d^3p_a} &= \frac{1}{16E_a^2\left(2\pi\right)^7}\int dE_b f_\pm(p_b)\int ds\int dt\overline{\left\vert\mathcal{M}\right\vert^2}(s,t)\nonumber\\
&\times\int_{\omega_\text{min}}^{\omega_\text{max}}d\omega\frac{\left[1 \mp f_\pm(p_2)\right] }{\sqrt{c+b\omega-a^2\omega^2}},
\end{align}
where the Jacobian from the $\delta$-functions gives
\begin{align}
a&=-s^2,\\
b&=2st\left[E_b-E_a\right],\\
c&=t\left[s^2+st-\left(E_a+E_b\right)^2 t-4E_aE_bs\right],
\end{align}
while the integration limits are given by the Jacobian's roots as
\begin{align}
\omega_\text{min}^\text{max}=\frac{b\pm\sqrt{b^2+4a^2c}}{2a^2},
\end{align}
which, as shown in Ref.~\cite{Peigne:2008nd}, follow from the last $\delta$-function constraints. To obtain a closed-form expression, the first term in Eq.~(\ref{eq:BE_geom}) is employed, thus giving
\begin{align}
\frac{d^3R}{d^3p_a} &\simeq \frac{1}{32E_a^2\left(2\pi\right)^6}\int_0^\infty dE_b f_\pm(E_b)\int_{s_-}^{s_+}\frac{ds}{s}\nonumber\\
&\times\int^{t_{\rm max}}_{t_{\rm min}}  dt \overline{\left\vert\mathcal{M}\right\vert^2}(s,t)+O\left(e^{-x_2}\right).    
\end{align}
The matrix element  
%
for $gg\to gg$ is now incorporated, along with the limits devised by Ref.~\cite{Peigne:2008nd}, namely
\begin{align}
    t_{\min}&=\begin{cases}
        m_D^2-s,\quad&\text{when necessary (for }u\text{-terms)}\\
        -s,\quad&\text{else}
    \end{cases}\\
    t_{\max}&=\begin{cases}
        -m_D^2,\quad&\text{when necessary (for }t\text{-terms)}\\
        0,\quad&\text{else}
    \end{cases}\\
    s_\pm&=\begin{cases}
        0,\quad&\text{for }s_-\\
        4E_aE_b,\quad&\text{for }s_+
    \end{cases}
\end{align}
where the divergence at $t_{\max}\to0$ is cut off by letting $t_{\max}\to-m_D^2$. Doing so generates
\begin{align}
\frac{d^3R_c}{d^3p_a}&\simeq\frac{9\alpha_s^2T}{8\pi^2x_a}+O\left(e^{-x_2}\right)\nonumber\\
\frac{d^3R_s}{d^3p_a}&\simeq-\frac{\alpha_s^2T}{16x_a\pi^2}+O\left(e^{-x_2}\right)\\
\frac{d^3R_t}{d^3p_a}&\simeq\frac{9\alpha_s^2T}{\pi^4}\bigg[\frac{\zeta(3)}{z^2}-\frac{\pi^2}{24x_a}\ln\left(\frac{4x_a}{z^2}\right)-\frac{\zeta'(2)}{4x_a}\nonumber\\
&\qquad\qquad-\frac{(1-\gamma)\pi^2}{24x_a}\bigg]+O\left(e^{-x_2}\right),\nonumber
\end{align}
for the non-HTL (hard) contribution, where $\gamma$ is the Euler–Mascheroni constant, and $\zeta'$ is the first derivative of the Riemann zeta function \cite{Cohl:2014drm}. Note that the $u$-channel gives the same expression as the $t$-channel. Together, the latter two give:
\begin{align}
\frac{d^3R_{t+u}}{d^3p_a}&\simeq\frac{9\alpha_s^2T}{\pi^4}\bigg[\frac{2\zeta(3)}{z^2}-\frac{\pi^2}{12x_a}\ln\left(\frac{4x_a}{z^2}\right)-\frac{\zeta'(2)}{2x_a}\nonumber\\
&\qquad\qquad-\frac{(1-\gamma)\pi^2}{12x_a}\bigg]+O\left(e^{-x_2}\right),
\end{align}
while the total rate, after adding four-point and $s$-term (which match our results in Eq.~(\ref{eq:rate_c_no_Bose_enh}) and Eq.~(\ref{eq:rate_s_approx})), is
\begin{align}
\frac{d^3R}{d^3p_a} &\simeq \frac{9\alpha_s^2T}{\pi^4}\bigg[\frac{2\zeta(3)}{z^2}-\frac{\pi^2}{12x_a}\ln\left(\frac{4x_a}{z^2}\right)-\frac{\zeta'(2)}{2x_a}\nonumber\\
&\qquad\qquad-\frac{(1-\gamma)\pi^2}{12x_a}\bigg]+\frac{17\alpha_s^2T}{16\pi^2x_a}+O\left(e^{-x_2}\right),
\label{eq:P&P_rate_w_c_s}
\end{align}
The Braaten \& Thoma limit \cite{Braaten:1991jj, Braaten:1991we} is obtained by solely keeping the double pole terms in Mandelstam variables present within the matrix element, yielding
\begin{align}
\frac{d^3R_{t+u}}{d^3 p_a}&\simeq\frac{9\alpha_s^2T}{\pi^4}\left[\frac{2\zeta(3)}{z^2}-\frac{\pi^2}{12x_a}\right].  
\end{align}
While the Braaten \& Thoma limit completely neglects four-point and $s$-term (since those don't contain any poles), adding our (non-Bose-enhanced) results from Eq.~(\ref{eq:rate_c_no_Bose_enh}) and Eq.~(\ref{eq:rate_s_approx}) gives
\begin{align}
\frac{d^3 R}{d^3 p_a}&\simeq\frac{9\alpha_s^2T}{\pi^4}\left[\frac{2\zeta(3)}{z^2}-\frac{\pi^2}{12x_a}\right]+\frac{17\alpha_s^2T}{16\pi^2x_a}.
\label{eq:BT_w_c_s}
\end{align}
%
\subsubsection{Comparisons between all approaches}
\begin{figure}[h]
    \centering
	\includegraphics[width=\linewidth]{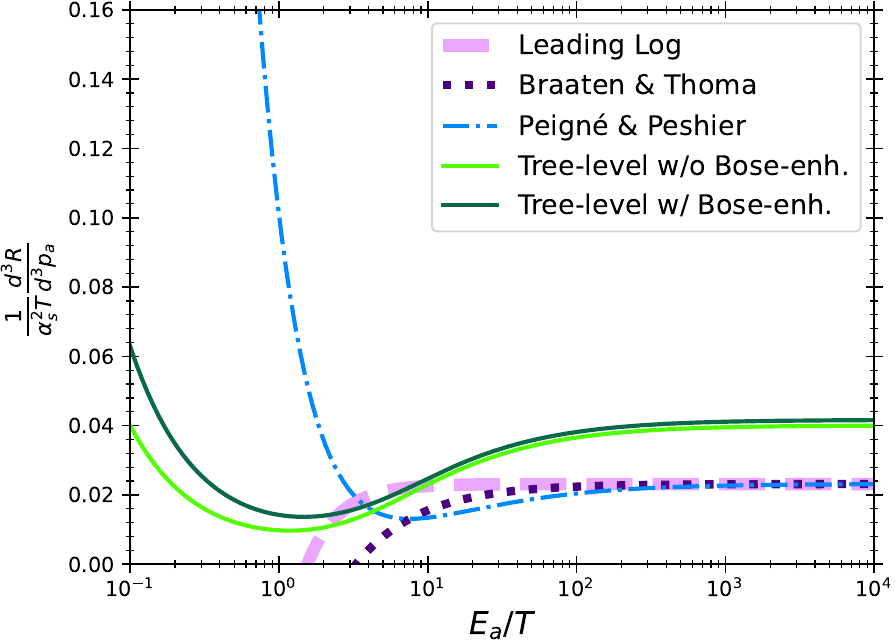}
    \caption{Comparison of different approximations to the tree-level rate from $t$ and $u$ scattering channels, with $m_D$ given by Eq.~(\ref{eq:mD_w_running}) and evaluated at $T=0.2$ GeV.}
    \label{fig:rate_approx_comp}
\end{figure}

A comparison of all approaches is given in Fig.~\ref{fig:rate_approx_comp} where solely the $t$- and $u$-channel contributions to the rate are included, while the cut-off momentum is given by Eq.~(\ref{eq:mD_w_running}). A comparison of the same approaches for a $z$ cut-off where $\alpha_s=0.3$ is used within Eq.~(\ref{eq:Debye_mass}) can be found in Ref.~\cite{short_paper}.

\begin{figure}[H]
    \centering
	\includegraphics[width=\linewidth]{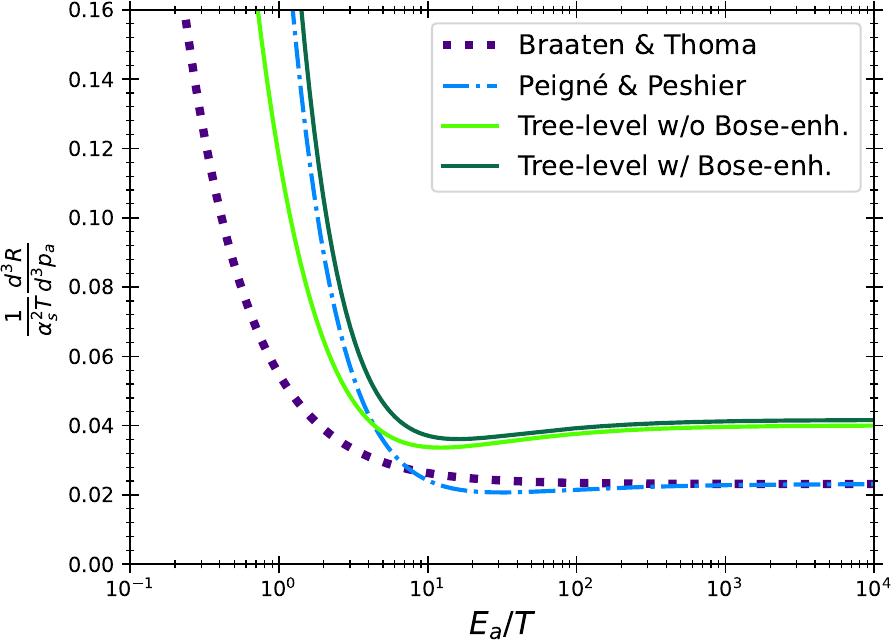}
    \caption{Comparison of different approximations to the tree-level rate including all four scattering channels, with $m_D$ given by Eq.~(\ref{eq:mD_w_running}) and evaluated at $T=0.2$ GeV.}
    \label{fig:all_channels_comp}
\end{figure}
For large $x_a$, the sizable discrepancy between our result and that of the leading-log result, the Braaten \& Thoma, as well as the  Peign\'e \& Peshier result, is therefore not stemming from the manner in which the Debye mass cut-off is computed. Instead, the kinematic approximation associated with evaluating $A^-_1(z)$ that other approaches use is the main reason for this discrepancy in the extreme UV regime, as shown in Fig.~\ref{fig:Struve}. Of course, approximations on $B^-_1(z)$ in the rate play less of a role in the extreme UV regime. Including the two remaining channels contributing to $gg\to gg$ scattering in Eqs.~(\ref{eq:P&P_rate_w_c_s}, \ref{eq:BT_w_c_s}), generates a rate whose shape resembles more the  Peign\'e \& Peshier result, as shown in Fig.~\ref{fig:all_channels_comp}, though there are still significant discrepancies. 

Computing the difference between various approaches explored herein constitutes a way of devising the theoretical systematic uncertainty to be included in jet-medium Bayesian analyses. Similarly, theoretical systematic uncertainties in Bayesian analysis can be included via model averaging as explained in Ref.~\cite{JETSCAPE:2020shq}.

\section{Jet-medium Transport Coefficients}
\label{sec:qhat_ehat}
Starting with the usual scattering rate
\begin{align}
\frac{d^3R}{d^3p_a}&=\frac{1}{16E_a\left(2\pi\right)^8}\prod_{i=b,1,2}\left[\int\frac{d^3p_i}{E_i}\right]\int d^4q\,\overline{\lvert\mathcal{M}\rvert^2}\,\nonumber\\
&\times \delta^{(4)}\left(p_a-p_1-q\right)\delta^{(4)}\left(p_b-p_2+q\right)\, \nonumber\\
&\times f_\pm(p_b)\left[1\mp f_\pm(p_2)\right],
\end{align}
the  non eikonalized jet-medium transport coefficients often used in jet-medium Monte Carlo simulations, such as Ref.~\cite{JETSCAPE:2024cqe}, take the form
\begin{align}
\hat{O}(p_a,h(q))&\equiv\int d^4q\frac{d^7R}{d^3p_a\,d^4q}h(q).
\end{align}
Mandelstam variables are included within the rate using the known \cite{Kapusta:2006pm} identities $1=\int ds\delta(s-(p_a+p_b)^2)$ and $1=\int dt\delta(t-(p_b-p_2)^2)$ to obtain
\begin{align}
\frac{d^7R}{d^3p_a\,d^4q}&=\frac{1}{16E_a\left(2\pi\right)^8}\int\frac{ds\;dt\;d^3p_b\;d^3p_1\;d^3p_2}{E_bE_1E_2}\,\overline{\lvert\mathcal{M}\rvert^2}\,\nonumber\\
&\times f_\pm(p_b)\left[1\mp f_\pm\left(p_2\right)\right]\delta^{(4)}\left(p_a-p_1-q\right)\nonumber\\
&\times\delta^{(4)}(p_b-p_2+q)\delta\left(s-\left(p_a+p_b\right)^2\right)\nonumber\\
&\times\delta\left(t-\left(p_b-p_2\right)^2\right).
\end{align}
This allows to simplify the computation of jet-medium transport coefficients, and after some algebra, one obtains
\begin{align}\label{OHat}
\hat{O}(p_a,h(q))&=\frac{1}{16\left(2\pi\right)^7\lvert\vec{p}_a\rvert E_a}\int ds\;dt\;dE_b\;dE_2\nonumber\\
&\times\frac{h(p_2-p_b)f_\pm(p_b)\left[1\mp f_\pm\left(p_2\right)\right]\overline{\lvert\mathcal{M}\rvert^2}(s,t)}{\sqrt{\Lambda(s,t,E_b,E_2)}}
\end{align}
where $h(q)=h(p_2-p_b)$. A jet-medium transport coefficients of interest is
\begin{align}
\hat{q}=\hat{O}(p_a,q_\perp^2)=\frac{1}{16\left(2\pi\right)^7\lvert\vec{p}_a\rvert E_a}\int ds\;dt\;dE_b\;dE_2\nonumber\\
\times\frac{q_\perp^2f_\pm(p_b)\left[1\mp f_\pm\left(p_2\right)\right]\overline{\lvert\mathcal{M}\rvert^2}(s,t)}{\sqrt{\Lambda(s,t,E_b,E_2)}}
\label{eq:Ohat_q}
\end{align}
describing transverse momentum broadening of the partons. The exchanged transverse momentum $q_\perp$ can be expressed in terms of Mandelstam variables via
\begin{align}
q_\perp^2&=-\frac{t^2+4E_aE_1t}{4E_a^2}
\end{align}
for gluon-gluon scattering. Another jet-medium transport coefficient of interest is
\begin{align}
\hat{e}=\hat{O}(p_a,\omega)&=\frac{1}{16\left(2\pi\right)^7\lvert\vec{p}_a\rvert E_a}\int ds\;dt\;dE_b\;dE_2\nonumber\\
&\times\frac{\omega f_\pm(p_b)\left[1\mp f_\pm\left(p_2\right)\right]\overline{\lvert\mathcal{M}\rvert^2}(s,t)}{\sqrt{\Lambda(s,t,E_b,E_2)}}.
\label{eq:Ohat_e}
\end{align}
For any other moment that involve powers of the components of $q^\mu$, one keeps in mind that
\begin{align}
q^\mu&=p^\mu_2-p^\mu_b=(E_2-E_b,\vec{p}_2-\vec{p}_b),\\
q^2&=\omega^2-\vert\vec{q}\vert\,^2=t,
\end{align}
which are enforced by energy-momentum conservation. The eikonalized version of jet-medium transport coefficients \cite{Ghiglieri:2015ala} are defined instead as 
\begin{align}
\hat{O}(h(q))&\equiv\int d^4q\lim_{\lvert p_a\rvert\to\infty}\left[\frac{d^7R}{d^3p_a\,d^4q}\right]h(q).
\end{align}
Unlike calculations based on the twist expansion \cite{Sirimanna:2021sqx,Kumar:2025egh,Kumar:2025asj} which compute transport coefficients using $t$-channel interactions when restricted to tree-level interactions, the eikonalized counterpart typically resums over all channels.

\subsection{Leading logarithm approximation}
Using the $t$-channel of the scattering rate, the jet broadening coefficient $\hat q_t$ is obtained via
\begin{align}
\hat q_t&\simeq \frac{288\alpha_s^2\left(2\pi\right)^2 }{\left(2\pi\right)^6 16E_a^2}\int^{\sqrt{2E_aT}}_{m_D} dq_{\perp} q_{\perp}  q^2_\perp \int^\infty_0 dE_b f(E_b)\nonumber\\
&\times\int^1_{-1} \frac{d[\cos(\theta_{ab})]}{\lvert 1-\cos\left(\theta_{ab}\right)\rvert}\left[\frac{\left\{2E_aE_b\left[1-\cos\left(\theta_{ba}\right)\right]\right\}^2}{q^4_\perp}\right]\nonumber\\
&\simeq \frac{9\alpha^2_s}{\pi^4}\ln\left[\frac{2x_a}{z^2}\right]\zeta(3)T^3.
\end{align}
\begin{widetext}
Now one sees why the boundaries of the integral over $q_\perp$ are so chosen: to obtain $\hat q\propto \ln\left[\frac{2E_a T}{m^2_D}\right]$. In this limit, the $u$-channel contribution to the matrix element is exactly the same, thus $\hat q_u=\hat q_t$ at leading logarithm accuracy.

Computing $\hat e$ in this limit yields
\begin{align}
\hat e &= \frac{1}{\left(2\pi\right)^8 2E_a}\int d^4q\, \omega\, \frac{d^3p_b}{2E_b}f\left(E_b\right)\left[1\pm f\left(E_{b} +\omega\right)\right]\overline{|\mathcal{M}|^2}\times\Theta\left(E_{a} - \omega\right)\Theta\left(E_{b} + \omega\right)\delta\left[\left(p_a - q\right)^2\right]\delta\left[\left(q + p_b\right)^2\right].
\end{align}

Using the first term in Eq.~(\ref{eq:BE_geom}) generates
\begin{align}
\hat e&\simeq\frac{1}{32E_a^2\left(2\pi\right)^{7}}\int_{0}^{\infty}\frac{dE_b}{E_b}f(E_b)\int_{m_D}^{\sqrt{2E_aT}} q_\perp dq_\perp \int_{-1}^{1}d(\cos\theta_{ab})\int_{0}^{2\pi}d(\Delta\phi)\,\overline{\left\vert\mathcal{M}\right\vert^2}\left[\frac{q_\perp^2 + 2E_b q_\perp\sin\theta_{ba}\cos(\Delta\phi)}{\left(1 - \cos\theta_{ba}\right)^2}\right]\nonumber
\end{align}
where $q^*_z\approx\omega^*$ has been used, as  $E_a\gg E_b\gg \omega$. After simplication, the above expression yields
\begin{align}
\hat e_t &\simeq \frac{72\alpha_s^2}{\left(2\pi\right)^4}\int_{0}^{\infty} dE_b E_b f(E_b) \int_{m_D}^{\sqrt{2E_aT}} \frac{dq_\perp}{q_\perp}=\frac{9 \alpha^2_s T^2}{4\pi^4}\left[\frac{\pi^2}{6}\right]\ln\left[\frac{2x_a}{z^2}\right]
\end{align}
%
\subsection{Complete tree-level transport coefficients}
Computation is facilitated (see Appendix~\ref{sec:Mandelstam_int}) by setting
\begin{align}
\lfloor E\rfloor\equiv&\min(E_a,E_b,E_1,E_2),\lfloor s\rfloor\equiv\min(E_aE_b,E_1E_2),\nonumber\\
\lfloor t\rfloor\equiv&\min(E_aE_1,E_bE_2), \,\,\,\,\,\lfloor u\rfloor\equiv\min(E_aE_2,E_bE_1),
\end{align}
where $E_1=E_a+E_b-E_2$. These definitions allow to perform the integral in Eq.~(\ref{eq:Ohat_q}) using the boundaries in Eq.~(\ref{eq:bdy_Lambda})  and Eq.~(\ref{eq:bdy_st}). Thus, one obtains for the four-point vertex channel of $gg\to gg$
\begin{align}
\hat{q}_c&=\frac{9\alpha_s^2}{2\pi^4E_a^4}\int_0^\infty\frac{dE_b}{e^{\beta E_b}-1}\int_0^{E_a+E_b}\frac{dE_2}{1-e^{-\beta E_2}}\lfloor E\rfloor\left[\lfloor E\rfloor^2\lfloor t\rfloor-2\lfloor t\rfloor^2-\frac{\lfloor E\rfloor^4}{5}+E_aE_1\left(3\lfloor t\rfloor-\lfloor E\rfloor^2\right)\right],\nonumber\\
&=\frac{9\alpha_s^2T^3}{2\pi^4x_a^4}\int_0^\infty\frac{dx_b}{e^{x_b}-1}\Bigg\{24\left[\polylog{6}\left(e^{-x_a}\right)-\polylog{6}\left(e^{-x_a-x_b}\right)\right]+\left[24\polylog{6}\left(e^{-x_b}\right)-\frac{8\pi^6}{315}\right]+24x_a\left[\zeta(5)-\polylog{5}\left(e^{-x_b}\right)\right]\nonumber\\
&+24x_b\left[\zeta(5)-\polylog{5}\left(e^{-x_a}\right)\right]+x_a^2\left[6\polylog{4}\left(e^{-x_b}\right)-\frac{\pi^4}{15}\right]+6x_a^2\left[\polylog{4}\left(e^{-x_a-x_b}\right)-\polylog{4}\left(e^{-x_a}\right)\right]+2x_b^2\left[6\polylog{4}\left(e^{-x_a}\right)-\frac{\pi^4}{15}\right]\nonumber\\
&+3x_ax_b\left(x_a+x_b\right)\left[\polylog{3}\left(e^{-x_a}\right)+2\zeta(3)\right]-\frac{4x_ax_b\pi^4}{15}+\frac{x_a^3x_b^2\left(3x_a+2x_b\right)}{6}\Bigg\},\nonumber\\
&=\frac{9\alpha_s^2T^3}{2\pi^4x_a^4}\left\{4\pi^2\left[2\zeta(5)+\zeta(3)\frac{\pi^2}{15}-\polylog{5}\left(e^{-x_a}\right)\right]-96\zeta(7)+4\zeta(3)\left[\polylog{4}\left(e^{-x_a}\right)-\frac{\pi^4}{15}\right]\right.\nonumber\\
&\qquad\qquad\quad+3x_a\left[\frac{\pi^2}{6}+2\zeta(3)\right]\polylog{3}\left(e^{-x_a}\right)+\frac{\pi^2 x_a}{2}+\Big[\pi^2\zeta(3)-18\zeta(5)\Big]x^2_a+\frac{\pi^4x_a^3}{45}+x_a^4\zeta(3)\nonumber\\
&\qquad\qquad\quad\left.+\int_0^\infty \frac{dx_b}{e^{x_b}-1}\Bigg[24\left[\polylog{6}\left(e^{-x_a}\right)-\polylog{6}\left(e^{-x_a-x_b}\right)\right]+\left[\polylog{4}\left(e^{-x_a-x_b}\right)-\polylog{4}\left(e^{-x_a}\right)\right]\Bigg]\right\}\nonumber\\
&=\frac{9\alpha_s^2T^3}{2\pi^4x_a^4}\left\{4\pi^2\left[2\zeta(5)+\zeta(3)\frac{\pi^2}{15}-\polylog{5}\left(e^{-x_a}\right)\right]-96\zeta(7)+4\zeta(3)\left[\polylog{4}\left(e^{-x_a}\right)-\frac{\pi^4}{15}\right]\right.\nonumber\\
&\qquad\qquad\quad+3x_a\left[\frac{\pi^2}{6}+2\zeta(3)\right]\polylog{3}\left(e^{-x_a}\right)+\frac{\pi^2 x_a}{2}+\Big[\pi^2\zeta(3)-18\zeta(5)-\polylog{5}\left(e^{-x_a}\right)-\polylog{4,1}\left(e^{-x_a},1\right)\Big]x^2_a\nonumber\\
&\qquad\qquad\quad\left.+\frac{\pi^4x_a^3}{45}+x_a^4\zeta(3)+24\Big[\polylog{7}\left(e^{-x_a}\right)+\polylog{6,1}\left(e^{-x_a},1\right)\Big]\right\},
\label{eq:qhat_c}
\end{align}
where the derivation of the term involving the multiple polylogarithms, e.g. $\polylog{6,1}\left(e^{-x_a},1\right)$, used to go from the second to the last step is given in Appendix~\ref{apnd:multi_Li_series}. The $s$-channel of $\hat q$ gives
\begin{align}
    \hat{q}_s&=\frac{3\alpha_s^2}{2\pi^4E_a^4}\int_0^\infty\frac{dE_b}{e^{\beta E_b}-1}\int_0^{E_a+E_b}\frac{dE_2}{1-e^{-\beta E_2}}\lfloor t\rfloor\frac{\lfloor t\rfloor^2(E_aE_1-\lfloor t\rfloor)+E_aE_bE_1E_2(2\lfloor t\rfloor-3E_aE_1)}{\left(E_a+E_b\right)^3}\nonumber\\
    &=\frac{3\alpha_s^2T^3}{2\pi^4x_a^4}\int_0^\infty\frac{dx_b}{e^{x_b}-1}\left[24x_b\polylog{5}(e^{-x_a})-\frac{24x_a^3x_b}{(x_a+x_b)^3}\polylog{5}(e^{-x_a-x_b})+6x_ax_b\polylog{4}(e^{-x_a})-\frac{6x_a^3x_b}{(x_a+x_b)^2}\polylog{4}(e^{-x_a-x_b})\right.\nonumber\\
    &\left.-\frac{24x_b^2\left(3x_a^2+3x_ax_b+x_b^2\right)\zeta(5)}{\left(x_a+x_b\right)^3}-\frac{6x_a^2x_b^2\zeta(3)}{x_a+x_b}+\frac{x_ax_b^2\left(2x_a^2+3x_ax_b+x_b^2\right)\pi^4}{5\left(x_a+x_b\right)^3}-\frac{x_a^3x_b^2\left(2x_a+x_b\right)}{20}\right]\\
    &=\frac{3\alpha_s^2T^3}{2\pi^4x_a^4}\left\{4\pi^2\polylog{5}\left(e^{-x_a}\right)+\pi^2x_a\polylog{4}\left(e^{-x_a}\right)-\frac{x_a^4\zeta(3)}{5}-\frac{\pi^4x_a^3}{300}-\int_0^\infty\frac{dx_b}{e^{x_b}-1}\left[\frac{24x_a^3x_b}{\left(x_a+x_b\right)^3}\polylog{5}\left(e^{-x_a-x_b}\right)\right.\right.\nonumber\\
    &\left.\left.+\frac{6x_a^3x_b}{\left(x_a+x_b\right)^2}\polylog{4}\left(e^{-x_a-x_b}\right)+\frac{24x_b^2\left(3x_a^2+3x_ax_b+x_b^2\right)\zeta(5)}{\left(x_a+x_b\right)^3}+\frac{6x_a^2x_b^2\zeta(3)}{x_a+x_b}-\frac{x_ax_b^2\left(2x_a^2+3x_ax_b+x_b^2\right)\pi^4}{5\left(x_a+x_b\right)^3}\right]\right\},\nonumber
\end{align}
where the remaining integral can be expressed in terms of a series involving $\zeta$-functions and multiple polylogarithms, following the procedure in Eq.~(\ref{eq:binomial_zeta_series}) for the former, with details in Appendix~\ref{apnd:multi_Li_series} for the latter. Neglecting Bose-enhancement by using the first term in Eq.~(\ref{eq:BE_geom}) yields
\begin{align}
    \left.\hat{q}\right|_c&\simeq\frac{9\alpha_s^2T^3}{2\pi^4}\left[\zeta(3)+\frac{\pi^4}{45x_a}\right]+O\left(e^{-x_2}\right),\\
    \left.\hat{q}\right|_s&\simeq-\frac{3\alpha_s^2T^3}{2\pi^4}\left[\frac{\zeta(3)}{5}+\frac{\pi^4}{300x_a}\right]+O\left(e^{-x_2}\right).
\end{align}
Finally, the $u$- and $t$-channel yield
\begin{align}
\hat{q}_t&=\frac{3\alpha_s^2}{2\pi^4E_a^4}\int_0^\infty\frac{dE_b}{e^{\beta E_b}-1}\int_0^{E_a+E_b}\frac{dE_2}{1-e^{-\beta E_2}}\left\{\frac{3E_aE_1\lfloor s\rfloor^2}{\sqrt{\left(E_b-E_2\right)^2+m_D^2}}\right.\nonumber\\
&\left.+\lfloor E\rfloor\left[2\lfloor s\rfloor\left(\lfloor t\rfloor-\lfloor s\rfloor-\lfloor E\rfloor^2\right)+\lfloor E\rfloor^2\left(\lfloor t\rfloor-E_aE_1\right)-\frac{3\lfloor E\rfloor^4}{5}-3E_aE_1\lfloor s\rfloor\right]\right\},\label{eq:qhat_t_BE}\\
\hat{q}_u&=\frac{3\alpha_s^2}{2\pi^4E_a^4}\int_0^\infty\frac{dE_b}{e^{\beta E_b}-1}\int_0^{E_a+E_b}\frac{dE_2}{1-e^{-\beta E_2}}\lfloor t\rfloor\frac{\lfloor t\rfloor^2\left(\lfloor t\rfloor-E_aE_1\right)+E_aE_bE_1E_2\left(3E_aE_1-2\lfloor t\rfloor\right)}{\left[\left(E_a-E_2\right)^2+m_D^2\right]^\frac{3}{2}}.
\label{eq:qhat_u_BE}
\end{align}
Unlike the previous two channels, the exact integral over $E_2$ in the $t$- and $u$-channels cannot be easily performed owing to the square root in the denominator, thus leaving the associated $\hat q_t$ and $\hat q_u$ as they stand. In the limit where Bose-enhancement effects can be neglected, one obtains
\begin{align}
\hat{q}_u&\simeq\frac{9\alpha_s^2T^3}{2\pi^4}\left\{\left[\frac{8\zeta(5)}{z^2x_a}-\frac{5\zeta(3)}{x_a}-\frac{\pi^4}{30x^2_a}-\frac{4\zeta(5)}{x^3_a}\right]\sqrt{x^2_a+z^2}+\left[2\zeta(3)-\frac{\pi^4}{15x_a}-\frac{3\zeta(3)z^2}{x^2_a}-\frac{\pi^4z^2}{10x^3_a}-\frac{12\zeta(5)z^2}{x^4_a}\right]\arcsinh\left(\frac{x_a}{z}\right)\right.\nonumber\\
&\left.+\frac{A^-_3(z)}{3z^2}-\frac{2}{3}A^-_1(z)+\frac{3}{2x_a}A^-_2(z)+z\left[\frac{\pi^2}{9}+\frac{4\zeta(3)}{x_a}+\frac{2\pi^4}{15x^2_a}+\frac{16\zeta(5)}{x^3_a}\right]+B^-_2(z)-\frac{B^-_3(z)}{x_a}+\frac{z^2B^-_1(z)}{2x_a}\right\}+O\left(e^{-x_2}\right).
\label{eq:qhat_u}
\end{align}
For the $t$-channel, important intermediate steps for obtaining $\hat q_t$ are presented in Appendix~\ref{sec:q_hat_ints}. Ultimately the $t$-channel contribution gives
\begin{align}
\hat{q}_t&\simeq\frac{9\alpha_s^2T^3}{2\pi^4}\left\{\left[\frac{\pi^2}{12}-\frac{11\zeta(3)}{3x_a}-\frac{13\pi^2z^2}{24x^2_a}+\frac{4\zeta(3)z^2}{3x^3_a}\right]\sqrt{x^2_a+z^2}+\left[2\zeta(3)+\frac{\pi^2z^2}{2x_a}-\frac{3\zeta(3)z^2}{x^2_a}-\frac{\pi^2 z^4}{8x^3_a}\right]\arcsinh\left(\frac{x_a}{z}\right)-\zeta(3)\right.\nonumber\\
&\left.-\frac{\pi^4}{18x_a}+B^-_2(z)+\frac{3z^2}{x_a}B^-_1(z)-\frac{3z^2}{2x^2_a}B^-_2(z)-\frac{3z^4}{4x^3_a}B^-_1(z)-\frac{3A^-_1(z)}{2}+\frac{A^-_2(z)}{x_a}+\frac{A^-_3(z)}{4x^2_a}+\frac{23z^2}{8x^2_a}A^-_1(z)+\frac{A^-_4(z)}{30x^3_a}-\frac{11z^2}{60x^3_a}A^-_2(z)\right.\nonumber\\
&\left.-\left[\frac{z^2}{2}-\frac{9z^4}{8x^2_a}\right]B^-_0(z)+z\left[2g^-_1(x_a)\left(1-\frac{2z^2}{x^2_a}\right)-\frac{2g^-_2(x_a)}{x_a}\left(1-\frac{z^2}{3x^2_a}\right)\right]-z^2\left[\frac{3}{x_a}\Delta\beta^-_1-\frac{3}{2x^2_a}\Delta\beta^-_2-\frac{3z^2}{4x^3_a}\Delta\beta^-_1\right]\right.\nonumber\\
&\left.-\frac{11}{20}\Delta\alpha^-_1+\frac{17}{60x_a}\Delta\alpha^-_2+\frac{17}{60x^2_a}\Delta\alpha^-_3+\frac{319z^2}{120x^2_a}\Delta\alpha^-_1+\frac{\Delta\alpha^-_4}{30x^3_a}-\frac{11z^2}{60x^3_a}\Delta\alpha^-_2+\left[\frac{z^2}{2}-\frac{9z^2}{8x^2_a}\right]\lim_{\epsilon\to0^+}\left[\Delta\beta^-_0(\epsilon)+\arcsinh\left(\frac{x_a}{z}\right)\ln(\epsilon)\right]\right.\nonumber\\
&\left.-\left[\frac{2z^2}{x_a}-\frac{8z^4}{15x^3_a}\right]\left[\lim_{\epsilon\to0^+}\Big[z\ln\left(\epsilon\right)+A^-_0\left(\epsilon;z\right)\Big]-2z\ln\left(1-e^{-x_a}\right)\right]-\left[\frac{x_a}{20}+\frac{137z^2}{120x_a}-\frac{8z^4}{15x^3_a}\right]\lim_{\epsilon\to0^+}\left[\Delta\alpha^-_0(\epsilon)+\sqrt{x^2_a+z^2}\ln\left(\epsilon\right)\right]\right\}\nonumber\\
&+\,\,O\left(e^{-x_2}\right),\label{eq:qhat_t}
\end{align}

where
\begin{align}
g^-_1\left(x_a\right)&=2x_a\ln\left(1-e^{-x_a}\right)-2\polylog{2}\left(e^{-x_a}\right)+\frac{\pi^2}{6},\\
g^-_2(x_a)&=2\zeta(3)+2x^2_a\ln\left(1-e^{-x_a}\right)-4x_a\polylog{2}\left(e^{-x_a}\right)-4\polylog{3}\left(e^{-x_a}\right).\nonumber
\end{align}
To obtain $\hat e$ in Eq.~(\ref{eq:Ohat_e}) is easier than $\hat q$, owing to the fact that $\omega=E_2-E_b$. The integrals over Mandelstam variables can immediately be performed, thus giving 
\begin{align}
\begin{split}
\hat{e}&=\frac{9\alpha_s^2}{4\pi^4E_a^2}\int_0^\infty\frac{dE_b}{e^{\beta E_b}-1}\int_0^{E_a+E_b}\frac{dE_2}{1-e^{-\beta E_2}}(E_2-E_b)\\
&\times\left[3\min(E_a,E_b,E_1,E_2)-\frac{E_aE_bE_1E_2}{\left(E_a+E_2\right)^3}+\frac{E_aE_bE_1E_2}{\left[\left(E_b-E_2\right)^2+m_D^2\right]^\frac{3}{2}}+\frac{E_aE_bE_1E_2}{\left[\left(E_a-E_2\right)^2+m_D^2\right]^\frac{3}{2}}\right],
\label{eq:ehat_BE}
\end{split}
\end{align}
where $E_1=E_a+E_b-E_2$. Except for the additional $E_2-E_b$ factor, all other terms are exactly the same as for the rate. Separating the above expression into individual channel contributions generates
\begin{align}
\hat{e}_t&=\frac{9\alpha_s^2}{4\pi^4E_a^2}\int_0^\infty\frac{dE_b}{e^{\beta E_b}-1}\int_0^{E_a+E_b}\frac{dE_2}{1-e^{-\beta E_2}}\frac{\left(E_2-E_b\right)E_aE_bE_1E_2}{\left[\left(E_b-E_2\right)^2+m_D^2\right]^\frac{3}{2}}\label{eq:ehat_t_BE}\\
\hat{e}_u&=\frac{9\alpha_s^2}{4\pi^4E_a^2}\int_0^\infty\frac{dE_b}{e^{\beta E_b}-1}\int_0^{E_a+E_b}\frac{dE_2}{1-e^{-\beta E_2}}\frac{(E_2-E_b)E_aE_bE_1E_2}{\left[\left(E_a-E_2\right)^2+m_D^2\right]^\frac{3}{2}}\label{eq:ehat_u_BE}
\end{align}
both of which do not admit simpler expressions. The Bose-Enhanced four-point vertex contribution, using Appendix~\ref{apnd:multi_Li_series}, gives
\begin{align}
\hat{e}_c&=\frac{27\alpha_s^2T^2}{4\pi^4x_a^2}\int_0^\infty\frac{dx_b}{e^{x_b}-1}\Bigg[2\left[\polylog{3}\left(e^{-x_a-x_b}\right)-\polylog{3}\left(e^{-x_a}\right)\right]+2\left[\zeta(3)-\polylog{3}\left(e^{-x_b}\right)\right]+x_a\left[\polylog{2}(e^{-x_a-x_b})-\polylog{2}\left(e^{-x_a}\right)\right]\nonumber\\
&\qquad\qquad\qquad\qquad\qquad\quad+x_b\left[\polylog{2}\left(e^{-x_a}\right)-\frac{\pi^2}{6}\right]+\frac{x_ax_b(x_a-x_b)}{2}\Bigg]\nonumber\\
&=\frac{27\alpha_s^2T^2}{4\pi^4x_a^2}\left\{\frac{x_a^2\pi^2}{12}+\frac{\pi^2}{6}\polylog{2}\left(e^{-x_a}\right)+x_a\left[\polylog{3}\left(e^{-x_a}\right)-\polylog{3}\left(1-e^{-x_a}\right)-2\zeta(3)-\polylog{2}\left(1-e^{-x_a}\right)+\frac{x_a}{2}\ln^2\left(1-e^{-x_a}\right)\right]\right.\nonumber\\
&\qquad\qquad\qquad\left.+2\int_0^\infty \frac{dx_b}{e^{x_b}-1}\Big[\polylog{3}\left(e^{-x_a-x_b}\right)-\polylog{3}\left(e^{-x_a}\right)\Big]\right\}\nonumber\\
&=\frac{27\alpha_s^2T^2}{4\pi^4x_a^2}\left\{\frac{x_a^2\pi^2}{12}+\frac{\pi^2}{6}\polylog{2}\left(e^{-x_a}\right)+x_a\left[\polylog{3}\left(e^{-x_a}\right)-\polylog{3}\left(1-e^{-x_a}\right)-2\zeta(3)-\polylog{2}\left(1-e^{-x_a}\right)+\frac{x_a}{2}\ln^2\left(1-e^{-x_a}\right)\right]\right.\nonumber\\
&\qquad\qquad\qquad\left.-2\Big[\polylog{4}\left(e^{-x_a}\right)+\polylog{3,1}\left(e^{-x_a},1\right)\Big]\right\}.
\label{eq:ehat_c}
\end{align}
Similarly, the $s$-channel is
\begin{align}
\hat{e}_s&=-\frac{9\alpha_s^2T^2}{4\pi^4x_a^2}\int_0^\infty\frac{dx_b}{e^{x_b}-1}\left[\frac{6x_ax_b}{\left(x_a+x_b\right)^3}\polylog{4}\left(e^{-x_a-x_b}\right)+\frac{2x_ax_b\left(2x_a+x_b\right)}{\left(x_a+x_b\right)^3}\polylog{3}\left(e^{-x_a-x_b}\right)+\frac{x_a^2x_b\left(x_a+x_b\right)}{\left(x_a+x_b\right)^3}\polylog{2}\left(e^{-x_a-x_b}\right)\right.\nonumber\\
&\left.\qquad\qquad\qquad\qquad\qquad\qquad-\frac{x_ax_b\pi^4}{15(x_a+x_b)^3}+\frac{x_ax_b(2x_a+4x_b)\zeta(3)}{(x_a+x_b)^3}-\frac{x_ax_b^2\pi^2}{6(x_a+x_b)^2}-\frac{x_ax_b(x_b-x_a)}{12}\right],\nonumber\\
&=\frac{9\alpha_s^2T^2}{4\pi^4x_a^2}\left\{\frac{x_a\zeta(3)}{6}-\frac{\pi^2x_a^2}{72}-\int_0^\infty\frac{dx_b}{e^{x_b}-1}\left[\frac{6x_ax_b}{\left(x_a+x_b\right)^3}\polylog{4}\left(e^{-x_a-x_b}\right)+\frac{2x_ax_b\left(2x_a+x_b\right)}{\left(x_a+x_b\right)^3}\polylog{3}\left(e^{-x_a-x_b}\right)\right.\right.\nonumber\\
&\left.\left.\qquad\qquad\qquad+\frac{x_a^2x_b\left(x_a+x_b\right)}{\left(x_a+x_b\right)^3}\polylog{2}\left(e^{-x_a-x_b}\right)-\frac{x_ax_b\pi^4}{15\left(x_a+x_b\right)^3}+\frac{x_ax_b\left(2x_a+4x_b\right)\zeta(3)}{\left(x_a+x_b\right)^3}-\frac{x_ax_b^2\pi^2}{6\left(x_a+x_b\right)^2}\right]\right\},
\end{align}
where, once more, the remaining integral can be expressed in terms of a series involving $\zeta$-functions and multiple polylogarithms, as shown in Appendix~\ref{apnd:multi_Li_series}. 
\end{widetext}
Neglecting Bose-enhancement these become
\begin{align}
\hat{e}_c&\simeq\frac{27\alpha_s^2T^2}{4\pi^4}\left[\frac{\pi^2}{12}-\frac{\zeta(3)}{x_a}\right]+O\left(e^{-x_2}\right),\\
\hat{e}_s&\simeq\frac{9\alpha_s^2T^2}{4\pi^4}\left[\frac{\zeta(3)}{6x_a}-\frac{\pi^2}{72}\right]+O\left(e^{-x_2}\right),\\
\hat{e}_t&\simeq\frac{9\alpha_s^2T^2}{4\pi^4}\Bigg[\left(\frac{\pi^2}{6}-\frac{2\zeta(3)}{x_a}\right)\arcsinh\left(\frac{x_a}{z}\right)+B^-_1(z)\nonumber\\
&-\frac{\pi^2\sqrt{x_a^2+z^2}}{3x_a}-\frac{B^-_2(z)}{x_a}+\frac{2A^-_1(z)}{x_a}\Bigg]+O\left(e^{-x_2}\right),\label{eq:ehat_t}\\
\hat{e}_u&\simeq\frac{9\alpha_s^2T^2}{4\pi^4}\Bigg[\left(\frac{2\zeta(3)}{z^2}-\frac{\pi^4}{15x_az^2}+\frac{\pi^2}{3x_a}\right)\sqrt{x_a^2+z^2}\nonumber\\
&+\frac{A^-_1(z)x_a}{z^2}+\left(\frac{4\zeta(3)}{x_a}-\frac{\pi^2}{3}\right)\arcsinh\left(\frac{x_a}{z}\right)-\frac{A^-_2(z)}{z^2}\nonumber\\
&-2B^-_1(z)-\frac{2A^-_1(z)}{x_1}+\frac{2B^-_2(z)}{x_a}\Bigg]+O\left(e^{-x_2}\right),\label{eq:ehat_u}
\end{align}
which can be readily used inside Monte Carlo simulations, owing to the simpler integrals involved in obtaining $A^-_i(z)$ as well as $B^-_i(z)$.
\subsection{Peign\'e \& Peshier approximation and the Braaten \& Thoma limit}
The Peign\'e \& Peshier approximation to transport coefficient is obtained by approximating the $\arcsinh$ and the square root using Eq.~(\ref{eq:approx_arcsinh_sqrt}) as was explained in the rate derivation, while the Braaten \& Thoma limit further neglects simple pole terms in Mandelstam variables within the matrix element, in addition to approximating the arcsinh and the square root. 

Following the approach by Peign\'e \& Peshier one obtains
\begin{align}
\hat{q}_c&\simeq\frac{9\alpha_s^2T^3}{2\pi^4}\left[\zeta(3)+\frac{\pi^4}{45x_a}\right]+O\left(z^2x^{-2}_a\right),\\
\hat{q}_s&\simeq\frac{9\alpha_s^2T^3}{2\pi^4}\left[-\frac{\zeta(3)}{15}-\frac{\pi^4}{900x_a}\right]+O\left(z^2x^{-2}_a\right),\\
\hat{q}_t&\simeq\frac{9\alpha_s^2T^3}{2\pi^4}\bigg[2\zeta(3)\ln\left(\frac{4x_a}{z^2}\right)+2\zeta'(3)+\frac{\pi^4}{90x_a}\nonumber\\
&\qquad\qquad\quad-\left(2\gamma+1\right)\zeta(3)\bigg]+O\left(z^2x^{-2}_a\right),\label{eq:qhat_t_PP}\\
\hat{q}_u&\simeq\frac{9\alpha_s^2T^3}{2\pi^4}\left[\left(2\zeta(3)-\frac{\pi^4}{15x_a}\right)\ln\left(\frac{4x_a}{z^2}\right)+2\zeta'(3)\right.\nonumber\\
&\left.\qquad\qquad\quad-\left(2\gamma+1\right)\zeta(3)+\frac{16\zeta(5)}{z^2}-\frac{6\zeta'(4)}{x_a}\right.\nonumber\\
&\left.\qquad\qquad\quad+\left(2\gamma-1\right)\frac{\pi^4}{30x_a}\right]+O\left(z^2x^{-2}_a\right),
\end{align}
whereas 
\begin{align}
\hat{e}_c&\simeq\frac{9\alpha_s^2}{\left(2\pi\right)^4}\left[\pi^2-\frac{12\zeta(3)}{x_a}\right]+O\left(z^2x^{-2}_a\right),\\
\hat{e}_s&\simeq\frac{9\alpha_s^2}{\left(2\pi\right)^4}\left[\frac{2\zeta(3)}{3x_a}-\frac{\pi^2}{18}\right]+O\left(z^2x^{-2}_a\right),\\
\hat{e}_t&\simeq\frac{9\alpha_s^2}{4\pi^4}\Bigg[\left(\frac{\pi^2}{6}-\frac{2\zeta(3)}{x_a}\right)\ln\left(\frac{4x_a}{z^2}\right)+\zeta'(2)\nonumber\\
&\qquad\qquad-\frac{(1+\gamma)\pi^2}{6}+\frac{(1+2\gamma)\zeta(3)}{x_a}-\frac{2\zeta'(3)}{x_a}\Bigg]\nonumber\\
&+O\left(z^2x^{-2}_a\right),\\
\hat{e}_u&\simeq\frac{9\alpha_s^2}{4\pi^4}\Bigg[\frac{4\zeta(3)x_a}{z^2}+\left(\frac{4\zeta(3)}{x_a}-\frac{\pi^2}{3}\right)\ln\left(\frac{4x_a}{z^2}\right)\nonumber\\
&-\frac{2\pi^4}{15z^2}+\frac{\gamma\pi^2}{3}-2\zeta'(2)+\frac{4\zeta'(3)}{x_a}+\frac{2(1-2\gamma)\zeta(3)}{x_a}\Bigg]\nonumber\\
&+O\left(z^2x^{-2}_a\right).
\end{align}
Keeping the most divergent double pole terms yields the Braaten \& Thoma limit
\begin{align}
\hat{q}_t&\simeq\frac{9\alpha_s^2T^3}{2\pi^4}\left[2\zeta(3)\ln\left(\frac{4x_a}{z^2}\right)+2\zeta'(3)-2\gamma\zeta(3)+\frac{\pi^4}{45x_a}\right]\nonumber\\
&\quad+O\left(z^2x^{-2}_a\right),\\
\hat{q}_u&\simeq\frac{9\alpha_s^2T^3}{2\pi^4}\left[\left(2\zeta(3)-\frac{2\pi^4}{45x_a}\right)\ln\left(\frac{4x_a}{z^2}\right)+2\zeta'(3)\right.\nonumber\\
&\left.\qquad\qquad-2\gamma\zeta(3)+\frac{16\zeta(5)}{z^2}-\frac{4\zeta'(4)}{x_a}+\left(\gamma-1\right)\frac{2\pi^4}{45x_a}\right]\nonumber\\
&\quad+O\left(z^2x^{-2}_a\right),
\end{align}
and 
\begin{align}
\hat{e}_t&\simeq\frac{9\alpha_s^2}{4\pi^4}\bigg[\left(\frac{\pi^2}{6}-\frac{2\zeta(3)}{x_a}\right)\ln\left(\frac{4x_a}{z^2}\right)+\zeta'(2)-\frac{\gamma\pi^2}{6}\nonumber\\
&\qquad\qquad+\frac{(2\gamma-1)\zeta(3)}{x_a}-\frac{2\zeta'(3)}{x_a}\bigg]+O\left(z^2x^{-2}_a\right),\\
\hat{e}_u&\simeq\frac{9\alpha_s^2}{4\pi^4}\bigg[\frac{4\zeta(3)x_a}{z^2}+\left(\frac{2\zeta(3)}{x_a}-\frac{\pi^2}{6}\right)\ln\left(\frac{4x_a}{z^2}\right)-\frac{2\pi^4}{15z^2}\nonumber\\
&\qquad\quad+\frac{(\gamma-1)\pi^2}{6}-\zeta'(2)+\frac{2\zeta'(3)}{x_a}+\frac{(3-2\gamma)\zeta(3)}{x_a}\bigg]\nonumber\\
&\quad+O\left(z^2x^{-2}_a\right).
\end{align}
It should be noted that four-point vertex and $s$-channel do not contribute in the Braaten \& Thoma limit, as they do not contain any poles. Therefore, removing anything but the highest order poles in $t$- and $u$-channel completely eliminates any contribution form other channels.
\begin{figure}[H]
    \centering
	\includegraphics[width=\linewidth]{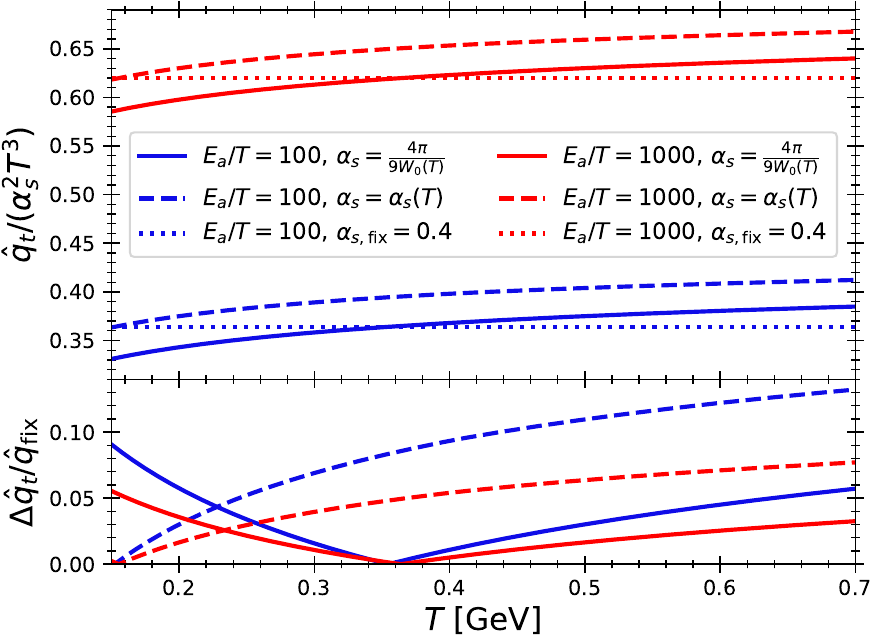}
    \caption{Top panel: $t$-channel contribution to $\hat{q}$ for fixed $E_a/T$ with running $m_D$ as a function of $T$. Bottom panel: comparison between changing and fixed Debye mass.}
    \label{fig:RunningAlpha-qhat-FixE}
\end{figure}
\subsection{Exploring $z$-indiced  cut-off effects on transport coefficients}
\subsubsection{The ultraviolet behavior of $\hat q$ and $\hat e$}

%
Recent Bayesian analysis of jet transverse momentum broadening in the QGP \cite{JETSCAPE:2024cqe}, which corresponds to $\hat q_t$ herein, has put constraints on its virtuality depend parameters. Translating that analysis to the situation considered here (where incoming and outgoing particles are on-shell) amounts to changing the normalization of $\hat q_t$, and Ref.~\cite{JETSCAPE:2024cqe} puts constraints on the overall allowed normalization of $\hat q_t$. However, neither the results in Ref.~\cite{JETSCAPE:2024cqe}, nor those of the previous Bayesian analysis in Ref.~\cite{JETSCAPE:2021ehl}, have accounted for the fact $z=m_D/T$ can change with temperature through $\alpha_s$, according to either Eq.~(\ref{eq:alpha_s_2piT}), or, equivalently, Eq.~(\ref{eq:alpha_s_run_W0}). 

The temperature-dependence of $\frac{\hat q_t}{\alpha^2_s T^3}$, owing to the running coupling inside $z$, is depicted in Fig.~\ref{fig:RunningAlpha-qhat-FixE}. The virtuality-independent portion of $\hat q$ used in Ref.~\cite{JETSCAPE:2024cqe}, labeled as $\hat q_{\rm fix}$, is given by
\begin{align}
    \frac{\hat{q}_\mathrm{fix}}{\alpha_s^2T^3}=\frac{9\zeta(3)}{\pi^4}\ln\left(\frac{2E_aT}{6\pi \alpha_{s,\mathrm{fix}}T^2}\right),
\end{align}
where $\alpha_{s,\mathrm{fix}}$ is a constant. The relative difference between our $\hat q_t$ in Eq.~(\ref{eq:qhat_t_BE}) and the one used in Ref.~\cite{JETSCAPE:2024cqe} is 
\begin{align}
    \Delta\hat{q}_t\equiv\lvert\hat{q}_t-\hat{q}_\mathrm{fix}\rvert.
\end{align}
This expression constitutes a theoretical systematic uncertainty, whose relative size is in the bottom panel of Fig.~\ref{fig:RunningAlpha-qhat-FixE}. The modest difference between the JETSCAPE expression for $\langle q_\perp^2\rangle/L$ and our $\hat q_t$ stems from both including Bose-enhancements, as explored in Ref.~\cite{short_paper}, as well as the running of $\alpha_s$ within $z$, according to Eq.~(\ref{eq:mD_w_running}). Such a theoretical systematic uncertainty should be accounted for in the future.

\begin{figure}[H]
    \centering
	\includegraphics[width=\linewidth]{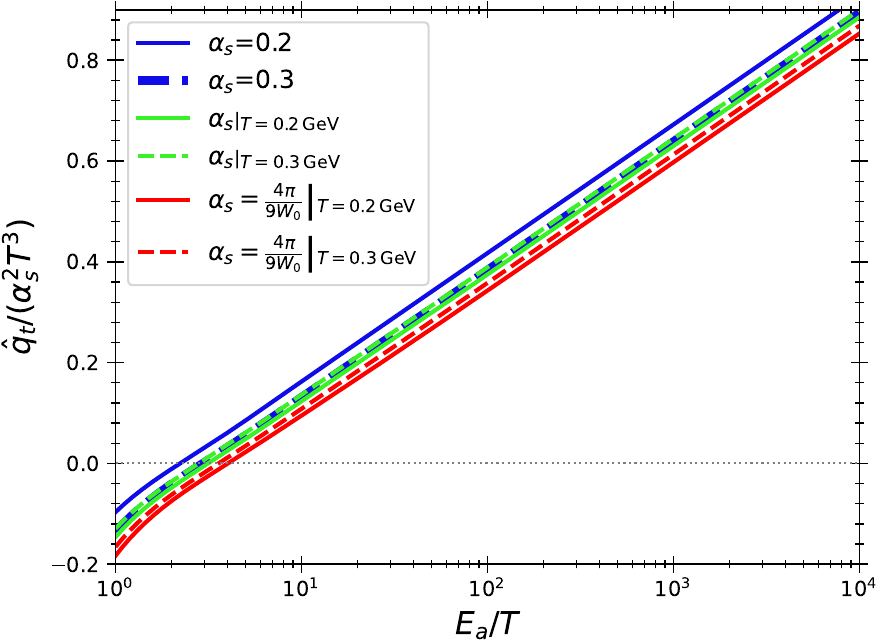}
    \caption{$t$-channel contribution to $\hat{q}$ for various ways of evaluating $\alpha_s$ in $z=m_D/T$.}
    \label{fig:RunningAlpha-qhat-t}
\end{figure}

Figure~\ref{fig:RunningAlpha-qhat-t} allows to appreciate how sensitive $\hat q_t(x_a)$ is to the definition of $z$. 

\begin{figure}[H]
    \centering
	\includegraphics[width=\linewidth]{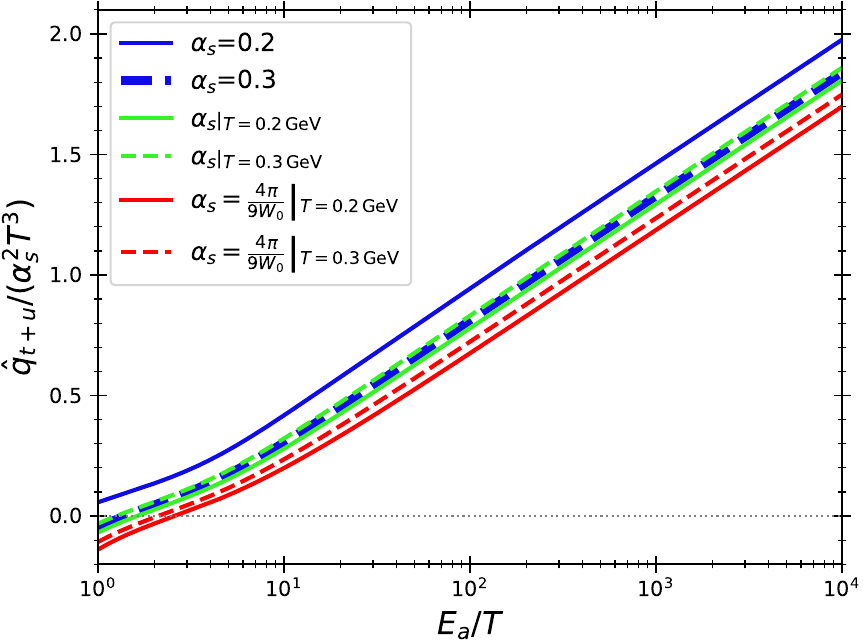}
    \caption{$t+u$-channel $\hat{q}$ for various ways of evaluating $\alpha_s$.}
    \label{fig:RunningAlpha-qhat-tu}
\end{figure}
\begin{figure}[H]
    \centering
	\includegraphics[width=\linewidth]{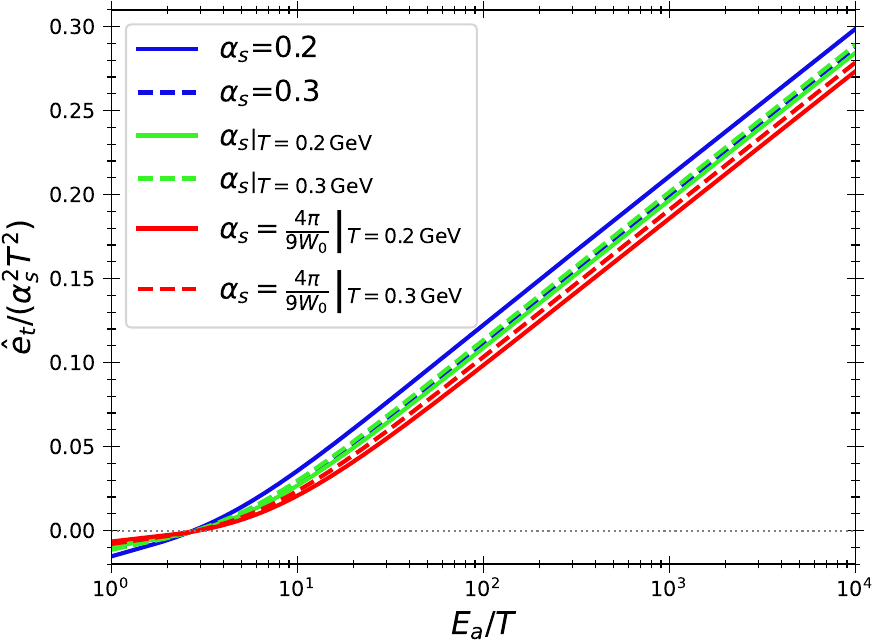}
    \includegraphics[width=\linewidth]{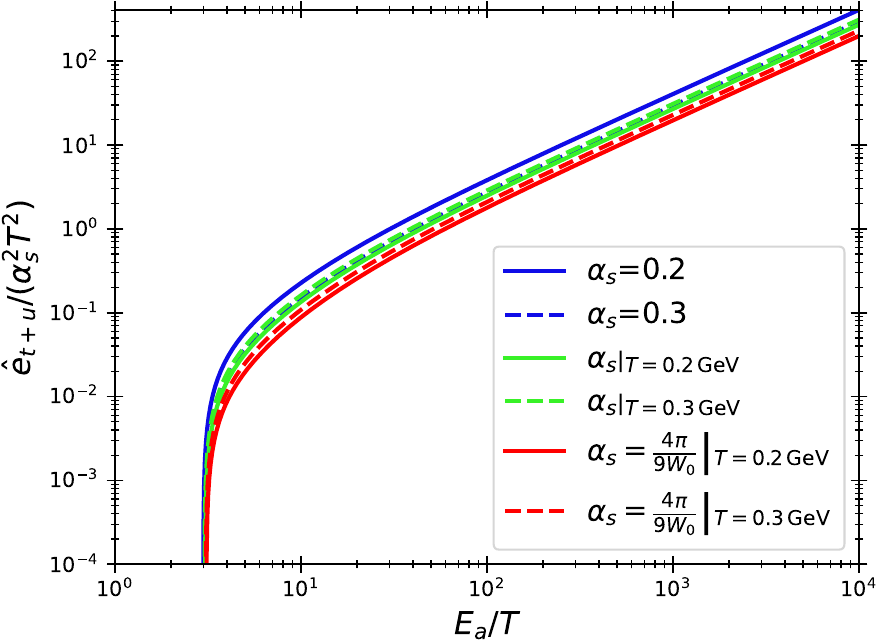}
    \caption{Top: $t$-channel $\hat{e}$ for various ways of evaluating $\alpha_s$. Bottom: Similar to top panel, but now combining $t$ and $u$ channels for $\hat e$.}
    \label{fig:RunningAlpha-ehat-tu}
\end{figure}

While the $u$-channel contribution to transverse momentum broadening $\langle q^2_\perp\rangle/L$ is not relevant within a higher-twist-based Monte Carlo evolution, we present the sum of $\hat q_u$ with $\hat q_t$ in Fig.~\ref{fig:RunningAlpha-qhat-tu} for completeness.

Figure~\ref{fig:RunningAlpha-ehat-tu} presents energy loss $\langle \omega\rangle/L$, using both the $t$-channel contribution relevant for higher-twist-based Monte Carlo simulations (top panel), along with the sum of $t$- and $u$-channel contributions, relevant for parton energy loss in the ultraviolet regime such as \cite{Ghiglieri:2015ala}.

\subsubsection{Complete tree-level results for $\hat q$ and $\hat e$}

\begin{figure}[h]
    \centering
	\includegraphics[width=\linewidth]{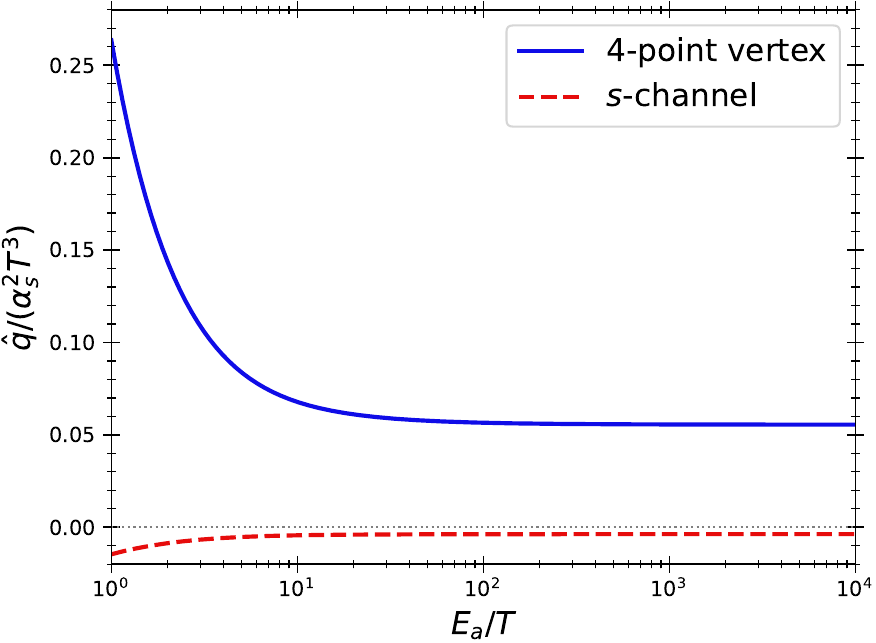}
    \includegraphics[width=\linewidth]{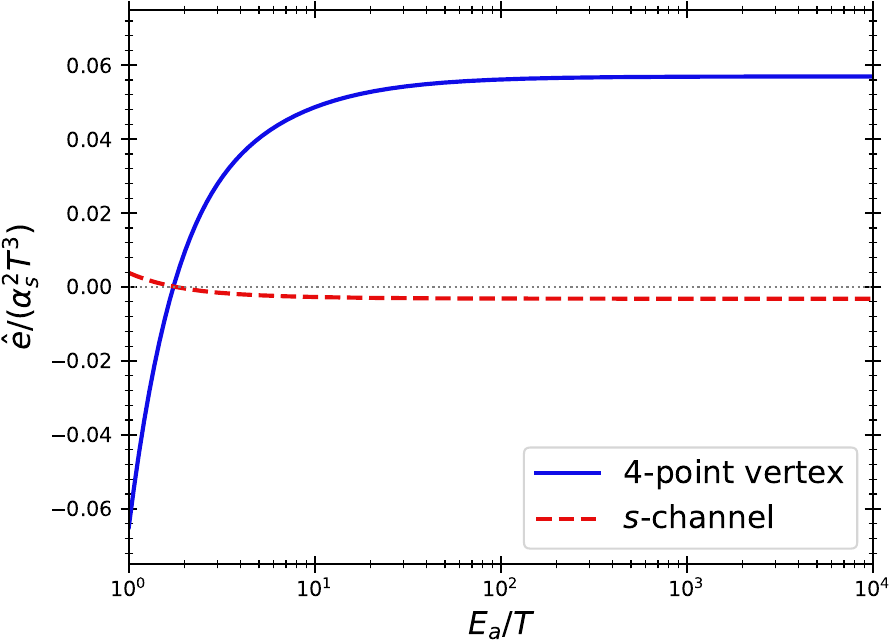}
    \caption{four-point vertex and $s$-channel contributions to $\hat{q}$ and $\hat e$.}
    \label{fig:qhat-ehat-cs}
\end{figure}

\begin{figure}[h]
    \centering
	\includegraphics[width=\linewidth]{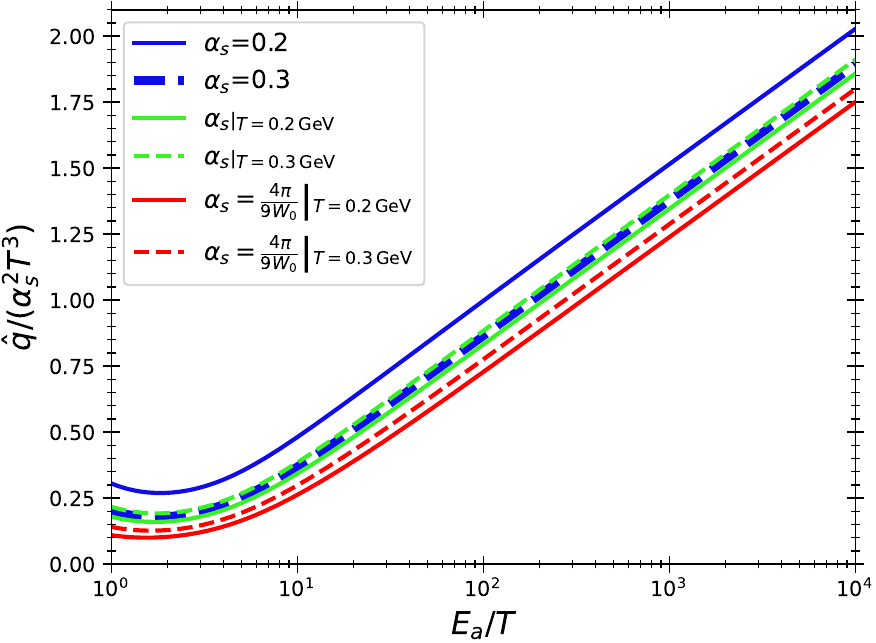}
    \includegraphics[width=\linewidth]{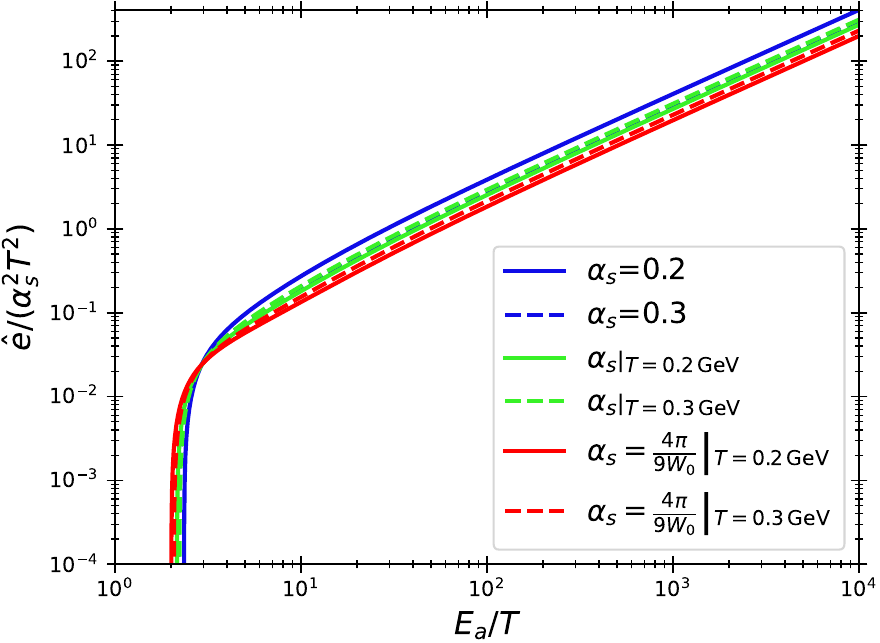}
    \caption{Top panel: $\hat{q}$ in Eq.~(\ref{eq:Ohat_q})  for various ways of evaluating the (Debye mass) $z$ cut-off. Bottom panel: same as top panel but for $\hat e$ in Eq.~(\ref{eq:ehat_BE}).}
    \label{fig:qhat-ehat-Full}
\end{figure}

To complete the picture at tree-level, the $s$-channel as well as the four-point vertex contributions are considered. These contributions do not require any cut-off, and their final $x_a$-dependence is presented in Fig.~\ref{fig:qhat-ehat-cs}. The sum from all four channels is given in Fig.~\ref{fig:qhat-ehat-Full}. For calculations of $\hat q$ that do not rely on the higher-twist formalism, e.g. \cite{Ghiglieri:2015ala}, all four scattering channels contribute to $\hat q$ and $\hat e$. 


\subsection{Finding a good approximation for $\hat q$ and $\hat e$}
\begin{figure}[H]
    \centering
	\includegraphics[width=\linewidth]{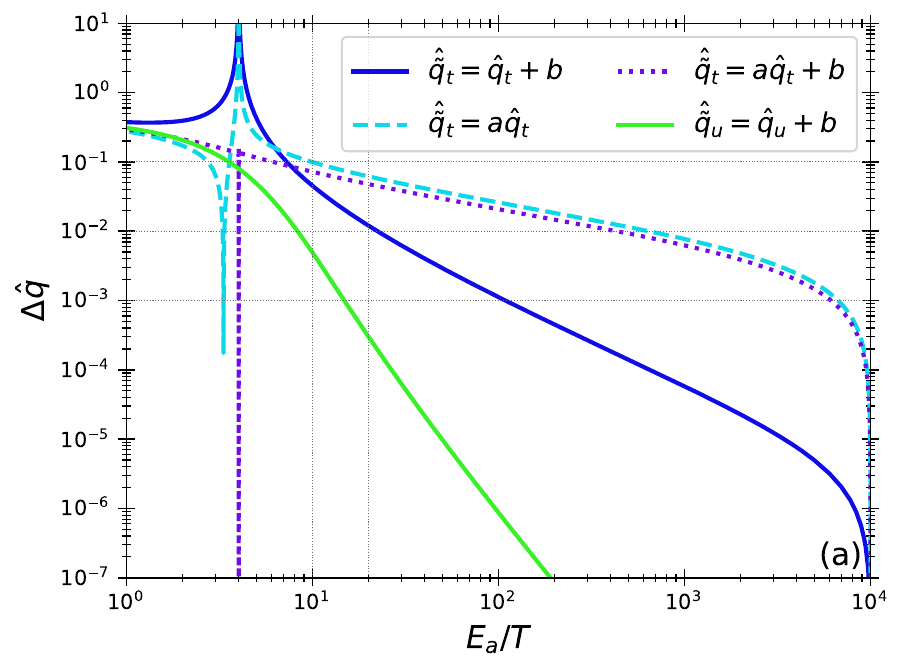}
    \includegraphics[width=\linewidth]{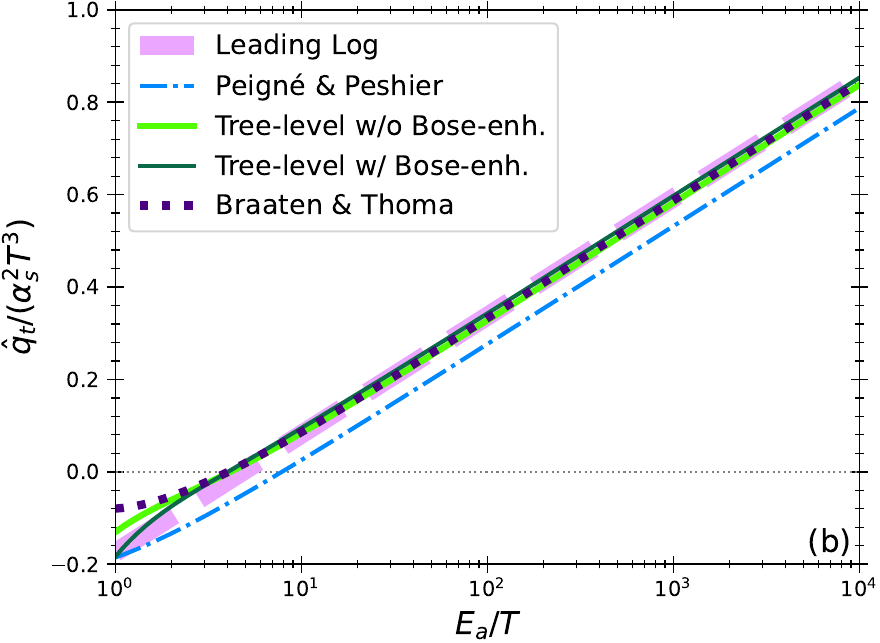}
    \includegraphics[width=\linewidth]{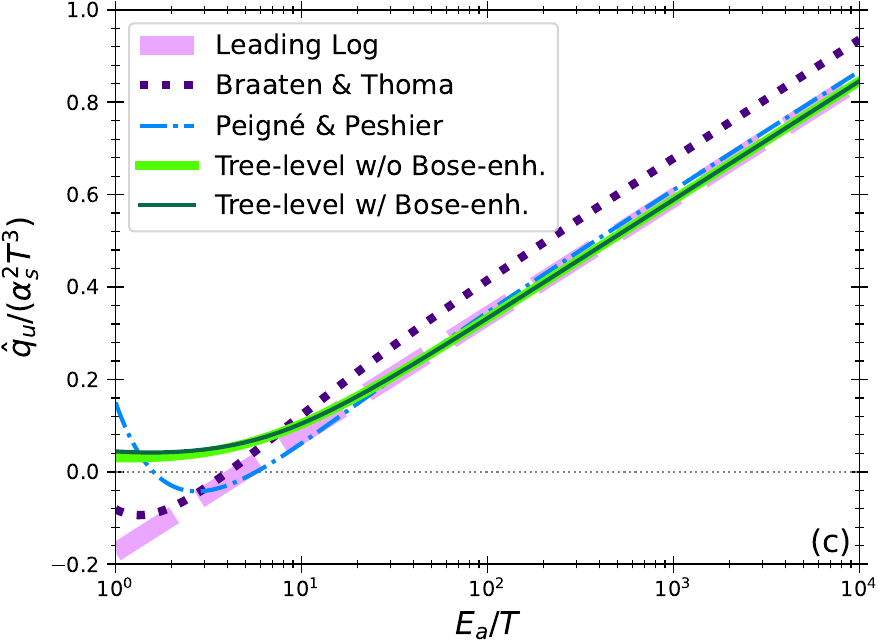}
    \caption{Comparison of various approximations to the full $\hat{q}_t$ and $\hat{q}_u$ in Eqs.~(\ref{eq:qhat_t_BE},\ref{eq:qhat_u_BE}), with the relative error of the non-Bose-enhanced $\hat{q}$ after linear transformation (a), while comparisons to other approximations to the $t$-channel are in Panel (b) and to the $u$-channel in (c). $m_D$ is taken to follow Eq.~(\ref{eq:mD_w_running}) with $T=0.2$ GeV.}
    \label{fig:qhatFit}
\end{figure}

\begin{figure}[H]
    \centering
	\includegraphics[width=\linewidth]{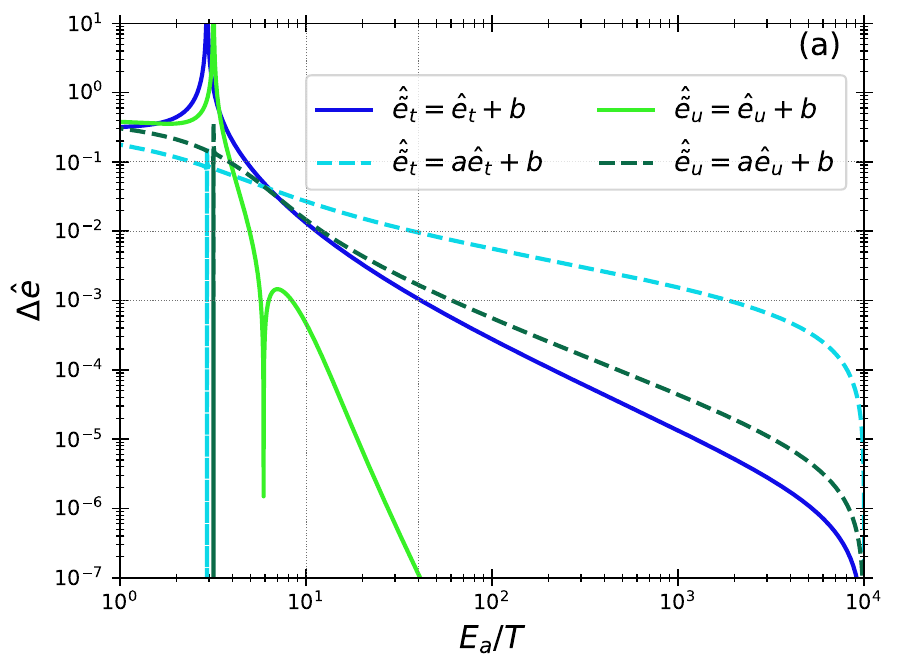}
    \includegraphics[width=\linewidth]{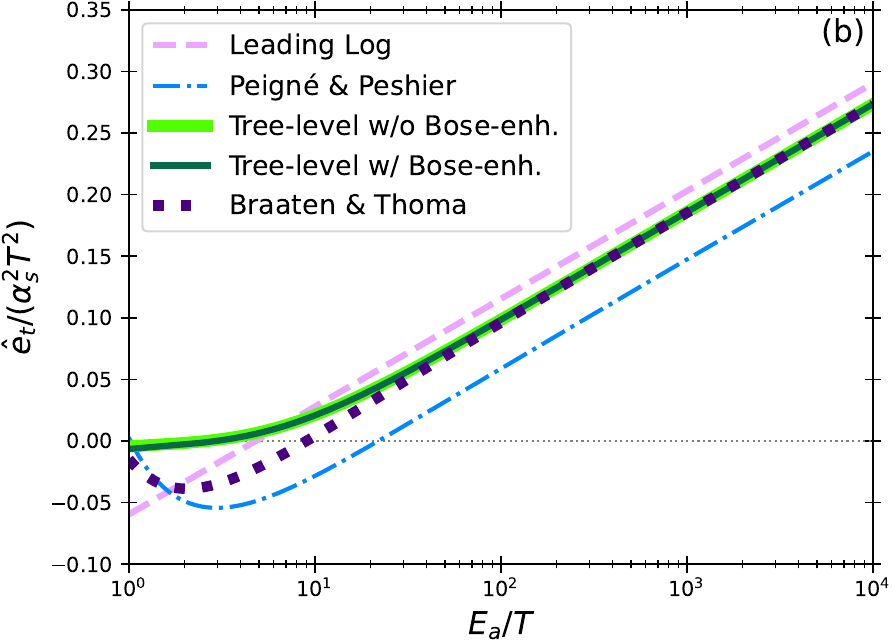}
    \includegraphics[width=\linewidth]{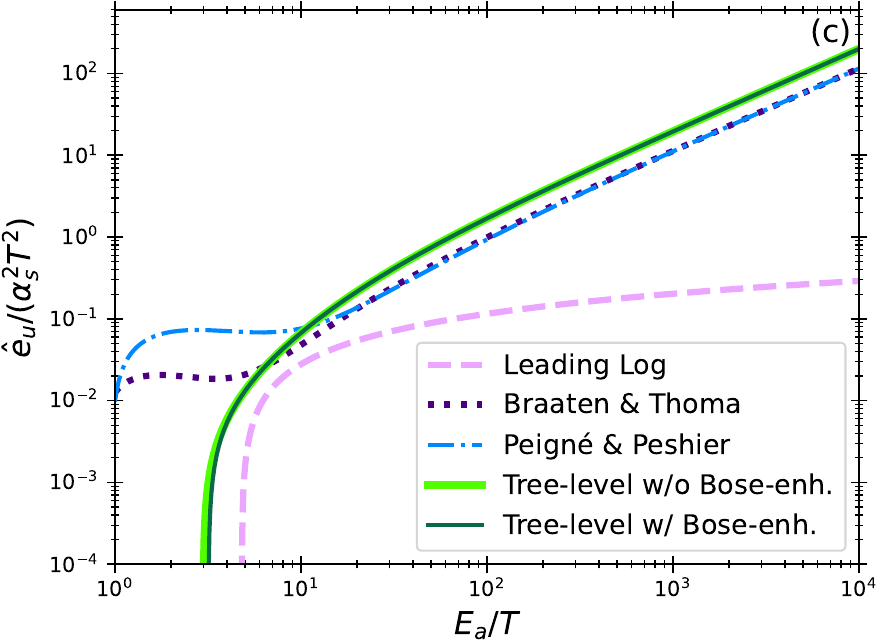}
    \caption{Comparison of various approximations to the full $\hat{e}_t$ and $\hat{e}_u$ in Eqs.~(\ref{eq:ehat_t_BE},\ref{eq:ehat_u_BE}), with the relative error of the non-Bose-enhanced $\hat{e}$ after linear transformation (a), while comparisons to other approximations to the $t$-channel are in Panel (b) and to the $u$-channel in (c). $m_D$ is taken to follow Eq.~(\ref{eq:mD_w_running}) with $T=0.2$ GeV.}
    \label{fig:ehatFit}
\end{figure}

As a detailed comparison between various approximations and our results has been done in Ref.~\cite{short_paper}, the main focus is now shifted towards finding a good approximation for $\hat q$ and $\hat e$, which is shown in Figs.~\ref{fig:qhatFit}(a),~\ref{fig:ehatFit}(a). The remaining panels in Figs.~\ref{fig:qhatFit},~\ref{fig:ehatFit} simply depict the comparison between various approaches already explored in Ref.~\cite{short_paper}, being presented here solely for reference.

The approximate transport coefficients $\hat{q}_{\rm i,\, approx}$ use Eq.~(\ref{eq:qhat_t}) and Eq.~(\ref{eq:qhat_u}) and, as done with the rate, linearly transform it according to
\begin{align}
    \hat{\tilde{q}}_{i}\equiv a\,\hat{q}_\mathrm{i,\,approx}+b
\label{eq:rescaled_transport_coeff}
\end{align}
where $i\in\{t,u\}$ for the specific channel's contribution. To compare the linearly transformed result with the full result, we define
\begin{align}
    \Delta\hat{q}\equiv\left|\frac{\hat{\tilde{q}}_{i}-\hat{q}_i}{\hat{q}_i}\right|,
\end{align}
where $\hat q_t$ and $\hat q_u$ are given by Eq.~(\ref{eq:qhat_t_BE}) and Eq.~(\ref{eq:qhat_u_BE}), respectively. The same procedure is applied to $\hat{e}$. 
The results of various transformations are shown in Fig.~\ref{fig:qhatFit}(a) for $\hat{q}$, and similarly in Fig.~\ref{fig:ehatFit}(a) for $\hat{e}$, using Eq.~(\ref{eq:ehat_t}) and Eq.~(\ref{eq:ehat_u}) for the latter. As was the case for the rate, at most two anchor points were used when obtaining the $a$ and $b$ coefficients within Eq.~(\ref{eq:rescaled_transport_coeff}): one at $x_a=10^{4}$ and the other satisfying $\hat q\left(x_a\right)=0=\hat e\left(x_a\right)$. If a simple shift or rescaling is applied, then $x_a=10^{4}$ is the sole anchor point. Simply shifting the transport coefficients in Eqs.~(\ref{eq:qhat_u}, \ref{eq:qhat_t}) as well as Eq.~(\ref{eq:ehat_t}, \ref{eq:ehat_u}) by a constant yields a very good approximation of the Bose-enhanced result, as was observed in Fig.~\ref{fig:rate_fit}. Adding a constant shift to Eq.~(\ref{eq:ehat_t}) approximates the Bose-enhanced $\hat e_t$ to within $\sim 1$\% or better for $x_a>10$, while doing so for $\hat e_u$ reaches better than $\sim 0.1$\% accuracy for $x_a>10$. For a shifted $\hat q_{\rm t,\, approx}$ to reach $\sim 1$\% accuracy, $x_a>20$ is needed. As the shifted $\hat q_{\rm u,approx}$ performs better than its $t$-channel counterpart, the precision admitted from a constant shift to Eqs.~(\ref{eq:qhat_u}, \ref{eq:qhat_t}) and Eq.~(\ref{eq:ehat_t}, \ref{eq:ehat_u}) enables them to be readily used within Monte Carlo simulations of jet-medium interactions in the QGP. From Fig.~\ref{fig:qhatFit}(b,c) and Fig.~\ref{fig:ehatFit}(b,c) theoretical systematic uncertainties can be computed by taking the difference between various approaches. This is particularly useful for Bayesian analyses.  

\subsection{Shift parameters used when approximating the rate and transport coefficients}
\begin{figure}[H]
    \centering
    \includegraphics[width=\linewidth]{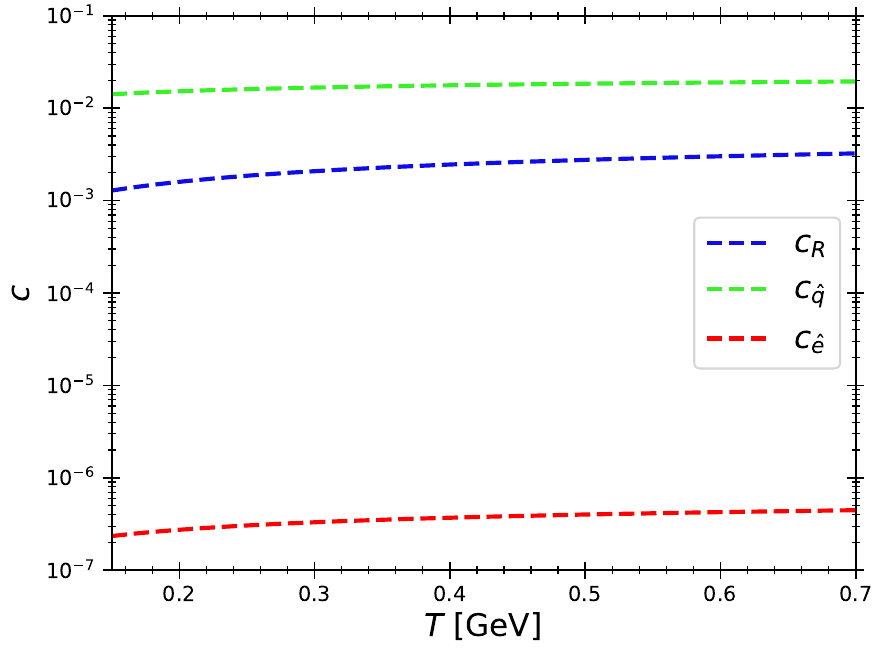}
    \caption{Difference between full tree-level result and non-Bose-enhanced approximations at $x_a=10^4$, for $m_D$ following Eq.~(\ref{eq:mD_w_running}). Note that the magnitude of $c_{\hat{e}}$ is depicted in this figure; its sign is negative, as in Table~\ref{tab:ShiftValues}.}
    \label{fig:Shifts}
\end{figure}

For all three quantities (the full scattering rate, $\hat{q}_t$ and $\hat{e}_t$), a simple shift of the non-Bose-enhanced expressions is found to best approximate the full, Bose-enhanced tree-level results. Since this makes the much simpler expressions (neglecting Bose-enhancement) useful for Monte Carlo simulations, the temperature dependence of the shift parameters is given in Fig.~\ref{fig:Shifts}. Plotted is the difference of Bose-enhanced and non-Bose-enhanced quantity in the UV at $x_a=10^4$. Taking the rate as an example, we have
\begin{align}
    c_R\equiv\frac{1}{\alpha^2_s T}\left[\left.\frac{d^3R}{d^3p_a}\right|_{x_a=10^4}-\left.\frac{d^3R_\mathrm{approx}}{d^3p_a}\right|_{x_a=10^4}\right]
\end{align}
where the approximate rate is taken from Eq.~(\ref{eq:rate_approx_stu_w_c_exact}), and the full rate, evaluated numerically, stems from Eq.~(\ref{eq:rate_full}). The difference is a constant, independent of $x_a$. However, as $z=m_D/T$ depends on $T$ through Eq.~(\ref{eq:mD_w_running}), so does $c_R$.

Similarly defined are $c_{\hat{q}}$, the difference between full Eq.~(\ref{eq:qhat_t_BE}) and non-Bose-enhanced approximate Eq.~(\ref{eq:qhat_t}), and likewise $c_{\hat{e}}$, difference between Eq.~(\ref{eq:ehat_t_BE}) and Eq.~(\ref{eq:ehat_t}). To be noted is the negative sign in front of $c_{\hat{e}}$, so that the red curve plotted has to be subtracted from the non-Bose-enhanced $\hat{e}$ to approximate the full result.

Table~\ref{tab:ShiftValues} lists these constant shifts for two common temperature values with $m_D$ following Eq.~(\ref{eq:mD_w_running}).

\begin{table}[h]
    \centering
    \renewcommand{\arraystretch}{3}
    \begin{tabular}{c|c|c}
        & $\left.c_i\right|_{T=0.2\,\mathrm{GeV}}$ & $\left.c_i\right|_{T=0.3\,\mathrm{GeV}}$\\
        \hline
         $\frac{1}{\alpha^{2}_s T}\Big[$Eq.~(\ref{eq:rate_approx_stu_w_c_exact})$\Big]\,+\,c_R$ & $1.598 4766\cdot10^{-3}$ & $2.084 8187\cdot10^{-3}$\\
         \hline
         $\frac{1}{\alpha^{2}_s T^{3}}\Big[$Eq.~(\ref{eq:qhat_t})$\Big]\,+\,c_{\hat q}$ & $1.527 5423\cdot10^{-2}$ & $1.674 0971\cdot10^{-2}$\\
         \hline
         $\frac{1}{\alpha^{2}_s T^{2}}\Big[$Eq.~(\ref{eq:ehat_t})$\Big]\,+\,c_{\hat e}$ & $-2.743 1881\cdot10^{-7}$ & $-3.307 6222\cdot10^{-7}$
    \end{tabular}
    \caption{Constant shifts used to approximate the Bose-enhanced quantities using the non-Bose-enhanced expressions, with $m_D$ following Eq.~(\ref{eq:mD_w_running})}
    \label{tab:ShiftValues}
\end{table}

\section{Conclusion}\label{sec:conclusion}
We have presented a first calculation where the scattering rate for the reaction $gg\to gg$ at tree-level leading order without any kinematic approximations, and have obtained jet-medium transport coefficients for that process. A few popular approximation used to obtain jet-medium transport coefficients were also explored, namely those in Refs.~\cite{Braaten:1991jj,Braaten:1991we,Peigne:2007sd,Peigne:2008nd}, as comparisons to guide our research. Future research will explore transport coefficients beyond those associated with gluon exchanges, by looking into transport coefficients where a fermion exchange is present. In the higher-twist formalism, those scattering kernels were recently computed \cite{Kumar:2025egh,Kumar:2025asj}, with the associated transport coefficients to be presented in an upcoming publication, that will leverage the theoretical formalism being presented here.     

The overall goal of this contribution was to focus on high-energy jet-medium interactions, where a tree-level calculation is a valid approximation, and obtain estimates of the theoretical systematic uncertainty associated with Bayesian constraints of jets in the QGP. Current Bayesian analysis, such as \cite{JETSCAPE:2024cqe}, do not quantify theoretical systematic uncertainties in jet-medium interactions, thus prompting our research. To illustrate how theoretical systematic uncertainties can be obtained in jet-medium interactions, we revisited tree-level $gg\to gg$ scatterings. Our estimates of these theoretical systematic uncertainties applicable within a dilute approximation of the QGP, which is also the limit used during the high virtuality stage of the multi-scale Monte Carlo simulations of jets in the QGP. Since jets partons decrease in energy as the shower develops, tree-level calculations will become less effective therein. In the lower virtuality portion of multi-scale jet-medium Monte Carlo simulations, infrared effects such as Landau Damping and Debye screening, become important. An upcoming publication will specifically address how thermal field theory self-energy corrections, and more specifically Hard Thermal Loops, modify our scattering rates and jet-medium transport coefficients. A better estimate of the theoretical uncertainty in the infrared region of jet-medium interactions will improve the reliability of Bayesian constraints of jets in the QGP. Indeed, the observed tension between leading hadrons and jet reported in Ref.~\cite{JETSCAPE:2024cqe} is affected by the theoretical systematic uncertainties not accounted for within those Bayesian constraints.
\newline

\textbf{\textit{Acknowledgments.}}---  The authors thank Charles Gale for valuable discussions and comments on this work. This work was supported by the Canada Research Chair under Grant Number CRC-2022-00146, the Natural Sciences and Engineering Research Council (NSERC) of Canada under Grant Number SAPIN-2023-00029, and the Canadian Foundation for Innovation John R. Evans Leaders Fund under Grant Number 44100.
\newline

\textbf{\textit{Data availability.}}--- No data were created or analyzed in this study. 
\appendix
\begin{appendices}
\section{Integration over Mandelstam variables}
\label{sec:Mandelstam_int}
Performing the integrals over Mandelstam variables is easier if they are slightly shifted, namely
\begin{align}
x&\equiv s-A_s, \quad A_s\equiv\frac{t^2-c_ut}{2C_s},\quad C_s\equiv a_t-t,\\
y&\equiv t-A_t, \quad A_t\equiv\frac{s^2-c_us}{2C_t}, \quad C_t\equiv a_s-s,
\end{align}
where $a_s,a_t$ and $c_u$ are defined in Eq.~(\ref{eq:abc_stu}). In that case, 
\begin{align}
\Lambda(s,t)&=(t^2-c_ut)s-a_st^2-(a_t-t)s^2,\\
&=C_s(2A_ss-B_s-s^2)=C_s(K_s-x^2)
\end{align}
where
\begin{align}
B_s&\equiv \frac{a_st^2}{C_s},\quad K_s\equiv A_s^2-B_s.
\end{align}
Equivalently, one has
\begin{align}
\Lambda(s,t)&=(s^2-c_us)t-a_ts^2-(a_s-s)t^2\\
&=C_t(2A_tt-B_t-t^2)=C_t(K_t-y^2)
\end{align}
where
\begin{align}
B_t&\equiv\frac{a_ts^2-b_ts-d_u}{C_t},\quad K_t\equiv A_t^2-B_t.
\end{align}
Since
\begin{align}
	C_t&=(E_a+E_b)^2-(p_a+p_b)^2=(\vec{p}_a+\vec{p}_b)^2>0,\\
	C_s&=(E_b-E_2)^2-(p_b-p_2)^2=(\vec{p}_b-\vec{p}_2)^2>0,
\end{align}
the Jacobian $\sqrt{\Lambda}$ in the scattering rate becomes
\begin{align}
\sqrt{\Lambda}&=\sqrt{C_s}\sqrt{K_s-x^2}, \,\,{\rm or}\label{eq:Lambda_x}\\
&=\sqrt{C_t}\sqrt{K_t-y^2}.\label{eq:Lambda_y}
\end{align}
Noting that the $\Lambda$-function vanishes at the integration boundaries, one obtains the boundaries of the innermost integral below
\begin{align}
	x_\pm&=\pm\sqrt{K_s}, \,\,{\rm or}\\
	y_\pm&=\pm\sqrt{K_t}.
\label{eq:xy_min_max}
\end{align}
The limits of the Mandelstam variable integrals are now determined, which, of course, depend on the order of integrations. The rate is thus expressed as
\begin{align}
\frac{d^3R}{d^3p_a}&=\frac{9\alpha_s^2}{16\pi^5E_a^2}\int_0^\infty\frac{dE_b}{e^{\beta E_b}-1}\int_0^{E_a+E_b}\frac{dE_2}{1-e^{-\beta E_2}} \nonumber\\
&\times\left\{\mathcal{G} [1]+\mathcal{G}\left[\frac{-tu}{s^2}\right]+\mathcal{G} \left[\frac{-su}{t^2}\right]+\mathcal{G}\left[\frac{-st}{u^2}\right]\right\},
\end{align}
where
\begin{align}
\begin{split}
\mathcal G [f(s,t)]&\equiv\int_{s_{\min}}^{s_{\max}}ds\int_{t_-}^{t_+}dt\frac{f(s,t)}{\sqrt{\Lambda(s,t)}}\\
&=\int_{t_{\min}}^{t_{\max}}dt\int_{s_-}^{s_+}ds\frac{f(s,t)}{\sqrt{\Lambda(s,t)}},  
\end{split}
\end{align}
where $t_\pm$ is obtained from requiring $\Lambda(s,t_\pm)=0$, while $s_{\rm max,min}$ is obtained from kinematics. A similar approach determines $s_\pm$ as well as $t_{\rm max,min}$. Integrating the matrix element term by term, starting with the four-point vertex, generates
\begin{align}
\mathcal G [1]&=\int_{s_{\min}}^{s_{\max}}ds\int_{t_-}^{t_+}\frac{dt}{\sqrt{C_t(2A_tt-B_t-t^2)}}\nonumber\\
&=\int_{s_{\min}}^{s_{\max}}\frac{ds}{\sqrt{C_t}}\int_{-\sqrt{K_t}}^{\sqrt{K_t}}\frac{dy}{\sqrt{K_t-y^2}}\\
&=\pi\int_{s_{\min}}^{s_{\max}}\frac{ds}{\sqrt{a_s-s}}\\
&=2\pi\left[\sqrt{(E_a+E_b)^2-s}\right]_{4\lfloor s \rfloor}^0\nonumber\\
&=2\pi\left[\sqrt{(E_a+E_b)^2-s}\right]_{4\min(E_aE_b,E_1E_2)}^0\nonumber\\
&=2\pi\left\{(E_a+E_b)-\max[(E_a-E_b),(E_1-E_2)]\right\}\nonumber\\
&=4\pi\min(E_a,E_b,E_1,E_2)\nonumber\\
&=4\pi\lfloor E\rfloor,
\end{align}
where $E_1=E_a+E_b-E_2$. The $s$-channel is given by
\begin{align}
\mathcal G\left[-\frac{tu}{s^2}\right]&=\int_{s_{\min}}^{s_{\max}}\frac{ds}{s^2\sqrt{C_t}}\int_{t_-}^{t_+}dt\frac{t(t+s)}{\sqrt{2A_tt-B_t-t^2}}\nonumber\\
&=\int_{s_{\min}}^{s_{\max}}\frac{ds}{s^2\sqrt{C_t}}\int_{-\sqrt{K_t}}^{\sqrt{K_t}}dy\frac{(y+A_t)(y+A_t+s)}{\sqrt{K_t-y^2}}\nonumber\\
&=\pi\int_{s_{\min}}^{s_{\max}}\frac{ds}{s^2\sqrt{C_t}}\left[\frac{K_t}{2}+A_t(A_t+s)\right]\nonumber\\
&=\pi\int_{s_{\min}}^{s_{\max}}ds\Bigg[\frac{(4a_sa_t+4a_sc_u-3c_u^2)}{8\left(a_s-s\right)^\frac{5}{2}}\nonumber\\
&\qquad\qquad\qquad+\frac{2(c_u-2a_s-2a_t)s-s^2}{8\left(a_s-s\right)^\frac{5}{2}}\Bigg]\nonumber\\
&=\pi\left[\frac{s^2+2(2a_t-c_u)s+c_u^2-4a_sa_t}{4\left(a_s-s\right)^\frac{3}{2}}\right]_0^{4\lfloor s\rfloor}\nonumber\\
&=-4\pi\frac{E_aE_bE_1E_2}{(E_a+E_2)^3},
\end{align}
where $\lfloor s\rfloor\equiv\min(E_aE_b,E_1E_2)$. For the $t$-channel one obtains
\begin{align}
\mathcal G\left[-\frac{su}{t^2}\right]&=\int_{t_{\min}}^{t_{\max}}\frac{dt}{t^2\sqrt{C_s}}\int_{s_-}^{s_+}ds\frac{s(s+t)}{\sqrt{2A_ss-B_s-s^2}},\nonumber\\
&=\pi\left[\frac{t^2+2(2a_s-c_u)t+c_u^2-4a_sa_t}{4(a_t-t)^\frac{3}{2}}\right]_{-4\lfloor t\rfloor}^{t_{\max}},
\end{align}
where $\lfloor t\rfloor\equiv {\rm min}(E_aE_1,E_bE_2)$. To avoid the $t\to0$ divergence, $t_{\rm max}=-m_D^2<0$ is set in the denominator. Finally,
\begin{align}
\begin{split}
\mathcal G\left[-\frac{su}{t^2}\right]&=4\pi\frac{E_aE_bE_1E_2}{\left[\left(E_b-E_2\right)^2+m_D^2\right]^\frac{3}{2}},
\end{split}
\end{align}
Similarly, after performing the relevant integrals one obtains
\begin{align}
\begin{split}
\mathcal G\left[-\frac{st}{u^2}\right]&=4\pi\frac{E_aE_bE_1E_2}{\left[\left(E_a-E_2\right)^2+m_D^2\right]^\frac{3}{2}},
\end{split}
\end{align}
which can also be obtained using crossing symmetry. Combining all four channels in $gg\to gg$ gives
\begin{align}
\begin{split}
\frac{d^3R}{d^3p_a}&=\frac{9\alpha_S^2}{4\pi^4E_a^2}\int_0^\infty\frac{dE_b}{e^{\beta E_b}-1}\int_0^{E_a+E_b}\frac{dE_2}{1-e^{-\beta E_2}}\\
&\left[3\min(E_a,E_b,E_1,E_2)-\frac{E_aE_bE_1E_2}{\left(E_a+E_2\right)^3}\right.\\
&\left.+\frac{E_aE_bE_1E_2}{\left[\left(E_b-E_2\right)^2+m_D^2\right]^\frac{3}{2}}+\frac{E_aE_bE_1E_2}{\left[\left(E_a-E_2\right)^2+m_D^2\right]^\frac{3}{2}}\right],
\end{split}
\end{align}
as claimed in Eq.~(\ref{eq:rate_I_IV}).
\section{About the convergence of the series expansions}
\label{apnd:zeta_series}
The (non-exponentially suppressed) contribution to the $s$-channel of the $gg\to gg$ scattering rate is expanded as a binomial series in Eq.~(\ref{eq:binomial_zeta_series}). There are some subtleties about its convergence that should be remarked. Starting from a general term on the left hand side of Eq.~(\ref{eq:binomial_zeta_series}), we first split the integral into two regions, and rewrite the integrand as
\begin{align}
\int_0^\infty\frac{dx_b}{e^{x_b}-1}\frac{x_b}{\left(x_a+x_b\right)^n}&=\int_0^{x_a}\frac{dx_b}{e^{x_b}-1}\frac{x_b}{x^n_a\left(1+\frac{x_b}{x_a}\right)^n}\nonumber\\
&+\int_{x_a}^\infty\frac{dx_b}{e^{x_b}-1}\frac{x_b}{x_b^n\left(\frac{x_a}{x_b}+1\right)^n}.
\end{align}
In this form, the usual binomial expansion
\begin{align}
    \frac{1}{(1+x)^n}=\sum_{k=0}^\infty\frac{(-1)^k(n+k-1)!}{k!(n-1)!}x^k,
\end{align}
which converges for $x<1$, allows to write
\begin{align}
\begin{split}
    &\int_0^\infty\frac{dx_b}{e^{x_b}-1}\frac{x_b}{(x_a+x_b)^n}\\
    &=\int_0^{x_a}\frac{dx_b}{e^{x_b}-1}\sum_{k=0}^\infty\frac{(-1)^k(n+k-1)!}{k!(n-1)!}\frac{x_b^{k+1}}{x_a^{n+k}}\\
    &+\int_{x_a}^\infty\frac{dx_b}{e^{x_b}-1}\sum_{k=0}^\infty\frac{(-1)^k(n+k-1)!}{k!(n-1)!}\frac{x_a^k}{x_b^{n+k-1}}.
\label{eq:split_int_binom}
\end{split}
\end{align}
Owing to negative integer powers in $x_b$, the second improper integral on the right hand side does not admit a known closed-form expression. For sufficiently large $x_a$ though, its contributions will be negligible compared to that of the first one. This follows from the fact that for $x_b>x_a\gg1$, the distribution function will behave as $e^{-x_b}$, thereby suppressing all contributions to the full integral from that region exponentially, while avoiding the simple pole $(e^{x_b}-1)^{-1}\sim x_b^{-1}\,\, \forall\,\, 0\simeq x_b\ll x_a$, where a large fraction of the area under the curve lies. The second integral on the right hand side of Eq.~(\ref{eq:split_int_binom}) therefore contributes a small fraction to the overall result, and one can approximate the left hand side using the limit $x_a\to\infty$ on the first term of the right hand side, thereby giving
\begin{align}
\begin{split}
    &\int_0^\infty\frac{dx_b}{e^{x_b}-1}\frac{x_b}{\left(x_a+x_b\right)^n}\\
    &\simeq\int_0^\infty\frac{dx_b}{e^{x_b}-1}\sum_{k=0}^\infty\frac{(-1)^k(n+k-1)!}{k!(n-1)!}\frac{x_b^{k+1}}{x_a^{n+k}}\\
    &=\sum_{k=0}^\infty(-1)^k\frac{(n+k-1)!(k+1)}{(n-1)!x_a^{n+k}}\zeta(k+2),
\end{split}
\end{align}
which, of course, amounts to naively expanding the binomial in $x_b/x_a$ across the whole integration range. This argument is therefore purely a numerical one, since formally this series does not converge when summing up infinitely many terms. We can confirm this via d'Alembert's ratio test
\begin{align}
    \left|\frac{a_{k+1}}{a_k}\right|=\frac{(k+2)\zeta(k+3)}{(k+1)\zeta(k+2)}\frac{(n+k)}{x_a}
\end{align}
which, for any finite $x_a$, diverges to $\infty$ for $k\to\infty$. But up to the point where $n+k\sim x_a$, the ratio is less than one. That means that at least up to that term, the partial sums of our series will get closer to the true value. Thus the $\zeta$-series is an asymptotic series and is quite useful when $x_a$ is sufficiently large, since as we saw $\sim 10$ terms are more than enough to give values within the precision required for numerical purposes. Indeed, for large enough $x_a$, the finite $\zeta$-series approximates the integral on the left hand side of Eq.~(\ref{eq:split_int_binom}) better compared to the Ei-series in Eq.~(\ref{eq:Ei-expansion}). Of course for small $x_a$ this approximation breaks down quickly, since we don't have enough terms to get anything useful until the crossover point of $k\sim x_a$ is reached, and we can only resort to the slower Ei-series.

While not being as quickly convergent as the $\zeta$-series (for sufficiently large $x_a$), the Ei-series is guaranteed to converge for any $x_a$, as can be shown by the integral test. To do so we have to integrate over the coefficients $a_k$ as follows
\begin{align}
\begin{split}
    &\int_1^\infty dk \;a_k=\int_1^\infty dk\int_0^\infty dx_b\frac{x_be^{-kx_b}}{(x_a+x_b)^n}\\
    &=\int_0^\infty dx_b\frac{x_b}{(x_a+x_b)^n}\int_1^\infty dke^{-kx_b},
\end{split}
\end{align}
where, interchanging the two integrals leads us to a simple result of
\begin{align}
    \int_1^\infty dk\;a_k&=\int_0^\infty \frac{dx_b}{(x_a+x_b)^n}e^{-x_b}\\
    &=e^{x_a}\int_{x_a}^\infty dt\frac{e^{-t}}{t^n}=e^{x_a}\Gamma(1-n,x_a)<\infty,\nonumber
\end{align}
where $t=x_a+x_b$ was substituted. Since we get a finite result, the Ei-series is actually convergent (albeit slower than the asymptotic $\zeta$-series).

Similarly we can show the convergence of the Struve expansions via the integral test. For the thermodynamic integral containing the square root, from Eq.~(\ref{eq:Struve-Expansion}) we had an expansion of the form
\begin{align}
    A^-_1(z)&=\sum^{\infty}_{k=1}\int_0^\infty dx \,x \sqrt{x^2+z^2}\,e^{-kx},
\end{align}
of which we just integrate the summand as
\begin{align}
\begin{split}
    \int_1^\infty dk\int_0^\infty dx \,x \sqrt{x^2+z^2}\,e^{-kx}&=\int_0^\infty dx\sqrt{x^2+z^2}\,e^{-x}\\
    &=\frac{z\pi}{2}K_1(z)<\infty.
\end{split}
\end{align}
Similarly for the arcsinh series
\begin{align}
    B^-_1(z)&=\sum^{\infty}_{k=1}\int_0^\infty dx\, x\arcsinh\left(\frac{x}{z}\right)e^{-kx},
\end{align}
we get
\begin{align}
\begin{split}
    &\int_1^\infty dk\int_0^\infty dx \,x \arcsinh\left(\frac{x}{z}\right)\,e^{-kx}\\
    &=\int_0^\infty dx\arcsinh\left(\frac{x}{z}\right)\,e^{-x}\\
    &=\int_0^\infty dy\frac{e^{-yz}}{\sqrt{1+y^2}}=\frac{\pi}{2}K_0(z)<\infty,
\end{split}
\end{align}
and thus the Struve-function series converge.

\section{Series approximation for the polylogarithmic terms for the $s$-channel contribution}
\label{apnd:multi_Li_series}
In the $s$-term in Eq.~(\ref{eq:rate_s}) we encounter polylogarithmic integrals of the general form
\begin{align}
    \int_0^\infty dx_b\frac{x_b^m\polylog{s}\left(e^{-x_a-x_b}\right)}{\left(x_a+x_b\right)^n\left(e^{x_b}-1\right)}.
\end{align}
Setting up a binomial series again using
\begin{align}
    \frac{1}{\left(x_a+x_b\right)^n}=\sum_{k=0}^\infty\frac{\left(-1\right)^k\left(n+k-1\right)!}{k!\left(n-1\right)!}\frac{x_b^k}{x_a^{n+k}},
\end{align}
we can write
\begin{align}
\begin{split}
    &\int_0^\infty dx_b\frac{x_b^m\polylog{s}\left(e^{-x_a-x_b}\right)}{\left(x_a+x_b\right)^n\left(e^{x_b}-1\right)}\\
    &=\sum_{k=0}^\infty\frac{\left(-1\right)^k\left(n+k-1\right)!}{k!\left(n-1\right)!x_a^{n+k}}\int_0^\infty dx_b\frac{x_b^{m+k}\polylog{s}\left(e^{-x_a-x_b}\right)}{e^{x_b}-1}.
\end{split}
\end{align}
Now each term of the series is integrable using the generalized version of the polylogarithm, the multiple polylogarithm defined as the nested series~\cite{Vollinga_2005}
\begin{align}
\begin{split}
    &\polylog{s_1,\dots,s_n}(z_1,\dots,z_n)\equiv\sum_{k_1>\dots>k_n>0}\frac{z^{k_1}\dots z^{k_n}}{k_1^{s_1}\dots k_n^{s_n}}\\
    &=\sum_{k_1=1}^\infty\dots\sum_{k_n=1}^\infty\frac{z_1^{k_1+\dots+k_n}z_2^{k_2+\dots+k_n}\dots z_n^{k_n}}{\left(k_1+\dots+k_n\right)^{s_1}\left(k_2+\dots+k_n\right)^{s_2}\dots k_n^{s_n}}.
\end{split}
\end{align}
Expanding the distribution function as well as the polylogarithm gives
\begin{align}
\begin{split}
    &\int_0^\infty dx_b\frac{x_b^{m+k}\polylog{s}\left(e^{-x_a-x_b}\right)}{e^{x_b}-1}\\
    &=\sum_{j_1=1}^\infty\sum_{j_2=1}^\infty e^{-j_2x_a}\int_0^\infty dx_b\frac{x_b^{m+k}e^{-{(j_1+j_2)x_b}}}{j_2^s}\\
    &=\sum_{j_1=1}^\infty\sum_{j_2=1}^\infty\frac{\left(m+k\right)!e^{-j_2x_a}}{\left(j_1+j_2\right)^{m+k+1}j_2^s}\\
    &=(m+k)!\polylog{m+k+1,s}(1,e^{-x_a}),
\end{split}
\end{align}
where we used the definition of the $\Gamma$-function from Eq.~(\ref{eq:PolyLog_Gamma}) to evaluate the integral. Then, after resumming our binomial series, the final result is
\begin{align}
    &\int_0^\infty dx_b\frac{x_b^m\polylog{s}\left(e^{-x_a-x_b}\right)}{\left(x_a+x_b\right)^n\left(e^{x_b}-1\right)}\label{eq:Multi_Li_Series}\\
    &=\sum_{k=0}^\infty\frac{(-1)^k\left(n+k-1\right)!\left(m+k\right)!}{k!\left(n-1\right)!x_a^{n+k}}\polylog{m+k+1,s}\left(1,e^{-x_a}\right).\nonumber
\end{align}
It should be noted that the series constructed in Eq.~(\ref{eq:Multi_Li_Series}) is an asymptotic one, like the $\zeta$-series in Eq.~(\ref{eq:binomial_zeta_series}), but gives numerically reasonable results for sufficiently large $x_a$, using the first few terms in the series.

\begin{widetext}
With that, we have everything we need to expand the $s$-terms of all relevant quantities. Evaluating the $s$-term of the rate in this manner, yields
\begin{align}
\begin{split}
    \frac{d^3R_s}{d^3p_a}&=-\frac{9\alpha_s^2T}{4\pi^4x_a}\left\{\frac{\pi^2}{36}-\frac{\pi^2}{6x_a}\sum_{n=1}^\infty(-1)^n\frac{n!\,n\,\zeta(n+1)}{x_a^n}+\frac{\zeta(3)}{x_a^2}\sum_{n=1}^\infty(-1)^n\frac{(n+1)!\,n\,\zeta(n+1)}{x_a^n}\right.\\
    &\left.\qquad\qquad\qquad-\frac{1}{x_a}\sum_{n=1}^\infty(-1)^n\frac{n!\,n}{x_a^n}\polylog{n+1,2}\left(1,e^{-x_a}\right)-\frac{1}{x_a^2}\sum_{n=1}^\infty(-1)^n\frac{(n+1)!\,n}{x_a^n}\polylog{n+1,3}\left(1,e^{-x_a}\right)\right\}
\end{split}
\end{align}
The equivalent result for the transport coefficients is given by
\begin{align}
\begin{split}
    \hat{q}_s&=\frac{3\alpha_s^2T^3}{2\pi^4x_a^4}\left\{4\pi^2\polylog{5}\left(e^{-x_a}\right)+\pi^2x_a\polylog{4}\left(e^{-x_a}\right)-\frac{x_a^4\zeta(3)}{5}-\frac{\pi^4x_a^3}{300}-12\sum_{n=0}^\infty(-1)^n\frac{(n+1)(n+2)!}{x_a^n}\polylog{n+2,5}\left(1,e^{-x_a}\right)\right.\\
    &-6x_a\sum_{n=0}^\infty(-1)^n\frac{(n+1)(n+1)!}{x_a^n}\polylog{n+2,4}\left(1,e^{-x_a}\right)-6\zeta(3)x_a\sum_{n=0}^\infty(-1)^n\frac{(n+2)!}{x_a^n}\zeta(n+3)\\
    &+6\zeta(5)\sum_{n=1}^\infty(-1)^n\frac{n(n+1)(n+1)!}{x_a^n}\left[6\zeta(n+2)+\frac{6(n+2)\zeta(n+3)}{x_a}+\frac{(n+2)(n+3)\zeta(n+4)}{x_a^2}\right]\\
    &\left.-\frac{\pi^4}{10}\sum_{n=1}^\infty(-1)^n\frac{n(n+1)(n+1)!}{x_a^n}\left[2x_a\zeta(n+2)+3(n+2)\zeta(n+3)+\frac{(n+2)(n+3)\zeta(n+4)}{x_a}\right]\right\}
\end{split}
\end{align}
for $\langle q_\perp^2\rangle/L$ and
\begin{align}
\begin{split}
    \hat{e}_s&=\frac{9\alpha_s^2T^2}{4\pi^4x_a^2}\left\{\frac{x_a\zeta(3)}{6}-\frac{\pi^2x_a^2}{72}+3\sum_{n=1}^\infty(-1)^n\frac{\left(n+1\right)!\,n}{x_a^{n+1}}\polylog{n+1,4}\left(1,e^{-x_a}\right)+2\sum_{n=1}^\infty(-1)^n\frac{\left(n+1\right)!\,n}{x_a^n}\polylog{n+1,3}\left(1,e^{-x_a}\right)\right.\\
    &+\sum_{n=1}^\infty(-1)^n\frac{\left(n+1\right)!\,n(n+1)}{x_a^{n+1}}\polylog{n+2,3}\left(1,e^{-x_a}\right)+\frac{x_a}{2}\sum_{n=1}^\infty(-1)^n\frac{\left(n+1\right)!\,n}{x_a^n}\polylog{n+1,2}\left(1,e^{-x_a}\right)\\
    &+\sum_{n=1}^\infty(-1)^n\frac{\left(n+1\right)!\,n\left(n+1\right)}{2x_a^n}\polylog{n+2,2}\left(1,e^{-x_a}\right)-\frac{\pi^4}{30}\sum_{n=1}^\infty(-1)^n\frac{(n+1)!\,n}{x_a^{n+1}}\zeta(n+1)+\zeta(3)\sum_{n=1}^\infty(-1)^n\frac{(n+1)!\,n}{x_a^n}\zeta(n+1)\\
    &\left.+2\zeta(3)\sum_{n=1}^\infty(-1)^n\frac{(n+1)!\,n(n+1)}{x_a^{n+1}}\zeta(n+2)+\frac{\pi^2}{6}\sum_{n=1}^\infty(-1)^n\frac{n!\,n(n+1)}{x_a^n}\zeta(n+2)\right\},
\end{split}
\end{align}
\end{widetext}
for $\langle \omega \rangle/L$. These multiple polylogarithms were also used to integrate all terms of the four-point vertex for $\hat{q}$ and $\hat{e}$ in Eqs.~(\ref{eq:qhat_c}, \ref{eq:ehat_c}) analytically. To do so, one has to group terms which exhibit a pole at $x_b=0$ with their respective bare pole terms, to generate expressions of the form
\begin{align}
    \int_0^\infty dx_b\frac{\polylog{s}\left(e^{-x_a}\right)-\polylog{s}\left(e^{-x_a-x_b}\right)}{e^{x_b}-1}
\end{align}
which are finite. Expanding the polylogarithm as a series gives
\begin{align}
\begin{split}
    &\int_0^\infty dx_b\frac{\polylog{s}\left(e^{-x_a}\right)-\polylog{s}\left(e^{-x_a-x_b}\right)}{e^{x_b}-1}\\
    &=\sum_{j=1}^\infty\int_0^\infty dx_b\frac{e^{-jx_a}(1-e^{-jx_b})}{j^s(e^{x_b}-1)}.
\end{split}
\end{align}
Recognizing the exponential terms as a partial sum of their geometric series $\frac{1-e^{-jx_b}}{e^{x_b}-1}=\sum_{k=1}^je^{-jx_b}$ and integrating over the remaining exponential gives
\begin{align}
\begin{split}
    &\int_0^\infty dx_b\frac{\polylog{s}\left(e^{-x_a}\right)-\polylog{s}\left(e^{-x_a-x_b}\right)}{e^{x_b}-1}\\
    &=\sum_{j=1}^\infty\sum_{k=1}^j\int_0^\infty dx_b\frac{e^{-jx_a}e^{-kx_b}}{j^s}=\sum_{j=1}^\infty\sum_{k=1}^j\frac{e^{-jx_a}}{kj^s}\\
    &=\sum_{k=1}^\infty\sum_{j=k}^\infty\frac{e^{-jx_a}}{kj^s}=\sum_{j=0}^\infty\sum_{k=1}^\infty\frac{e^{-(j+k)x_a}}{k\left(j+k\right)^s},
\end{split}
\end{align}
where we also changed the order of summation and relabeled the $j$-index, so that it starts from 0. Then all that is left to do is to separate out the $j=0$ term, so that the double series matches the definition of the multiple polylogarithm, to get our result
\begin{align}
\begin{split}
    &\int_0^\infty dx_b\frac{\polylog{s}\left(e^{-x_a}\right)-\polylog{s}\left(e^{-x_a-x_b}\right)}{e^{x_b}-1}\\
    &=\sum_{k=1}^\infty\frac{e^{-kx_a}}{k^{s+1}}+\sum_{j=1}^\infty\sum_{k=1}^\infty\frac{e^{-(j+k)x_a}}{k\left(j+k\right)^s}\\
    &=\polylog{s+1}\left(e^{-x_a}\right)+\polylog{s,1}\left(e^{-x_a},1\right).
\end{split}
\end{align}

\section{Thermodynamic integrals}
The most common integrals involving thermodynamics distributions that do not have a well-known closed form are defined in terms of functions below
\begin{align}
\alpha^{-}_n\left(c,\epsilon;x_a,z\right)&\equiv\int^c_\epsilon dx_b\frac{x^n_b\sqrt{\left(x_a-x_b\right)^2+z^2}}{e^{x_b}-1}\quad\forall n\in\mathbb{N}_0,\nonumber\\
A^{-}_n\left(\epsilon;z\right)&=\alpha^-_n(\infty,\epsilon;0,z)\equiv\lim_{c\to\infty}[\alpha^{-}_n(c,\epsilon;0,z)]\nonumber\\
A^{-}_n\left(z\right)&=\lim_{\epsilon\to0^+}\left[A^-_n(\epsilon;z)\right]\nonumber\\
\beta^{-}_n(c,\epsilon;x_a,z)&\equiv\int^c_\epsilon dx_b\frac{x^n_b\,\,{\rm arcsinh}\left(\frac{x_a-x_b}{z}\right)}{e^{x_b}-1}\quad\forall n\in\mathbb{N}_0,\nonumber\\
-B^{-}_n(\epsilon;z)&=\beta^{-}_n(\infty,\epsilon;0,z)\equiv\lim_{c\to\infty}[\beta^-_n(c,\epsilon;0,z)],\nonumber\\
-B^{-}_n(z)&=\lim_{\epsilon\to0^+}[\beta^{-}_n(\infty,\epsilon;0,z)],\nonumber\\
\Delta\alpha^-_n(\epsilon)&=2\alpha^-_n\left(x_a,\epsilon;x_a,z\right)-\alpha^-_n\left(\infty,\epsilon;x_a,z\right),\nonumber\\
\Delta\alpha^-_n&=\lim_{\epsilon\to 0^+}\left[\Delta\alpha^-_n\left(\epsilon\right)\right],\nonumber\\
\Delta\beta^-_n(\epsilon)&=2\beta^-_n\left(x_a,\epsilon;x_a,z\right)-\beta^-_n\left(\infty,\epsilon;x_a,z\right)\nonumber\\
\Delta\beta^-_n&=\lim_{\epsilon\to 0^+}\left[\Delta\beta^-_n\left(\epsilon\right)\right]
\label{eq:thermo_int}
\end{align}

\section{Performing $\hat q$ integrals}
\label{sec:q_hat_ints}
Performing the integral in Eq.~(\ref{eq:qhat_t_BE}) is non trivial even in the limit where Bose-enhancement is neglected. Starting in that limit allows to perform the inner integral analytically, yielding
\begin{align}
\hat{q}_t&\simeq\frac{3\alpha_s^2}{2\pi^4E_a^4}\int_0^\infty\frac{dE_b}{e^{\beta E_b}-1}\int_0^{E_a+E_b}dE_2\Bigg\{\frac{3E_aE_1\lfloor s\rfloor^2}{\sqrt{\left(E_b-E_2\right)^2+m_D^2}}\nonumber\\
&+\lfloor E\rfloor\Bigg[2\lfloor s\rfloor\left(\lfloor t\rfloor-\lfloor s\rfloor-\lfloor E\rfloor^2\right)+\lfloor E\rfloor^2\left(\lfloor t\rfloor-E_aE_1\right)\nonumber\\
&\qquad\qquad-\frac{3\lfloor E\rfloor^4}{5}-3E_aE_1\lfloor s\rfloor\Bigg]\Bigg\}.
\label{eq:q_t_no_BE}
\end{align}
The second term in Eq.~(\ref{eq:q_t_no_BE}) is straightforward to evaluate, since it doesn't depend on the momentum cut-off, and the integrand solely consists of polynomials in $E_2$ and $E_b$, thus one obtains
\begin{align}
    \hat{q}_{t,2}&\equiv\frac{3\alpha_s^2}{2\pi^4E_a^4}\int_0^\infty\frac{dE_b}{e^{\beta E_b}-1}\int_0^{E_a+E_b}dE_2\, \lfloor E\rfloor\nonumber\\
    &\times\Bigg[2\lfloor s\rfloor\left(\lfloor t\rfloor-\lfloor s\rfloor-\lfloor E\rfloor^2\right)+\lfloor E\rfloor^2\left(\lfloor t\rfloor-E_aE_1\right)\nonumber\\
    &\qquad-\frac{3\lfloor E\rfloor^4}{5}-3E_aE_1\lfloor s\rfloor\Bigg]\nonumber\\
    &=-\frac{9\alpha_s^2T^3}{2\pi^4}\left[\zeta(3)+\frac{\pi^4}{18x_a}\right].
\end{align}

The first, square-root-dependent, term in Eq.~(\ref{eq:q_t_no_BE}) is more involved, as following the integral over $E_2$ integral, pairs of poles will appear that mutually cancel, that is to say that each divergence has a counter-term. To ensure that this cancellation is maintained throughout the derivation, so that the answer can be expressed using Eq.~(\ref{eq:thermo_int}), we proceed as follows
\begin{widetext}
\begin{align}
\begin{split}
    \hat{q}_{t,1}&\equiv\frac{3\alpha_s^2}{2\pi^4E_a^4}\int_0^\infty\frac{dE_b}{e^{\beta E_b}-1}\int_0^{E_a+E_b}dE_2\frac{3E_aE_1\lfloor s\rfloor^2}{\sqrt{\left(E_b-E_2\right)^2+m_D^2}}\\
    &=\frac{9\alpha_s^2T^3}{2\pi^4}\int_0^\infty\frac{dx_b}{e^{x_b}-1}\left\{\sqrt{x_a^2+z^2}\left(\frac{x_a}{20}+\frac{x_b}{2}-\frac{11x_b^2}{6x_a}+\frac{137z^2}{120x_a}-\frac{13z^2x_b}{4x_a^2}+\frac{2z^2x_b^2}{3x_a^3}-\frac{8z^4}{15x_a^3}\right)\right.\\
    &+\left(x_b^2-\frac{z^2}{2}+\frac{3z^2x_b}{x_a}-\frac{3z^2x_b^2}{2x_a^2}+\frac{9z^4}{8x_a^2}-\frac{3z^4x_b}{4x_a^3}\right)\left[\arcsinh\left(\frac{x_a}{z}\right)+\arcsinh\left(\frac{x_b}{z}\right)\right]\\
    &-\sqrt{x_b^2+z^2}\left(\frac{3x_b}{2}-\frac{x_b^2}{x_a}+\frac{2z^2}{x_a}-\frac{x_b^3}{4x_a^2}-\frac{23z^2x_b}{8x_a^2}-\frac{x_b^4}{30x_a^3}+\frac{11z^2x_b^2}{60x_a^3}-\frac{8z^4}{15x_a^3}\right)\\
    &+\text{sgn}(x_a-x_b)\left[z\left(2x_b-\frac{2x_b^2}{x_a}+\frac{2z^2}{x_a}-\frac{4z^2x_b}{x_a^2}+\frac{2z^2x_b^2}{3x_a^3}-\frac{8z^4}{15x_a^3}\right)\right.\\
    &+z^2\left(\frac{1}{2}-\frac{3x_b}{x_a}+\frac{3x_b^2}{2x_a^2}-\frac{9z^2}{8x_a^2}+\frac{3z^2x_b}{4x_a^3}\right)\arcsinh\left(\frac{x_a-x_b}{z}\right)-\sqrt{\left(x_a-x_b\right)^2+z^2}\\
    &\left.\left.\times\left(\frac{x_a}{20}+\frac{11x_b}{20}-\frac{17x_b^2}{60x_a}+\frac{137z^2}{120x_a}-\frac{17x_b^3}{60x_a^2}-\frac{319z^2x_b}{120x_a^2}-\frac{x_b^4}{30x_a^3}+\frac{11z^2x_b^2}{60x_a^3}-\frac{8z^4}{15x_a^3}\right)\right]\right\}.
\end{split}
\end{align}
Naively some of the terms appear to be divergent when evaluating the remaining integral; in particular, all terms without any powers of $x_b$ in the numerator to suppress the Bose-Einstein pole at $x_b=0$. However, careful analysis determines that the expression is in fact finite, as all diverging terms have counter-terms. To see this, we start by separating the expression into two parts: one solely containing well-behaved terms, and one consisting of just the divergent terms and counter-terms. Explicitly, this gives
\begin{align}
\hat{q}_{t,1}=\mathfrak{q}_1+\mathfrak{q}_2,
\end{align}
where
\begin{align}
\mathfrak{q}_1\equiv&\frac{9\alpha_s^2T^3}{2\pi^4}\int_0^\infty\frac{dx_b}{e^{x_b}-1}\left\{x_b\sqrt{x_a^2+z^2}\left(\frac{1}{2}-\frac{11x_b}{6x_a}-\frac{13z^2}{4x_a^2}+\frac{2z^2x_b}{3x_a^3}\right)+x_b\left(x_b+\frac{3z^2}{x_a}-\frac{3z^2x_b}{2x_a^2}-\frac{3z^4}{4x_a^3}\right)\arcsinh\left(\frac{x_a}{z}\right)\right.\nonumber\\
&+\left(x_b^2-\frac{z^2}{2}+\frac{3z^2x_b}{x_a}-\frac{3z^2x_b^2}{2x_a^2}+\frac{9z^4}{8x_a^2}-\frac{3z^4x_b}{4x_a^3}\right)\arcsinh\left(\frac{x_b}{z}\right)-x_b\sqrt{x_b^2+z^2}\left(\frac{3}{2}-\frac{x_b}{x_a}-\frac{x_b^2}{4x_a^2}-\frac{23z^2}{8x_a^2}-\frac{x_b^3}{30x_a^3}+\frac{11z^2x_b}{60x_a^3}\right)\nonumber\\
&+\text{sgn}(x_a-x_b)\left[zx_b\left(2-\frac{2x_b}{x_a}-\frac{4z^2}{x_a^2}+\frac{2z^2x_b}{3x_a^3}\right)-z^2x_b\left(\frac{3}{x_a}-\frac{3x_b}{2x_a^2}-\frac{3z^2}{4x_a^3}\right)\arcsinh\left(\frac{x_a-x_b}{z}\right)\right.\nonumber\\
&\left.\left.\qquad\qquad\qquad\quad-x_b\sqrt{\left(x_a-x_b\right)^2+z^2}\left(\frac{11}{20}-\frac{17x_b}{60x_a}-\frac{17x_b^2}{60x_a^2}-\frac{319z^2}{120x_a^2}-\frac{x_b^3}{30x_a^3}+\frac{11z^2x_b}{60x_a^3}\right)\right]\right\}
\end{align}
contains no poles at $x_b=0$. For the remaining terms, containing these poles, we start by splitting the integral at $x_b=x_a$. Then we can explicitly evaluate ${\rm sgn}(x_a-x_b)$, which yields
\begin{align}
    \mathfrak{q}_2&=\frac{9\alpha_s^2T^3}{2\pi^4}\int_0^{x_a}\frac{dx_b}{e^{x_b}-1}\left\{\left(\frac{2z^2}{x_a}-\frac{8z^4}{15x_a^3}\right)\left(z-\sqrt{x_b^2+z^2}\right)+\left(\frac{9z^4}{8x_a^2}-\frac{z^2}{2}\right)\left[\arcsinh\left(\frac{x_a}{z}\right)-\arcsinh\left(\frac{x_a-x_b}{z}\right)\right]\right.\nonumber\\
    &\left.\qquad\qquad\qquad\qquad\qquad+\left(\frac{x_a}{20}+\frac{137z^2}{120x_a}-\frac{8z^4}{15x_a^3}\right)\left(\sqrt{x_a^2+z^2}-\sqrt{\left(x_a-x_b\right)^2+z^2}\right)\right\}\nonumber\\
    &+\frac{9\alpha_s^2T^3}{2\pi^4}\int_{x_a}^\infty\frac{dx_b}{e^{x_b}-1}\left\{\left(\frac{8z^4}{15x_a^3}-\frac{2z^2}{x_a}\right)\left(z+\sqrt{x_b^2+z^2}\right)+\left(\frac{9z^4}{8x_a^2}-\frac{z^2}{2}\right)\left[\arcsinh\left(\frac{x_a}{z}\right)+\arcsinh\left(\frac{x_a-x_b}{z}\right)\right]\right.\nonumber\\
    &\left.\qquad\qquad\qquad\qquad\qquad+\left(\frac{x_a}{20}+\frac{137z^2}{120x_a}-\frac{8z^4}{15x_a^3}\right)\left(\sqrt{x_a^2+z^2}+\sqrt{\left(x_a-x_b\right)^2+z^2}\right)\right\}.
\end{align}
Grouping the terms together by their leading coefficients and focusing on the region $0\leq x_b< x_a$, one notices that each divergence is canceled by its counterpart. Specifically, expanding around $x_b=0$ yields
\begin{align}
\begin{split}
&\frac{\sqrt{x_b^2+z^2}-z}{e^{x_b}-1}\simeq\frac{x_b}{2z}+O(x^2_b),\\
&\frac{\arcsinh\left(\frac{x_a-x_b}{z}\right)-\arcsinh\left(\frac{x_a}{z}\right)}{e^{x_b}-1}\simeq-\frac{1}{\sqrt{x^2_a+z^2}}+O(x_b),\\
&\frac{\sqrt{\left(x_a-x_b\right)^2+z^2}-\sqrt{x_a^2+z^2}}{e^{x_b}-1}\simeq-\frac{x_b}{\sqrt{x^2_a+z^2}}+O(x^2_b),
\end{split}
\label{eq:divergent_cancel}
\end{align}
\end{widetext}
which are well-behaved as $x_b\to 0$. The region $x_a\leq x_b<\infty$ has no poles and thus generates a finite result. We write down this cancellation in terms of limits of our special functions defined in Eq.~(\ref{eq:thermo_int}). Starting with the simple square root, we have terms of the form
\begin{align}
\int_\epsilon^{x_a}\frac{dx_b}{e^{x_b}-1}\left(\sqrt{x_b^2+z^2}-z\right)\nonumber\\
+\int_{x_a}^\infty\frac{dx_b}{e^{x_b}-1}\left(\sqrt{x_b^2+z^2}+z\right)
\label{eq:1st_pole}
\end{align}
with the implication that $\epsilon\to0^+$. Here we recombine the square root integrals as
\begin{align}
\int_\epsilon^{\infty}\frac{dx_b}{e^{x_b}-1}\sqrt{x_b^2+z^2}=A_0^-(\epsilon,z).
\label{eq:1st_pole_fin}
\end{align}
As the second term does change sign, we evaluate both pieces separately, giving
\begin{align}
\int_{x_a}^\infty\frac{dx_b}{e^{x_b}-1}-\int_\epsilon^{x_a}\frac{dx_b}{e^{x_b}-1}&=\left[\ln\left(1-e^{-x_b}\right)\right]_{x_a}^\infty\nonumber\\
    &-\left[\ln\left(1-e^{-x_b}\right)\right]_\epsilon^{x_a}\nonumber\\
    &=\ln\left(1-e^{-\epsilon}\right)\nonumber\\
    &-2\ln\left(1-e^{-x_a}\right)\nonumber\\
    &\simeq \ln(\epsilon)-2\ln\left(1-e^{-x_a}\right)\nonumber\\
    &+O(\epsilon).
\label{eq:1st_pole_div}
\end{align}
Combining Eq.~(\ref{eq:1st_pole_fin}) and Eq.~(\ref{eq:1st_pole_div}) in Eq.~(\ref{eq:1st_pole}), and taking the limit $\epsilon\to0^+$, we get the finite  expression below
\begin{align}
&\int_\epsilon^{x_a}\frac{dx_b}{e^{x_b}-1}\left(\sqrt{x_b^2+z^2}-z\right)\nonumber\\
&+\int_{x_a}^\infty\frac{dx_b}{e^{x_b}-1}\left(\sqrt{x_b^2+z^2}+z\right)\nonumber\\
&=\lim_{\epsilon\to0^+}\left[z\ln\left(\epsilon\right)+A^-_0\left(\epsilon;z\right)\right]-2z\ln\left(1-e^{-x_a}\right).
\end{align}
The generalized version of the above expression encountered in our calculation is the integral of the form
\begin{align}
\int_\epsilon^{x_a}\frac{dx_b}{e^{x_b}-1}\left(\sqrt{\left(x_a-x_b\right)^2+z^2}-\sqrt{x_a^2+z^2}\right)\nonumber\\
+\int_{x_a}^\infty\frac{dx_b}{e^{x_b}-1}\left(\sqrt{\left(x_a-x_b\right)^2+z^2}+\sqrt{x_a^2+z^2}\right).
\end{align}
In this case the simple pole doesn't change sign, so we just recombine both parts to get
\begin{align}
\int_\epsilon^{\infty}\frac{dx_b}{e^{x_b}-1}&=\left[\ln\left(1-e^{-x_b}\right)\right]_\epsilon^\infty\nonumber\\
&=-\ln\left(1-e^{-\epsilon}\right)\nonumber\\
&\simeq-\ln(\epsilon)+O(\epsilon).
\label{eq:bose_pole}
\end{align}
For the root, we start by first adding and subtracting the lower region to be able to express it solely in terms of functions defined in Eq.~(\ref{eq:thermo_int}), giving
\begin{align}
&\int_\epsilon^{x_a}\frac{dx_b}{e^{x_b}-1}\sqrt{\left(x_a-x_b\right)^2+z^2}\nonumber\\
&-\int_{x_a}^\infty\frac{dx_b}{e^{x_b}-1}\sqrt{\left(x_a-x_b\right)^2+z^2}\nonumber\\
&=2\int_\epsilon^{x_a}\frac{dx_b}{e^{x_b}-1}\sqrt{\left(x_a-x_b\right)^2+z^2}\nonumber\\
&\quad-\int_{\epsilon}^\infty\frac{dx_b}{e^{x_b}-1}\sqrt{\left(x_a-x_b\right)^2+z^2}\nonumber\\
&=2\alpha_0^-(x_a,\epsilon;x_a,z)-\alpha_0^-(\infty,\epsilon;x_a,z)\nonumber\\
&=\Delta\alpha_0^-(\epsilon,z).
\end{align}
Combining the two similar divergences (with different sign) again yields
\begin{align}
&\int_0^{x_a}\frac{dx_b}{e^{x_b}-1}\left(\sqrt{\left(x_a-x_b\right)^2+z^2}-\sqrt{x_a^2+z^2}\right)\nonumber\\
&+\int_{x_a}^\infty\frac{dx_b}{e^{x_b}-1}\left(\sqrt{\left(x_a-x_b\right)^2+z^2}+\sqrt{x_a^2+z^2}\right)\nonumber\\
&=\lim_{\epsilon\to0^+}\left[\Delta\alpha_0^-(\epsilon,z)+\sqrt{x_a^2+z^2}\ln(\epsilon)\right].
\end{align}
For the inverse hyperbolic sine terms, the same procedure is applied to terms of the form
\begin{align}
\int_\epsilon^{x_a}\frac{dx_b}{e^{x_b}-1}\left[\arcsinh\left(\frac{x_a-x_b}{z}\right)-\arcsinh\left(\frac{x_a}{z}\right)\right]\nonumber\\
+\int_{x_a}^\infty\frac{dx_b}{e^{x_b}-1}\left[\arcsinh\left(\frac{x_a-x_b}{z}\right)+\arcsinh\left(\frac{x_a}{z}\right)\right],
\end{align}
resulting in
\begin{align}
&\int_\epsilon^{x_a}\frac{dx_b}{e^{x_b}-1}\arcsinh\left(\frac{x_a-x_b}{z}\right)\nonumber\\
&-\int_{x_a}^\infty\frac{dx_b}{e^{x_b}-1}\arcsinh\left(\frac{x_a-x_b}{z}\right)\nonumber\\
&=2\int_\epsilon^{x_a}\frac{dx_b}{e^{x_b}-1}\arcsinh\left(\frac{x_a-x_b}{z}\right)\nonumber\\
&\quad-\int_{\epsilon}^\infty\frac{dx_b}{e^{x_b}-1}\arcsinh\left(\frac{x_a-x_b}{z}\right)\nonumber\\
&=2\beta_0^-(x_a,\epsilon;x_a,z)-\beta_0^-(\infty,\epsilon;x_a,z)\nonumber\\
&=\Delta\beta_0^-(\epsilon,z),
\end{align}
which, combined with the other pole given by Eq.~(\ref{eq:bose_pole}), ultimately yields
\begin{align}
&\int_0^{x_a}\frac{dx_b}{e^{x_b}-1}\left[\arcsinh\left(\frac{x_a-x_b}{z}\right)-\arcsinh\left(\frac{x_a}{z}\right)\right]\nonumber\\
&+\int_{x_a}^\infty\frac{dx_b}{e^{x_b}-1}\left[\arcsinh\left(\frac{x_a-x_b}{z}\right)+\arcsinh\left(\frac{x_a}{z}\right)\right]\nonumber\\
&=\lim_{\epsilon\to0^+}\left[\Delta\beta_0^-(\epsilon,z)+\arcsinh\left(\frac{x_a}{z}\right)\ln(\epsilon)\right].
\end{align}
\end{appendices}
\bibliography{references}
\end{document}